\documentclass[12pt, preprint]{aastex}
\usepackage{epsfig}
\usepackage{natbib}

\newcommand{\bm}[1]{\mbox{\boldmath{$#1$}}}
\newcommand{\Hm}{\rm{H}^{-}}
\newcommand{\Hp}{\rm{H}^{+}}
\newcommand{\me}{\rm{e}}
\newcommand{\mHtp}{\rm{H}_{2}^{+}}
\newcommand{\mH}{\rm{H}}
\newcommand{\mHt}{\rm{H}_{2}}
\newcommand{\He}{\rm{He}}
\newcommand{\xhteq}{x_{\rm H_{2}, eq}}
\newcommand{\Htvol}{\langle x_{\rm{H}_{2}}\rangle_{\rm V}}
\newcommand{\Htmass}{\langle x_{\rm{H}_{2}}\rangle_{\rm M}}

\newcommand{\cii}{\hbox{C\,{\sc ii}}\,}
\newcommand{\sii}{\hbox{Si\,{\sc ii}}\,}
\newcommand{\oi}{\hbox{O\,{\sc i}}\,} 

\newcommand{\pderiv}[2]{\frac{\partial #1}{\partial #2}}
\newcommand{\tderiv}[2]{\frac{{\rm D} #1}{{\rm D} #2}}
\def\simless{\mathbin{\lower 3pt\hbox
   {$\rlap{\raise 5pt\hbox{$\char'074$}}\mathchar"7218$}}}   
\def\simgreat{\mathbin{\lower 3pt\hbox  
   {$\rlap{\raise 5pt\hbox{$\char'076$}}\mathchar"7218$}}} 

\begin{document}

\title{Simulating the formation of molecular clouds. I. 
Slow formation by gravitational collapse from static initial
conditions}

\author{Simon~C.~O. Glover$^{1,2}$ \& Mordecai-Mark {Mac Low}$^1$}
\affil{$^1$Department of Astrophysics, American Museum of Natural History, \\
       Central Park West at 79th Street, New York, NY 10024}
\affil{$^2$Astrophysikalisches Institut Potsdam,\\An der Sternwarte 16, 14482
  Potsdam, Germany}
\email{sglover@aip.de, mordecai@amnh.org}

\begin{abstract}
We study the formation of $\mHt$ in the ISM, using a modified version of the 
astrophysical magnetohydrodynamical code ZEUS-MP that includes a 
non-equilibrium treatment of the formation and destruction of $\mHt$. We examine two 
different approximations to treat the shielding of $\mHt$ against photodissociation: 
a local approximation, which gives us a solid lower bound on the amount of 
shielding, and a method based on ray-tracing that is considerably more accurate
in some circumstances but that produces results that are harder to clearly interpret.
In both cases, the computational cost of determining $\mHt$ photodissociation
rates is reduced by enough to make three-dimensional high-resolution
simulations of cloud formation feasible with modest computational
resources. Our modification to ZEUS-MP also includes a detailed treatment of 
the thermal behaviour of the gas.

In this paper, we focus on the problem of molecular cloud formation in 
gravitationally unstable, initially static gas. (In a subsequent paper, we 
consider turbulent flow).  We show that in these conditions, and for
initial densities consistent with those observed in the cold, neutral 
atomic phase of the interstellar medium, $\mHt$ formation occurs on 
a timescale $t \geq 10 \: {\rm Myr}$, comparable to or longer than the 
gravitational free-fall timescale of the cloud. We also show that the collapsing 
gas very quickly reaches thermal equilibrium and that the equation of state 
of the thermal equilibrium gas is generally softer than isothermal.

Finally, we demonstrate that although these results show little sensitivity to 
variations in most of our simulation parameters, they are highly sensitive to the 
assumed initial density $n_{\rm i}$. Reducing $n_{\rm i}$
significantly increases the cloud formation timescale and decreases the amount 
of hydrogen ultimately converted to $\mHt$.
\end{abstract}


\section{Introduction}
Since essentially all observed Galactic star formation occurs within dense,
self-gravitating molecular clouds, developing an understanding of the origin 
of these clouds is an important goal of research into star formation.
Research in this area has typically focused on trying to answer a few basic
questions:
\begin{enumerate}
\item[(i)] How does the molecular gas form? In other words, what are the chemical
processes involved?
\item[(ii)] Where does the molecular gas form? Does it form before or after the 
assembly of the gas into dense clouds?
\item[(iii)] How quickly does the molecular gas form?
\end{enumerate}
By far the largest constituent of the molecular gas is molecular hydrogen, 
$\mHt$, with other molecules such as ${\rm CO}$ being present only in small
amounts, and so in practice the study of the formation of molecular gas is usually
simply the study of the formation of $\mHt$.

Although the chemistry of $\mHt$ formation in space remains an active field
of study, the basic principles have been understood for some time. Gas-phase
formation of $\mHt$ by direct radiative association is highly forbidden and 
proceeds at a negligible rate, while gas-phase formation via intermediate
molecular ions such as $\Hm$ or $\mHtp$ is strongly suppressed by the 
interstellar radiation field \citep{gl03} and in any case cannot produce 
molecular fractions much higher than $x_{\mHt} \simeq 10^{-3}$. Consequently, 
most Galactic $\mHt$ cannot have formed in the gas phase, leaving 
grain-surface reactions as the only viable option. The pioneering work of 
\citet{gs63} and \citet{hs70,hs71} showed that $\mHt$ molecules could form on 
the surface of idealized dust  grains with an effective rate coefficient compatible with 
that inferred from UV observations of $\mHt$ in the local interstellar medium, 
$R_{\mHt} \sim 10^{-17} \:  \: {\rm s}^{-1}$ \citep{jura74}. This remains widely 
accepted, at least for cold dust, although there is ongoing debate about the 
efficacy of $\mHt$ formation on warm dust  \citep[see e.g.][]{katz99,ct04}.

Answers to the other questions remain far more uncertain. One school of
thought argues that $\mHt$ forms {\em in situ}, in the locations presently 
occupied by the observed clouds. According to this picture, gas accumulates
due to the action of large scale flows, which may be driven by large-scale
gravitational instability \citep{kenn89}, magnetic instabilities such as the Parker 
instability (\citeauthor{park66}~1966, although see \citealt{kos02} for a recent 
view of its importance), or may simply be part of the general turbulent velocity field, 
which itself is probably driven primarily by some combination of energy input from 
supernovae and from the magnetorotational instability \citep{mk04}. However, others have argued 
that the $\mHt$ forms long before the molecular clouds themselves are assembled, residing 
in the interstellar medium in a diffuse state or in the 
form of small cloudlets that eventually coalesce to form observable clouds
(see \citealt{elm90} and references therein, or \citealt{pal01} for a more 
recent version of this model). Since coalescence will happen at a much faster 
rate in regions of  converging flow, such as spiral arms, this model can be used 
to explain the enhanced star formation rates found within spiral arms.

One way to discriminate between these explanations is by determining the ages of
observed molecular clouds. If most molecular clouds are young, with ages comparable 
to their dynamical timescales, then this suggests that they are transient objects, 
and argues for a dynamical origin. On the other hand, if clouds are old, with 
lifetimes that are significantly greater than their dynamical timescale, then this is
much easier to explain within a model in which clouds build up slowly and are 
dispersed slowly.

In recent years, evidence that clouds are young has been accumulating. 
For instance, \citet{bhv99} 
argue that the absence of post-T Tauri stars with ages greater than $3 \: {\rm Myr}$ in 
the Taurus-Auriga molecular cloud complex implies that the age of this cloud complex
can be no more than a few million years. \citet{hbpb01} elaborate on this idea
and show that similarly young ages are implied for most local star-forming regions. 
Additionally, the age dispersion of stars in open clusters suggests that 
the molecular cloud complexes that give rise to them must have lifetimes
of the order of their dynamical timescales, which are typically no more than 
a few million years \citep{elm00}. Short cloud lifetimes also make it easier 
to understand the presence of supersonic magnetohydrodynamical
turbulence within the 
molecular gas: in the absence of forcing, this will decay away within a
few turbulent crossing times \citep{sog98,mac99}, and so its presence in 
long-lived clouds {\em requires} there to be some form of external or internal 
driving, whereas its presence in short-lived clouds does not.

Another key piece of evidence for youthful clouds is discussed by 
\citet{fukui99} and \citet{bli06}, who show that giant molecular clouds in the Large
Magellanic Cloud (LMC) are well-correlated with young stellar clusters, but do
not correlate well with older clusters or with supernova remnants. 
They find that these very large clouds can only last $\sim 6$~Myr
before the onset of OB star formation, although they may last another
20~Myr, supported by internal H~{\sc ii} region
expansion \citep[e.g.][]{mat02,kmm06} or external driving by background 
supernovae \citep[e.g.][]{jm06}, before their final dissolution by the same agents.  

However, for a model involving rapid cloud formation to be viable, it must be 
possible to produce the required quantity of $\mHt$ within a few million 
years. Given the relatively slow rate at which $\mHt$ forms,
it is natural to ask whether it is possible to satisfy this requirement.
Simple back-of-the-envelope estimates made using the $\mHt$ formation
rate quoted above suggest that it is possible, provided that the mean density 
of the material making up the cloud exceeds $10^{3} \: {\rm cm}^{-3}$ 
\citep{hbpb01}, but we would ideally like to be able to confirm this result with 
more detailed numerical modeling. Various efforts in this direction have been 
made by a number of groups \citep{hp99,hp00,ki00,ki02,berg04}, but to date 
this modeling has generally been restricted to one or two dimensions, and has 
assumed an initially ordered, large-scale velocity field, such as a convergent 
flow, despite the observational evidence that the velocity field of the neutral 
interstellar medium (ISM)
is dominated by disordered, turbulent motions \citep[see e.g.][]{lp00}.

On the other hand, existing three-dimensional simulations of the neutral ISM,
which do properly treat the velocity field and often also include an approximate
treatment of the thermal balance of the gas (e.g.\ \citealt{kor99}; \citealt{da00};
\citealt{wada01}; \citeauthor{kn02a}~2002a,b,~2004; \citealt{bkmm04};
\citealt{dab04};  \citealt{sdbs05}; \citealt{jm06}) have not previously included 
sufficient chemistry to follow the formation of  $\mHt$ and so have been unable to 
directly address the questions posed here.

To bridge this gap, we have performed simulations of the neutral ISM
using a hydrodynamical code capable of following both the thermal balance of the gas 
and the formation and destruction of $\mHt$ within it. Our goal is to determine
how quickly significant quantities of $\mHt$ can form in the dense neutral ISM,
and to use this information to assess whether models in which cloud formation is assumed 
to be rapid are likely to work in practice. In this paper, we discuss in detail the 
numerical method used to follow the coupled chemical, thermal and dynamical evolution 
of the gas, and the tests used to verify the correctness of our implementation. We also 
present results from an application of our method to the problem of $\mHt$ formation
in gas that is gravitationally unstable, but not turbulent. In a companion paper 
(\citealt{gm06}; hereafter, paper II) we present results from a large suite of simulations 
that include the effects of supersonic turbulence. Although highly simplified, and
probably not representative of real clouds, the non-turbulent models examined in
the second half of the present paper allow us to place an upper limit on the time
required to form a molecular cloud, given neutral atomic gas with the density assumed 
in our initial conditions. By comparing the results of these simulations with the
results from the turbulent models examined in paper II, we can more clearly identify
the effects of the turbulence, allowing us to demonstrate that supersonic turbulence
significantly reduces the time required to form large quantities of $\mHt$.

The structure of the current paper is as follows. In \S~\ref{num_section} we 
describe the methods used to solve the equations of fluid flow, with a focus on our 
treatment of the thermal and chemical evolution of the gas, and in \S~\ref{tests} 
we present the results of some basic tests of our approach. In \S~\ref{init_conds}, 
we describe and motivate the initial conditions used for our simulations. 
In \S~\ref{static_results}, we present results from simulations of gravitationally 
unstable gas which is initially at rest, paying particular attention to the rate of 
$\mHt$ formation and the spatial distribution of the resulting molecular gas. 
Finally, in \S~\ref{summary} we summarize our main results.

\section{Numerical method}
\label{num_section}
\subsection{Magnetohydrodynamical equations}
\label{mhd_eq}
The governing equations for the flow of an inviscid, magnetized, 
self-gravitating gas can be written as \citep{sn92b}:
\begin{equation}
\tderiv{\rho}{t}  = - \rho \nabla \cdot \bm{v}, 
\end{equation}
\begin{equation}
\rho \tderiv{\bm{v}}{t}  =  - \nabla p - \rho \nabla \Phi + \frac{1}{4\pi} 
(\nabla \times \bm{B}) \times \bm{B}, 
\end{equation}
\begin{equation}
\rho \tderiv{}{t} \left(\frac{e}{\rho} \right) =  - p \nabla \cdot \bm{v}  
- \Lambda,  \label{energy_eq} 
\end{equation}
\begin{equation}
\pderiv{\bm{B}}{t}  = \nabla \times (\bm{v} \times \bm{B}),
\end{equation}
\begin{equation}
\nabla^{2}\Phi   =   4 \pi G \rho,
\end{equation}
where $\rho$ $e$, $p$, $\bm{v}$, $\bm{B}$ and $\Phi$ are, respectively, 
the mass density, internal energy density, pressure, velocity, magnetic 
field and gravitational potential of the gas, where ${\rm D}/{\rm D}t$ denotes
the comoving derivative
\begin{equation}
\tderiv{}{t} = \pderiv{}{t} + \bm{v} \cdot \nabla,
\end{equation}
and where $\Lambda$ denotes the net rate at which the gas gains or loses 
internal energy due to radiative and chemical heating and 
cooling.\footnote{Note that $\Lambda > 0$ corresponds to a net loss of 
energy and $\Lambda < 0$ to a net gain.}
Additionally, in a chemically reactive flow, each chemical species satisfies
an equation of the form
\begin{equation}
\tderiv{\rho_{i}}{t}  = - \rho_{i} \nabla \cdot \bm{v} + {\rm C}_{i} 
 - {\rm D}_{i}, \label{chem_eq}
\end{equation}
where $\rho_{i}$ is the mass density of species $i$, and where ${\rm C}_{i}$ 
and ${\rm D}_{i}$ represent its creation and destruction by chemical reactions.
Finally, to close the set of equations, it is necessary to specify an 
equation of state relating the internal energy and the pressure. For an 
ideal gas we can write this as
\begin{equation}
 p = (\gamma - 1) e,
\end{equation}
where $\gamma$, the ratio of specific heats, depends upon the composition of
the gas. For a gas with a number density $n$ of hydrogen nuclei, a number 
density $n_{\He} = 0.1n$ of helium nuclei, and with a molecular hydrogen 
abundance $x_{\mHt} = 2 n_{\mHt} / n$ and an electron abundance 
$x_{\me} =  n_{\me} / n$, we can write $\gamma$ as
\begin{equation}
 \gamma = \frac{5.5 + 5x_{\me} - 1.5x_{\mHt}}{3.3 + 3x_{\me} - 0.5x_{\mHt}},
\end{equation} 
where we have assumed that the rotational degrees of freedom of $\mHt$
are populated, but that the vibrational degrees of freedom are not. With the 
definition of $x_{\mHt}$ that we have adopted here, a value of $x_{\mHt} = 1.0$ 
corresponds to gas in which all of the hydrogen is in molecular form, in
which case $\gamma = 10/7$. (Note that the more familiar $\gamma = 7/5$
is for a gas which is pure $\mHt$; the presence of the helium in our simulations
causes a slight hardening of the equation of state).

To solve this set of equations, we used a modified version of the 
publicly available ZEUS-MP hydrodynamical code. ZEUS-MP is a multi-physics, 
massively-parallel, message-passing code for astrophysical fluid dynamics 
\citep{norman00}, developed by the Laboratory for Computational Astrophysics 
at UC San Diego, which solves the equations of self-gravitating 
magnetohydrodynamics (MHD) in three dimensions. The algorithms used in the ZEUS 
family of hydrocodes are described in detail in \citeauthor{sn92a}~(1992a,b) 
and \citet{hs95}. Their implementation within ZEUS-MP is discussed 
in \citet{fied97} and \citet{norman00}. The Poisson solver used is a Fourier 
space solver that utilizes the FFTW library \citep{fftw}. Our modified version
of ZEUS-MP is derived from version 1.0b of the code. (For details of the more
recently released version 2, see \citealt{zmp2}).

We have modified ZEUS-MP in two main respects. Firstly, in order to follow 
non-equilibrium chemistry within the gas it is necessary to add an extra 
field variable for each chemical species that we wish to track. A natural 
choice of variable is the mass density of each species, as in that case in a 
medium with $N$ non-equilibrium chemical species, we will have $N$ equations 
of the form of equation~\ref{chem_eq} to solve. As discussed below in 
\S~\ref{chem_sec}, we follow only two non-equilibrium species in the 
simulations presented here -- $\mHt$ and $\Hp$ -- and so have two such 
equations to solve. To solve these equations, we use operator splitting to
separate the effects of advection (which is treated in the same fashion as
advection of the total mass density; see \citealt{sn92a}) from those of the
chemical creation and destruction terms. In other words, during the reaction 
step we solve the equations
\begin{eqnarray}
 \pderiv{\rho_{\mHt}}{t} & = & {\rm C}_{\mHt}(\rho_{\mHt}, \rho_{\Hp}, T)
  - {\rm D}_{\mHt}(\rho_{\mHt}, \rho_{\Hp}, T), \label{chem_eq1} \\
  \pderiv{\rho_{\Hp}}{t } & = & {\rm C}_{\Hp}(\rho_{\mHt}, \rho_{\Hp}, T) 
   - {\rm D}_{\Hp}(\rho_{\mHt}, \rho_{\Hp}, T). \label{chem_eq2}
\end{eqnarray}
The method of solution that we adopt is implicit finite differencing: we 
approximate equations~\ref{chem_eq1} \& \ref{chem_eq2} as
\begin{eqnarray}
 \rho_{\mHt}^{i+1} & = & \rho_{\mHt}^{i} + \Delta t 
  \left[ {\rm C}_{\rm \mHt}^{i+1} - {\rm D}_{\rm \mHt}^{i+1}  \right], 
  \label{chem_eq3} \\
  \rho_{\Hp}^{i+1} & = & \rho_{\Hp}^{i} + \Delta t 
  \left[{\rm C}_{\rm \Hp}^{i+1} - {\rm D}_{\rm \Hp}^{i+1} \right], 
  \label{chem_eq4}
\end{eqnarray}
where the superscripts indicate values at the beginning and end of the timestep. 
The advantage of using a first-order implicit method is that we can guarantee 
that the abundances will remain non-negative and that the solution will remain
stable regardless of the size of the timestep chosen (although the 
requirement that we solve equations~\ref{chem_eq3} \& \ref{chem_eq4}
accurately still places some limits on the size of the timestep). The 
disadvantage of using an implicit method is that the resulting finite 
difference equations are coupled and must be solved iteratively. Moreover, 
the fact that the creation and destruction terms also depend strongly on the 
internal energy of the gas (through the temperature $T$) suggests that we 
should solve these equations simultaneously with the energy equation. We 
therefore defer discussion of the solution of equations~\ref{chem_eq3} \& 
\ref{chem_eq4} until after we have discussed the modifications that we have
 made to the treatment of the internal energy equation.

This has been modified to include a term representing the combined effects of 
radiative and chemical heating or cooling, i.e.\ the $\Lambda$ term of 
equation~\ref{energy_eq}. Details of the processes included 
are given in \S~\ref{cool_sec} below and are summarized in 
Table~\ref{cool_model}. Solution of the resulting equation 
proceeds much as it does in the unmodified version of ZEUS-MP: the equation is
operator split, with the effects of artificial viscosity, compressional 
heating and advection treated separately. Our treatment of the artificial 
viscosity and advection steps mirrors that in the unmodified version of the 
code, as discussed in detail in \citet{sn92a}; we will not discuss it 
here. However, during the compressional heating step, instead of solving the 
equation
\begin{equation}
 \label{pdv}
 \pderiv{e}{t} = - p \nabla \cdot \bm{v},
\end{equation}
we solve
\begin{equation}
 \label{pdvcool}
 \pderiv{e}{t} = - p \nabla \cdot \bm{v} - \Lambda.  
\end{equation}
To solve equation~\ref{pdvcool}, we use an algorithm based on a 
combination of  the ZEUS-3D {\bf pdvcool} algorithm (originally 
implemented by M.~Norman and subsequently modified by M.-M.~Mac~Low,
J.~Stone and D.~Clarke) with the implicit algorithm used by \citet{sutt97} 
and further developed by \citet{pav02} and \citet{sr03}. We construct the following 
implicit approximation to equation~\ref{pdvcool}
\begin{equation}
\frac{e^{i+1} - e^{i}}{\Delta t} = - \tilde{p}
\nabla \cdot \bm{v}^{i} - \Lambda^{i+1},
\end{equation}
where $\tilde{p}$ represents the time-centered pressure, which we 
approximate as $\tilde{p} \simeq 0.5 [p(e^{i+1}) + p(e^{i})]$,
and where $\Lambda^{i+1}$ is the cooling rate at the end of the
timestep. We rearrange this equation to give
\begin{equation}
 e^{i+1} =  \frac{(1 - q) e^{i} - \Lambda^{i+1}
 \Delta t}{1 + q}, \label{fd_energy}
\end{equation}
where
\begin{equation}
 q =  \frac{\Delta t}{2} (\gamma - 1) (\nabla \cdot \bm{v}^{i}),
\end{equation}
and then solve equation~\ref{fd_energy} together with 
equations~\ref{chem_eq3} \& \ref{chem_eq4} using a form of Gauss-Seidel
iteration. This works as follows:
\begin{enumerate}
\item Update $\rho_{\mHt}$, using old values for $\rho_{\Hp}$ and $e$.
\item Update $\rho_{\Hp}$, using new value for $\rho_{\mHt}$ but old value 
for $e$.
\item Update $e$, using new values for $\rho_{\mHt}$ and $\rho_{\Hp}$.
\item Test for convergence. If not converged, return to step 1.
\end{enumerate}
On our first pass through the loop, we take the old values for $\rho_{\Hp}$ 
and $e$ to be those at the start of the current timestep. On subsequent 
passes through the loop, we instead use the values from the previous 
iteration. 

Since the internal energy converges more slowly than either of the chemical
abundances, we monitor its value to determine when to stop the iteration,
halting once the relative difference between updated and old values 
becomes less than $10^{-7}$. On rare occasions, the iteration may fail to 
converge. In this case, we switch to solving for $e$ using a more expensive
bisection algorithm that is sure to find a solution. In this case, we lose the 
benefits of the iteration for refining our initial guesses for $\rho_{\mHt}$ and
$\rho_{\Hp}$. Fortunately, the abundances of $\mHt$ and $\Hp$ generally
do not vary much during a single timestep and so the error that this
introduces is probably not significant, particularly since the iteration almost
always succeeds.

To compute the cooling timestep, $\Delta t_{\rm cool}$, we use the same 
procedure as in \citet{sutt97} and \citet{pav02}: we compute $\Lambda$
and then require that the timestep be such that the internal energy will
not vary by more than 30\% given this value of $\Lambda$, i.e.\
\begin{equation}
\Delta t_{\rm cool} = 0.3 \frac{e}{|\Lambda|}.
\end{equation}

To improve the efficiency of the code, we also make use of subcycling. 
Rather than constraining the global timestep of the 
simulation $\Delta t$ to be less than $\Delta t_{\rm cool}$, we instead 
use the same global timestep as we would within the unmodified version 
of the code, but solve the chemistry and cooling substeps alone over the 
shorter cooling timescale, repeating the procedure for as many times as 
is necessary to make the total elapsed time equal to $\Delta t$. Thus, 
if $\Delta t_{\rm cool} \geq \Delta t$, we proceed through the chemistry 
and cooling substep only once per simulation timestep, while if 
$\Delta t_{\rm cool} \ll \Delta t$, we do so many times, using the updated 
values of the chemical abundances and internal energy from the end of 
one substep as the input to the next. We subcycle at the level of the
individual grid zones, so only zones for which $\Delta t_{\rm cool} < \Delta t$
take multiple chemistry and cooling substeps  per hydrodynamical timestep.
The gain in computational efficiency from this subcycling procedure is difficult 
to quantify, as it depends on the physical state of the gas, but tests indicate 
that it can easily be as much as an order of magnitude.

\subsection{Chemistry}
\label{chem_sec}
The chemical composition of the ISM is complex. Over 120 different 
molecular species have been detected in interstellar space \citep{wik04}
and while many of these are found in detectable amounts only in dense,
well-shielded gas, there remain a significant number that have been 
detected in diffuse, unshielded gas 
\citep[see, for instance][and references therein]{ovw02}. 
A full chemical model of the ISM can easily
involve several hundred different atomic and molecular species and several
thousand different reactions, even if reactions on grain surfaces are 
neglected \citep[e.g.][]{rate99}.

It is currently impractical to incorporate this amount of chemistry into a 3D
hydrodynamical code such as ZEUS-MP, due to the extreme impact it
would have on the code's performance. Fortunately, much of this chemistry
is not relevant to our current study, and can be neglected without significantly
compromising our results. Recall that the main goals of this paper are to 
determine the timescale on which molecular clouds form in gravitationally
unstable atomic gas, and the fraction of this gas that is converted to 
molecular form. To achieve these goals, we clearly need to be able to 
follow the formation and destruction of $\mHt$ with a reasonable degree 
of accuracy. Beyond this, however, the only chemistry that we really need to be 
concerned with is that which plays a role in determining the thermal 
balance of the gas. In other words, we need only follow the chemistry
of $\mHt$, and of a few other major coolants such as ${\rm C}^{+}$ or ${\rm O}$;
the chemistry of other molecules such as ${\rm CH}$, while undeniably of 
interest, is not central to the task at hand and so can be sacrificed in pursuit 
of efficiency. 

To identify the major coolants in the diffuse ISM, we can use the results of 
\citet{w95,w03}, who studied its thermal evolution in some detail. They show
that in the warm neutral medium (WNM), which has a characteristic 
temperature $T \sim 8000 \: {\rm K}$, most of the cooling comes from 
Lyman-$\alpha$ emission from atomic hydrogen, electron recombination
with small grains and polycyclic aromatic hydrocarbons (PAHs), and fine 
structure emission from atomic oxygen. Cooling in the cold neutral medium 
(CNM), which has a characteristic temperature $T < 300 \: {\rm K}$, is dominated 
by fine structure emission from ionized carbon, $\cii$, with fine structure 
emission from oxygen also contributing significantly in the warmer parts of
the CNM. These two coolants also dominate in the thermally 
unstable temperature regime $300 < T < 8000 \: {\rm K}$. Heating in all three 
regimes is dominated by photoelectric emission from dust grains and PAHs;
the heating rate is therefore primarily determined by the strength of the 
ultraviolet background and by the dust to gas ratio, but it also has a
dependence on the electron abundance (see equation~\ref{pe_eff} in 
\S~\ref{cool_sec}). We also include fine structure cooling from 
${\rm Si^{+}}$. Although never the dominant coolant, ${\rm Si^{+}}$ does
produce $\sim 10\%$ of the total fine structure emission at temperatures
$T > 200 \: {\rm K}$ at all densities, and can produce as much as 
30\% of the emission at high densities, owing to the relatively large
size of Einstein coefficient for the ${\rm Si^{+}}$ $^{2}P_{3/2} \rightarrow
 ^{2}P_{1/2}$ transition \citep{sv02}.

In denser gas that is shielded from the ultraviolet background, other coolants
such as ${\rm CO}$ and ${\rm H_{2}O}$ become important, as demonstrated 
by \citet{gl78}, \citet{nk93} and \citet{nlm95}. However, since we are primarily 
interested in the transition from atomic to molecular gas, and since these 
coolants will not become important until after the gas is already fully molecular, 
we have chosen to make another major simplification in the chemistry: we 
assume that carbon, oxygen and silicon remain primarily in the form of \cii, \oi
and \sii respectively, throughout the simulation. By making this assumption, we 
essentially reduce the chemistry that must be followed to that of only 
three species -- electrons, neutral hydrogen and $\mHt$ -- at the cost of 
computing an incorrect cooling rate in dense molecular gas. The dense
gas in our simulations is also typically rather cold, with a temperature of no
more than a few tens of Kelvin, and so we will probably underestimate the
cooling rate within it, since in these conditions molecular coolants such as 
${\rm CO}$ are more effective than the \cii, \oi and \sii included in our
simulations. This means that we probably overestimate the temperature of
this gas, but the fact that the dense gas is so cold shows that we do not 
overestimate its temperature by a large amount, and we therefore do not
expect this approximation to have a significant impact on the $\mHt$ 
formation rate. 

Having discussed the underlying assumptions, we now present our chemical
model. We adopt standard solar abundances of hydrogen and helium, along
with abundances of carbon, oxygen and silicon taken from \citet{sem00}: 
$x_{\rm C} = 1.4 \times 10^{-4}$, $x_{\rm O} = 3.2 \times 10^{-4}$, 
$x_{\rm Si} = 1.5 \times 10^{-5}$. We 
assume that helium remains neutral and plays no direct role in the chemistry,
and that carbon and silicon remain singly ionized throughout the simulation.
The ionization state of oxygen is assumed to track that of hydrogen, due to
the influence of the charge transfer reactions
\begin{eqnarray}
{\rm O}^{+} + \mH & \rightarrow & {\rm O} + \Hp, \\
\Hp + {\rm O} & \rightarrow & \mH + {\rm O^{+}}.
\end{eqnarray}

As previously noted in \S~\ref{mhd_eq}, we follow directly the abundances 
of two chemical species, namely $\Hp$ and $\mHt$. The abundances of the other
major species -- atomic hydrogen and electrons -- are computed from the 
conservation laws:
\begin{equation}
 x_{\mH} = x_{\rm H, tot} - x_{\Hp} - x_{\mHt},
\end{equation}
and
\begin{equation}
 x_{\me} = x_{\Hp} + x_{\rm C^{+}} + x_{\rm Si^{+}},
\end{equation}
where $x_{\rm H, tot}$ is the total abundance of hydrogen nuclei, in all forms,
and where $x_{\rm C^{+}}$ and $x_{\rm Si^{+}}$ represent the abundances of
ionized carbon and silicon respectively. We neglect the small contributions to
$x_{\me}$ made by electrons from other ionized metals, such as ${\rm Mg}$
or ${\rm S}$, since these are small compared to the contribution from carbon.
We also neglect any contribution from ${\rm O^{+}}$, as this will be negligible
compared to the contribution from hydrogen. 

Our assumption that carbon and silicon remain ionized throughout the 
simulation becomes inaccurate in dense, well-shielded gas, where UV
photoionization of neutral carbon and silicon becomes ineffective. We
therefore overestimate the fractional ionization of gas in these regions.
However, our main motivation for tracking the fractional ionization is
to compute the photoelectric heating rate accurately, and 
since this is unimportant in dense, shielded gas, the inaccuracy in the
fractional ionization is likely unimportant there.

The reactions included in our chemical model are summarized in 
Table~\ref{chem_model}. In most cases, we also list the source or sources
from which we took the rate coefficient used in our model. The exception is
$\mHt$ photodissociation, the treatment of which is discussed in detail in 
\S~\ref{h2_phdis} below.  Two other reactions deserve further comment:
the collisional dissociation of $\mHt$ by atomic hydrogen
\begin{equation}
\mHt + \mH \rightarrow 3\mH,
\end{equation}
and molecular hydrogen
\begin{equation}
\mHt + \mHt \rightarrow \mHt + 2\mH.
\end{equation}
The reaction coefficients for both of these reactions are density-dependent,
since they are sensitive to the population of the vibrational and rotational 
levels of $\mHt$. To treat the former, we use a rate coefficient
\begin{equation}
\log k_{\mH}  = \left( \frac{n/n_{\rm cr}}{1 + n/n_{\rm cr}} \right)
\log k_{\rm \mH, h} + \left(\frac{1}{1 + n/n_{\rm cr}} \right) \log k_{\rm \mH, l},
\end{equation}
where $k_{\rm \mH, l}$ is the low density limit of the collisional dissociation 
rate and is taken from \citet{ms86}, while $k_{\rm \mH, h}$ is the high density 
limit, taken from \citet{ls83}. The critical density, $n_{\rm cr}$, is given by
\begin{equation}
\frac{1}{n_{\rm cr}} = \frac{x_{\mH}}{n_{\rm cr, \mH}} + 
\frac{x_{\mHt}}{n_{\rm cr, \mHt}},
\end{equation}
where $n_{\rm cr, \mH}$ and $n_{\rm cr, \mHt}$ are the critical densities in pure
atomic gas with an infinitesimally dilute quantity of $\mHt$ and in pure molecular 
gas respectively. The first of these values is taken from \citet{ls83}, but has been
decreased by an order of magnitude, as recommended by \citet{msm96}; the 
other value comes from \citet{sk87}. To treat the collisional dissociation of
$\mHt$ by $\mHt$ we use a similar expression
\begin{equation}
\log k_{\mHt}  = \left( \frac{n/n_{\rm cr}}{1 + n/n_{\rm cr}} \right)
\log k_{\rm \mHt, h} + \left(\frac{1}{1 + n/n_{\rm cr}} \right) \log k_{\rm \mHt, l},
\end{equation}
where the low density limit, $k_{\rm \mHt, l}$, is taken from \citet{mkm98} and the 
high density limit, $k_{\rm \mHt, h}$, is taken from \citet{sk87}.  The collisional 
dissociation rates computed in this way are acceptably accurate when 
$n_{\mH} \gg n_{\mHt}$ or $n_{\mH} \ll n_{\mHt}$,  but may be less accurate in 
gas with $n_{\mH} \sim n_{\mHt}$; further study of the collisional dissociation of 
$\mHt$ in gas which is a mixture of $\mH$ and $\mHt$ would be desirable to 
help remedy this.

We do not include the gas phase formation of $\mHt$ via the $\Hm$ or 
$\mHtp$ ions, as in the typical conditions of the Galactic ISM,  this is 
unimportant compared to the formation of $\mHt$ on dust grains 
\citep{gl03}.

As can be seen from Table~\ref{chem_model}, our $\Hp$ chemistry is 
straightforward. However, a couple of reactions are deserving of comment.
One is gas-phase recombination, where we note that we use the case 
B value for the recombination coefficient, as this is the most appropriate choice 
in all but the most highly ionized gas. The other is the cosmic ray ionization of
$\mH$. In the majority of our simulations, we use a value 
of the cosmic ray ionization rate $\zeta = 10^{-17} \: {\rm s^{-1}}$. This is 
consistent (within the error bars) with 
recent determinations of the ionization rate in dense cores within molecular
clouds \citep{cwth98,bpwm99,vv00}. However, measurements of the 
${\rm H_{3}^{+}}$ column density along sight-lines passing through diffuse gas 
are only explained by a much larger value of $\zeta$, within the range 
$10^{-16} < \zeta < 10^{-15} \: {\rm s^{-1}}$ \citep{mac03,liszt03,lrh04}, where 
the uncertainty is due primarily to the uncertain distribution of gas along the 
observed lines of sight. The cause of this discrepancy is currently unknown 
(although \citealt{ps05} suggest one possible mechanism), and so for simplicity 
we take $\zeta$ to be constant, independent of the gas density. We briefly
examine in \S~\ref{static-temp} the consequences of adopting a larger value 
for $\zeta$.

\subsubsection{$\mHt$ photodissociation}
\label{h2_phdis}
Following \citet{db96}, we can write the photodissociation rate of $\mHt$ in
optically thin gas as
\begin{equation}
 k_{\rm ph, 0} = 3.3 \times 10^{-11} \chi \:\: {\rm s}^{-1}, \label{kph_thin}
\end{equation}
where we have assumed that the ultraviolet field has the same spectral shape
as the \citet{d78} field, and where $\chi$ is a dimensionless factor which
characterizes the intensity of the field at $1000\:$\AA~relative
to the \citet{habing68} field; note that for the original \citet{d78} field,
$\chi = 1.7$.

By balancing this dissociation rate against the rate at which $\mHt$ forms on 
dust grains, we can easily show that the equilibrium $\mHt$ fraction in cold
gas is given by
\begin{equation}
 x_{\mHt} \sim 10^{-6} \chi^{-1}  \left(\frac{T}{100}\right)^{1/2} n_{\mH}, 
\end{equation}
which is clearly far less than one unless $n_{\mH}$ is very large. Since 
observations indicate that significant molecular gas is present at densities 
below $n_{\mHt} < 10^{4} \: {\rm cm}^{-3}$ \citep[see e.g.][]{falg98},
it is clear that some shielding of interstellar $\mHt$ molecules from the 
effects of UV photodissociation must occur, and that simulations of
molecular cloud formation which do not take this shielding into account 
are going to be tremendously inaccurate.

$\mHt$ can be shielded against photodissociation in two main ways: by
line absorption due to other $\mHt$ molecules (i.e.\ self-shielding), and by
continuous absorption due to dust. The effects of both processes have 
been treated in some detail by \citet{db96}. They show that in shielded 
gas the photodissociation rate can be written approximately as
\begin{equation}
\label{h2_kph}
k_{\rm ph} = f_{\rm shield}(N_{\mHt}) e^{-\tau_{\rm d, 1000}} k_{\rm ph,0},
\end{equation}
where $f_{\rm shield}$ is a numerical factor accounting for the effects of
self-shielding, $\tau_{\rm d, 1000}$ is the optical depth due to dust 
at $1000 $\AA, and $k_{\rm ph, 0}$ is the unshielded photodissociation
rate, given by equation~\ref{kph_thin} above. 

\citeauthor{db96} show that for a static, plane parallel slab of gas, 
$f_{\rm shield}$ is well approximated by
\begin{equation}
f_{\rm shield} = \frac{0.965}{(1 + x/b_{5})^{2}} + \frac{0.035}{(1 + x)^{1/2}}
 \exp\left[-8.5 \times 10^{-4} (1+x)^{1/2}\right], \label{fsh}
\end{equation}
where $x = N_{\mHt} / 5 \times 10^{14} \: {\rm cm}^{-2}$ with $N_{\mHt}$ 
being the $\mHt$ column density, and $b_{5} = b / 10^{5} \: {\rm cm}
\:{\rm s}^{-1}$, where $b$ is the Doppler broadening parameter.

If the gas is not static or in uniform motion, but instead has a spatially 
varying velocity, then equation~\ref{fsh} will overestimate the amount
of self-shielding that occurs. The reason for this is that if the velocity
field is non-uniform, then the relative velocity between any two fluid 
elements will, in general, be non-zero. Consequently, the contribution to 
the total absorption coming from the first fluid element will be Doppler 
shifted when viewed from the rest frame of the second fluid element. 
If this Doppler shift is large compared to the line widths of the 
Lyman-Werner band transitions, then the effect is to significantly reduce 
the extent to which the absorption coming from the first fluid element 
contributes to the self-shielding seen by the second fluid element. 
For $\mHt$ column densities $N_{\mHt} < 10^{17} \: {\rm cm^{-2}}$,
the intrinsic widths of even the strongest Lyman-Werner band
transitions are unimportant and the line profiles are dominated by
Doppler broadening. In this regime, the neglect of changes in the velocity
along the line of sight is justified if the differences in velocity are
much smaller than the sound speed of the gas, and equation~\ref{fsh}
remains a good estimator for $f_{\rm shield}$. On the other hand, if the
differences in the velocity are much greater than the sound speed, as
will generally be the case in a supersonically turbulent flow, then
equation~\ref{fsh} will significantly overestimate the degree of
self-shielding that will actually occur.

For $N_{\mHt} > 10^{17} \: {\rm cm^{-2}}$, the line widths of the strongest
transitions are dominated by Lorentz broadening, and the degree of
self-shielding in these lines becomes insensitive to the velocity
distribution of the gas, unless the range of velocities is extremely
large. However, the total $\mHt$ photodissociation rate remains
sensitive to the velocity dispersion, as a large fraction
of the total rate is due to absorption in weaker lines, whose widths are
still dominated by Doppler broadening. Only for $N_{\mHt} > 10^{18} \:
{\rm cm^{-2}}$ does the total photodissociation rate become relatively
insensitive to the velocity distribution of the gas.

An additional problem with using equation~\ref{fsh} to compute the self-shielding
factor is that it requires us to compute the $\mHt$ column densities along a 
large number of lines of sight for every zone in our simulation. Unfortunately, 
the computational cost of doing so is very large: for a simulation with $N$
fluid elements, the cost of solving for the column densities scales as 
$O(N^{5/3})$ (assuming that we require an angular resolution that is well 
matched with the spatial resolution of the code), compared to scalings of 
$O(N)$ for the hydrodynamical algorithms and $O(N \log N)$ for the self-gravity.
Consequently, solving for the column densities would quickly come to dominate
the cost of the simulation, and would prevent us from performing any high 
resolution simulations.  Moreover, while parallelization of the code would 
help to some extent, efficient parallelization of the radiative transfer is 
difficult, owing to the way in which it couples together widely separated 
regions in the computational volume.

Computation of the optical depth due to dust, $\tau_{\rm d, 1000}$, is in 
principle considerably simpler than computation of $f_{\rm shield}$, since
it involves continuous absorption rather than line absorption, and so Doppler
effects can generally be neglected (at least for the velocities of interest 
here). \citet{ccm89} showed that it is possible to approximate the extinction 
observed along many lines of sight in the Galaxy by a one-parameter family of 
curves, where the controlling parameter is $R_{V}$, defined as
\begin{equation}
R_{V} = \frac{A_{V}}{A_{B} - A_{V}},
\end{equation}
where $A_{B}$ and $A_{V}$ are the extinctions in the B and V bands 
respectively.  For $R_{V} = 3.1$, which is typical for many lines of sight 
in the diffuse ISM, \citet{db96} quote an effective attenuation cross-section 
for the dust of $\sigma_{\rm d, 1000} = 2 \times 10^{-21} \: {\rm cm^{2}}$.
The corresponding dust opacity is given by
\begin{equation}
\label{tau_d}
\tau_{\rm d, 1000} = 2 \times 10^{-21} N_{\rm \mH, tot},
\end{equation}
where $N_{\rm \mH, tot} = N_{\mH} + N_{\Hp} + 2 N_{\mHt}$ is the total
column density of hydrogen nuclei between the point of interest and the 
source  of the radiation. Therefore, to compute the effects of the dust
shielding, we again need to compute appropriate column densities,
although in this case the effect of velocity differences along the line of
sight is unimportant. As in the case of self-shielding, the cost of solving
for the column densities is $O(N^{5/3})$ for a simulation with $N$ fluid
elements, and so a high resolution treatment of both the fluid flow and
the radiation field is not computationally feasible.

To overcome these difficulties, we have performed simulations using two
different approximations. Our first approximation is very simple. We assume 
that the dominant contribution to the shielding of a given fluid element comes 
from gas in the immediate vicinity of that element. In the case of $\mHt$ 
self-shielding, this assumption can be justified to some extent by the fact 
that the relative velocity between this gas and the point of interest will typically 
be small, whereas gas at larger distances will typically possess a much larger relative 
velocity, particularly in supersonically turbulent gas. Additionally, in gravitationally 
collapsing regions, the local gas density will be substantially larger than the 
density in most regions of the flow, further increasing the importance of the 
spatially local contribution. In the case of dust shielding, the former justification
no longer holds true, although the latter remains valid. To give a quantitative form 
to this approximation, we note that we continue to use equation~\ref{fsh} to 
compute the self-shielding factor, but use an $\mHt$ column density given by
\begin{equation}
 \tilde{N}_{\mHt} = \frac{\Delta x}{2} n_{\mHt},
\end{equation}
where $n_{\mHt}$ is the $\mHt$ number density in the zone of interest and 
$\Delta x$ is the width of the zone, measured parallel to one of the coordinate
axes. Similarly, we continue to use equation~\ref{tau_d} to compute the dust
opacity, but replace $N_{\rm \mH, tot}$ with the local approximation:
\begin{equation}
\tilde{N}_{\rm \mH, tot} = \frac{\Delta x}{2} \left( n_{\mH} + n_{\Hp} + 
2 n_{\mHt} \right),
\end{equation}
where $n_{\mH}$, $n_{\Hp}$ and $n_{\mHt}$ are the local number densities of
atomic hydrogen, $\Hp$ and $\mHt$ respectively. In the remainder of this
paper, we refer to this approximation as the local shielding approximation.

The main advantage of the local shielding approximation is that we can be 
certain that we are {\em underestimating} the true amount of shielding and 
hence {\em overestimating} the $\mHt$ photodissociation rate.  Consequently, 
$\mHt$ fractions computed in simulations which use the local shielding approximation 
will be strong lower limits on the true values, and the corresponding $\mHt$
formation timescales will be solid upper limits. Moreover, in paper II we
show that this very simple approximation proves to be surprisingly accurate 
in supersonically turbulent gas, although it is significantly less effective in 
the simulations of smooth collapse presented in this paper.

A significant disadvantage of the local shielding approximation is that it makes the 
photodissociation rate dependent on $\Delta x$ and hence on the resolution 
of the simulation. Consequently, the abundance of $\mHt$ that we expect to 
find in gas in our simulations that is in photodissociation equilibrium also 
becomes resolution-dependent. To see this, consider the equation for the 
equilibrium $\mHt$ abundance:
\begin{equation}
 \frac{\xhteq}{1 - \xhteq} = \frac{2 R_{\rm form}}{R_{\rm ph}} n, \label{eq:xhteq}
\end{equation}
where $R_{\rm form}$ is the formation rate of $\mHt$ on dust grain surfaces,
taken from \citet{hm79}, and $R_{\rm ph} = k_{\rm ph} n_{\mHt}$. Since 
$R_{\rm ph}$ depends on  $\Delta x$ but $R_{\rm form}$ does not, the equilibrium
$\mHt$ abundance $\xhteq$ must inevitably depend on $\Delta x$. How significant 
this is for any given grid zone depends in large part on the total hydrogen column 
density and the $\mHt$ column density for that zone. If both are sufficiently small 
that the zone is optically thin to UV radiation, then changing the size of the zone 
will have little effect on $R_{\rm ph}$, provided that the zone remains optically 
thin. Conversely, if either $N_{\rm \mH, tot}$ or $N_{\mHt}$ for that zone 
are large enough to make the zone highly optically thick to $\mHt$ photodissociation, 
then changing the zone size will have a large effect on $R_{\rm ph}$, but only a 
very small effect on $\xhteq$, which will remain close to one. However,
for zones that lie in the intermediate regime between these two limiting cases, the
effect on $\xhteq$ can be significant and it is here that our results will be least 
accurate.

The second approximate technique that we have used to compute the 
photodissociation rate again makes use of equations~\ref{fsh} and \ref{tau_d}.
However, in this approach we compute the $\mHt$ and dust column densities
seen by each grid zone by averaging over a small number of lines of sight.
Specifically, we compute column densities in both the positive and negative
directions along each of the three coordinate axes of the simulation, compute
the associated values of $f_{\rm shield}$ and $\tau_{\rm d, 1000}$ for each
of these lines of sight, and then compute the total photodissociation rate  
by means of a suitably weighted average of these values. The main 
advantage of this approach is that its computational cost should not be 
very much greater than the cost of solving the hydrodynamic equations, 
as in both cases the cost of the algorithm is $O(N)$. Moreover, by 
restricting the lines of sight that are treated to be those parallel to the
coordinate axes, we also limit the amount of communication between 
different processors that is required, and so limit the adverse impact 
on the scalability of the code. An approach of this type has previously 
been used to study the stability and dynamics of low mass
molecular clouds and Bok globules \citep{nl97,nl99}, while a very
similar approach has been used to study the formation of $\mHt$ in
early protogalaxies illuminated by an intergalactic UV background
\citep{yahs03}. We refer to this approximation in the remainder of 
this paper as the six-ray shielding approximation.

An obvious disadvantage of this approach is the extremely coarse 
angular resolution of the radiation field that it provides. This poor
angular resolution will cause us to overestimate the amount of 
shielding in some regions, and underestimate it in others: the
precise details will depend on the particular form of the density 
field, but in general we will tend to underestimate the amount of
shielding whenever the volume filling factor of dense gas is 
small. Another
significant problem is that this approach does not take account of 
velocity structure along any of the lines of sight. It therefore may 
significantly overestimate $f_{\rm shield}$ in a supersonic flow,
particularly if the gas is turbulent. For the main problem that we are 
interested in investigating -- the determination of the $\mHt$ 
formation rate in dynamically evolving, cold atomic gas -- this is
problematic, as it may lead us to derive an artificially short timescale 
for $\mHt$ formation. Nevertheless, in many scenarios (including
the simulations of smooth collapse presented later in this paper),
the six-ray shielding approximation produces significantly more 
accurate results than the local shielding approximation. 

Another possible approach to treating the effects of shielding
which we considered but discarded makes use of the fact that the 
contribution towards the total $\mHt$ photodissociation rate made 
by any individual Lyman-Werner line is directly proportional to the 
penetration probability for that line. For a line of sight ${\mathbf n}$, 
this is given by
\begin{equation}
P_{\rm p}({\mathbf n}) = \int_{0}^{\infty} \phi(\nu) e^{-\tau(\nu, {\mathbf n})} d\nu,
\end{equation}
where $\phi(\nu)$ is the line profile function for the line in question
and $\tau(\nu, {\mathbf n})$ is the optical depth at frequency $\nu$ 
in the direction ${\mathbf n}$. This penetration probability is analogous 
to the more commonly encountered escape probability, and this similarity 
can be fruitfully exploited. In particular, if the gas has a large monotonic 
velocity gradient in the direction ${\mathbf n}$, then the Sobolev approximation 
\citep{sob57} can be used to compute $P_{\rm p}({\mathbf n})$ given 
only the local $\mHt$ number density and the magnitude of the velocity 
gradient. If the velocity gradient is large in all directions, then we can repeat
this procedure for many lines of sight, and with suitable averaging can
derive a mean penetration probability for the line. Finally, by repeating
this for each of the lines which contribute to the total photodissociation rate, 
we can compute the rate itself, using only local quantities.

There are, however, two major drawbacks to this approach. First, it can
only be relied upon to give accurate answers in conditions where the
Sobolev approximation applies. This formally limits the applicability of
this method to regions where the velocity gradient is monotonic and
where any variations in density, temperature or $\mHt$ abundance 
occur far more slowly than variations in velocity. These conditions are
not satisfied in turbulent molecular clouds, and so although the 
Sobolev approximation can sometime be fruitfully applied 
\citep[see, for instance][]{oss97},
in general it will not give accurate results. Moreover, in contrast
to our local shielding approximation, we cannot be confident that we know the 
sense of the inaccuracy, i.e.\ whether we produce too much or too little $\mHt$,
as this will depend on the physical conditions within a given simulation,
and may also change over time within that simulation. Therefore,
although this approach avoids the resolution dependence of our local shielding
approximation, it does so at the cost of producing results that have no
clear interpretation, being neither lower nor upper limits. The other major 
drawback to this approach is that it cannot be used to model dust absorption, as 
in this case the Sobolev approximation simply does not apply. 

To sum up, our local shielding approximation has the advantages of simplicity 
and a straightforward interpretation, while our six ray shielding
approximation will often be more accurate but has a less clear interpretation. 
Although neither approximation is ideal, by comparing and contrasting
the results from both, we can draw important conclusions about the 
formation timescale of  $\mHt$ in static and in turbulent gas.

\subsection{Heating and cooling}
\label{cool_sec}
In order to solve equation~\ref{pdvcool}, we need to calculate 
$\Lambda$, the net rate at which the gas gains or loses energy due to 
radiative and/or chemical heating and cooling. We can write $\Lambda$ 
as the sum of a heating and a cooling term:
\begin{equation}
\Lambda = \Lambda_{\rm cool} - \Gamma_{\rm heat}.
\end{equation}
As discussed in \S~\ref{chem_sec}, a number of different processes 
contribute to $\Lambda_{\rm cool}$. At high temperatures 
($T \simgreat 8000 \: {\rm K}$), much of the cooling is provided by excitation 
of the resonance lines of hydrogen, 
helium or heavier elements. To treat cooling in this temperature range, we 
adopted a tabulated cooling function from \citet{sd93}: specifically, the cooling 
function listed in Table~10 of their paper, which assumes [Fe/H] = -0.5. 
Significant cooling also comes from the recombination of electrons with small 
dust grains and PAHs. We incorporate this using a cooling rate taken from 
\citet{w03}, based on an original formulation by \citet{bt94}:
\begin{equation}
 \Lambda_{\rm rec} = 4.65 \times 10^{-30} \phi_{\rm pah} T^{0.94} \psi^{\beta} 
 n_{\me} n  \: {\rm ergs} \: {\rm cm^{-3}} \: {\rm s}^{-1},
\end{equation}
where $\beta = 0.74 / T^{0.068}$, $\psi$ is given by
\begin{equation}
\label{psi_eq}
\psi = \frac{\chi_{\rm eff} T^{1/2}}{n_{\rm e} \phi_{\rm pah}},
\end{equation}
where $\chi_{\rm eff} = e^{-2.5A_{\rm V}} \chi$ represents the strength of the UV 
background in the gas after dust shielding is taken into account
\citep[see][]{berg04}, and where $\phi_{\rm pah}$ is an adjustable 
parameter introduced by \citet{w03} to make the heating rate consistent 
with the values of the electron attachment and electron recombination 
rates that are inferred observationally for PAHs; the original 
\citeauthor{bt94} treatment corresponds to $\phi_{\rm pah} = 1$.

At lower temperatures, the contribution of these coolants becomes negligible,
and cooling by \cii and \oi fine structure lines dominates. To compute the cooling
from \cii, we used atomic data from \citet{sv02}, together with collisional 
de-excitation rates from \citet{fl77}  for collisions with $\mHt$, from
\citet{hm89}  for collisions with atomic hydrogen at $T < 2000 \: {\rm K}$, 
from \citet{k86}  for collisions with atomic hydrogen at 
$T > 2000 \: {\rm K}$, and from \citet{wb02}  for collisions with electrons. 

For \oi fine structure cooling, we used atomic data from \citet{sv02}, 
together with collisional rates provided by D.~Flower (priv.\ comm.) 
for collisions with $\mH$ and $\mHt$, as well as rates from \citet{bbt98}
for collisions with electrons and \citeauthor{p90}~(1990,~1996) for 
collisions with protons. 

In addition to cooling from \cii and \oi, we also included contributions 
from \sii fine structure line emission -- which can be more effective than
\cii cooling if the temperature and density are both large -- and from the 
rotational and vibrational lines of $\mHt$. 

To compute the \sii cooling rate, we again used atomic
data from \citet{sv02}, and collisional rates from \citet{r90} for collisions 
with atomic hydrogen and from \citet{dk91} for collisions with electrons. 
De-excitation rate coefficients for collisions between $\sii$ and $\mHt$ 
were unavailable, and so were arbitrarily set to zero; however, this is
unlikely to introduce a significant error into the computed cooling rate in 
our simulations as gas with a significant molecular fraction is typically far 
too cold for \sii cooling to be effective (see, for instance, 
\S~\ref{static-temp}).
 
For cooling due to $\mHt$, we use the cooling function from \citet{lpf99},
which we have extended to temperatures below $100 \: {\rm K}$ by 
assuming that only the $J = 2 \rightarrow 0$ and  $J = 3 \rightarrow 1$ 
transitions contribute significantly to the cooling rate. For simplicity, we
also fix the ortho:para ratio at 3:1. However, variations in this ratio are
unlikely to significantly affect the $\mHt$ cooling rate at temperatures
at which it contributes significantly to the total cooling rate (see, for
instance, Figure~5 in \citealt{lpf99}).

In each case, we assumed that cooling occurs in the optically thin limit.
This is a reasonable assumption in diffuse gas, but breaks down in dense,
high column density cores. However, in these conditions of high density
and high optical depth, we would expect all of the hydrogen to already be
in molecular form (and indeed much of the carbon and oxygen to be in the 
form of ${\rm CO}$, rather than \cii and \oi as assumed here) and so errors in 
the gas temperature in these dense cores are unlikely to have any significant
effect on our results.

In addition to these processes, we also include the effects of collisional 
transfer of energy between gas and dust, following the prescription of
\citet{hm89}. This acts to cool the gas whenever $T_{\rm gas} > 
T_{\rm dust}$, and to heat it when $T_{\rm dust} > T_{\rm gas}$. 
However, this is not an important process at the gas densities studied
here, although it does become increasingly important in higher density
gas.

Finally, we also allow for the effects of cooling due to the collisional 
dissociation of $\mHt$ and collisional ionization of $\mH$, although in
practice neither process is of much importance in our simulations.

As far as heating is concerned, the most important contribution to 
$\Gamma_{\rm heat}$ comes from the heating produced by photoelectric 
emission from UV-irradiated dust grains  and PAHs.
Following \citet{bt94} and \citet{w03}, we write 
the photoelectric heating rate as
\begin{equation}
 \Gamma_{\rm pe} = 1.3 \times 10^{-24} n \epsilon \chi_{\rm eff} 
 \: {\rm ergs} \: {\rm s}^{-1} \: {\rm cm^{-3}},
\end{equation}
where $\epsilon$ is the heating efficiency, given by
\begin{equation}
\label{pe_eff}
\epsilon = \frac{4.9 \times 10^{-2}}{1 + 4.0 \times 10^{-3} \psi^{0.73}} 
+ \frac{3.7 \times 10^{-2} \left(T / 10^{4}\right)^{0.7}}{1 + 2.0 \times 
10^{-4} \psi},
\end{equation}
with $\psi$ as given by Equation~\ref{psi_eq} above.

Additional contributions to $\Gamma_{\rm heat}$ come from
several other processes:
\begin{itemize}
\item Photodissociation of $\mHt$ by far-ultraviolet (FUV) radiation

Following \citet{bd77}, we assume that each photodissociation 
deposits $0.4 \: {\rm eV}$ of heat into the gas, giving us a heating
rate 
\begin{equation}
\Gamma_{\rm ph} = 6.4 \times 10^{-13} k_{\rm ph} n_{\mHt} 
 \: {\rm ergs} \: {\rm s}^{-1} \: {\rm cm}^{-3},
\end{equation}
where $k_{\rm ph}$ is given by Equation~\ref{h2_kph} above.

\item Excitation of $\mHt$ by FUV radiation

As well as dissociating some of the $\mHt$, the FUV radiation
also produces vibrationally excited $\mHt$ via radiative 
pumping. At high densities, this leads to heating of the gas,
as most of the excited molecules undergo collisional 
de-excitation. We adopt a radiative pumping rate that is
8.5 times larger than the photodissociation rate 
\citep{db96}, and assume that each excitation transfers an
average of $2 \, (1 + n_{\rm cr}/n)^{-1} \: {\rm eV}$ to the gas
\citep{bht90}, where $n_{\rm cr}$ is the critical density at 
which collisional de-excitation of vibrationally excited 
$\mHt$ occurs at the same rate as radiative de-excitation. 
As previously noted,
our value for $n_{\rm cr}$ is a weighted harmonic mean of the value for
$\mHt$-$\mH$ collisions given by \citet{ls83} and the value for 
$\mHt$-$\mHt$ collisions given by \citet{sk87}.

\item $\mHt$ formation on dust grains

The formation of an $\mHt$ molecule from two hydrogen atoms
releases approximately $4.5 \: {\rm eV}$ of energy. In all likelihood,
some fraction of this energy will go into the rotational and vibrational 
excitation of the newly-formed $\mHt$ molecule, with the remainder 
going into the translational energy of the molecule or into heating 
the grain \citep{dw93}. The fraction of the total energy going into 
rotational and vibrational excitation remains a subject of investigation
\citep[see e.g.][]{ros03}, but in our simulations we assume that this fraction is
close to one. As with FUV pumping, most of this energy will be lost
through radiative de-excitation of the $\mHt$ molecule if  
$n \ll n_{\rm cr}$, but will be converted to heat if $n \gg n_{\rm cr}$. 
We therefore adopt an $\mHt$ formation heating rate of
\begin{equation}
\Gamma_{\mHt} = 7.2 \times 10^{-12} \frac{R_{\mHt}}{(1 + n_{\rm cr}/n)} 
\: {\rm ergs} \: {\rm s}^{-1} \: {\rm cm}^{-3},
\end{equation}
where $R_{\mHt}$ is the rate of $\mHt$ formation on dust grains 
and $n_{\rm cr}$ is the critical density, calculated as described 
above.

\item Cosmic ray ionization

Following \citet{gl78}, we assume that each ionization deposits 
$20 \: {\rm eV}$ of energy into the gas, which gives us a heating rate 
\begin{equation}
 \Gamma_{\rm cr} = 3.2 \times 10^{-28} \left( \frac{\zeta}{10^{-17} \: 
 {\rm s}^{-1}} \right) n \: {\rm ergs} \: {\rm s}^{-1} \: {\rm cm}^{-3}.
\end{equation}
\end{itemize}

We do not include the effects of heating by soft X-rays, as this is 
of little importance compared to photoelectric heating in low density gas, 
and is negligible in high density, optically thick gas. Other potential heat 
sources, such as the photoionization of carbon or silicon by the UV 
background, are also insignificant and can be neglected.

A full list of all of the thermal processes included in our model can be found in
Table~\ref{cool_model}.

\section{Tests}
\label{tests}
Tests of the hydrodynamical and MHD algorithms used in ZEUS are presented
in \citeauthor{sn92a}~(1992a,b) and \citet{hs95} and therefore are not 
discussed here. Some potential problems relating to the treatment of rarefaction 
waves and adiabatic MHD shocks in ZEUS have been pointed out by \citet{falle02}, 
but we do not anticipate that these problems will invalidate the results presented 
in this paper. In the case of the rarefaction errors, our confidence that
they will not significantly affect our results rests on the fact that most of the 
relevant chemistry for $\mHt$ formation occurs in regions of compression while 
regions of rarefaction are relatively inactive, and so errors in the treatment of 
the latter will have very little effect. As for the shock errors, these vanish
if an isothermal equation of state is used \citep{falle02}, and since the effective
equation of state in our simulations is much closer to the isothermal case than 
the adiabatic case (see \S~\ref{static-temp} below), it is reasonable to expect 
that any errors will be small.

However, it is necessary to ensure that our modifications to the basic ZEUS
code operate as intended. To verify the modified code, we have performed 
tests of the advection of the species mass densities in non-reacting flows
(discussed in \S~\ref{advect_test} below) and of the solution of the
chemistry and cooling substep in static gas (discussed in 
\S~\ref{static_test}). 

\subsection{Advection tests}
\label{advect_test}
In non-reacting flow, Equation~\ref{chem_eq} reduces to
\begin{equation}
\tderiv{\rho_{i}}{t}  = - \rho_{i} \nabla \cdot \bm{v},
\end{equation}
and so the species mass densities should be advected in precisely the same manner
as the physical density of the gas. To verify that this is indeed the case, we 
performed a series of tests of the advection algorithm. Our test suite was based
on the advection tests used by \citet{sn92a} to validate the advection algorithms 
used in ZEUS-2D, a predecessor of the current ZEUS-MP code. These tests include 
the advection of a square pulse of high density material by a uniform flow, the
classic Sod shock tube test \citep{sod78}, and various tests taken from \citet{cw84}
that involve the interaction of strong shocks.  To use these test problems to 
verify that our species mass densities are advected correctly, we modified them
to include two extra field variables, $\rho_{1}$ and $\rho_{2}$, representing 
arbitrary chemical species, both  of which evolve according to Equation~\ref{chem_eq}. 
Since we wished to simulate non-reacting flow, the chemical creation and destruction
terms for these variables ($C_{1}$, $C_{2}$, $D_{1}$ and $D_{2}$) were set to zero.
The fields were initialized such that $\rho_{1} = \rho$ and $\rho_{2} = 10^{-8} \rho$,
the tests were run, and then the resulting values of $\rho_{1}$ and $\rho_{2}$ were
examined. Correct operation of the code implies that $\rho_{1}/\rho$ and 
$\rho_{2}/\rho$ should be conserved. In our tests, we found that these ratios were
generally conserved to within 0.01\%.

\subsection{Chemistry and cooling tests}
\label{static_test}
Our initial tests of the chemistry and cooling substeps focused on verifying that
the correct reaction rates and heating and cooling rates were computed by the code, 
given various different sets of input parameters. To do this, we constructed a 
simple test harness, using a mixture of Fortran and Perl, which allowed us to run 
the main chemistry and cooling subroutines separately from the remainder of the 
hydrodynamical code. We also inserted debugging statements into these subroutines
which caused them to write out the values of the various rates into a log file.
The values produced were then compared with those produced for the same set of 
input parameters by an independent test implementation of the chemical and cooling 
rates. Disagreements between the two sets of rates could then be easily identified and 
investigated, and the whole process could be repeated for many different sets of input
parameters. By eliminating the overhead of running the full hydrodynamical code, 
we were able to quickly explore a wide range of input parameters and verify that 
the code operated correctly in each case. This approach proved especially useful for 
catching simple bugs during the code development process, particularly those caused by 
typographical errors in the software, as these were generally present in either the
test implementation or the real implementation, but were very unlikely to be present
in both. 

We also tested the operation of the chemistry and cooling code at a much higher level
by performing several sets of simulations of static gas using the full hydrodynamical
code. In the first set of simulations, we artificially disabled the operation of the 
cooling subroutines and set $\Lambda = 0$, so that the gas would remain at its initial 
temperature. We then ran a large number of simulations of static gas with different 
initial temperatures, densities, UV fields and dust-to-gas ratios. In each case, we 
ran the simulation for $3 \times 10^{15} \: {\rm s}$, or almost 100~Myr, which even at the lowest 
densities that we examined greatly exceeded the time required for the chemical abundances 
to reach equilibrium. We then compared the value of the $\mHt$ and $\Hp$ abundances 
produced by the simulation with the predicted equilibrium values, which are easy to compute 
for a fixed temperature and density. In every case that we examined, we found good agreement 
between the predicted and the actual values.

In our second set of simulations, we performed a similar series of tests of the 
cooling routines: we artificially disabled the chemistry, equivalent to setting 
${\rm C}_{i} = {\rm D}_{i} = 0$ in Equation~\ref{chem_eq} above, and solved for the 
equilibrium temperature of the gas for various different initial densities, temperatures, 
UV fields and $\mHt$ fractions. Again, the resulting values could be easily compared with 
the predicted equilibrium values. Good agreement was again found in every case.

Finally, we performed a set of simulations without artificial constraints on either 
the chemical or the cooling terms. The aim of this set of simulations was to reproduce 
the phase diagram for interstellar atomic gas calculated by \citet{w03}. To do this,
we ran simulations for a wide range of initial densities, but used the same initial 
temperature, $T_{\rm i} = 1000 \: {\rm K}$, in each case. The other free parameters of our 
simulations were set up to correspond as closely as possible to those used by \citet{w03} 
in computing their $R = 8.5 \: {\rm kpc}$ Galactrocentric radius model. Specifically, we
adopted gas phase carbon and oxygen abundances of $x_{\rm C} = 1.4 \times 10^{-4}$
and $x_{\rm O} = 3.2 \times 10^{-4}$ respectively, a cosmic ray ionization rate 
$\zeta = 1.8 \times 10^{-17} \: {\rm s^{-1}}$ and an ultraviolet field strength $\chi = 1.7$
(or in other words, the standard \citet{d78} field).
 
\begin{figure}  
\centering
\epsfig{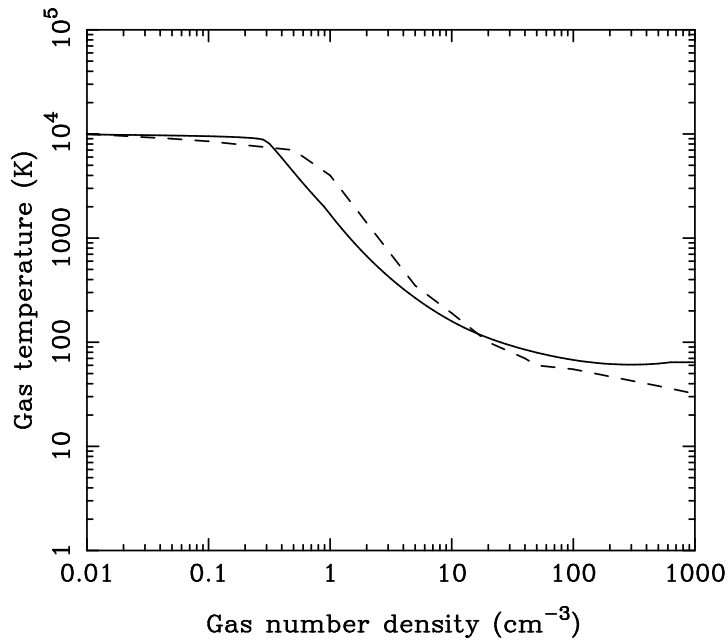}
\caption{Gas temperature of the ISM as a function of number density, 
computed for a Galactrocentric radius of $8.5 \: {\rm kpc}$ using
our modified version of ZEUS-MP (solid line). For comparison, the
results of \citet{w03} are plotted (dashed line). Note that since our 
modelling of the gas-phase chemistry and gas-grain interactions is considerably 
less detailed than that of Wolfire et~al., we do not expect complete
agreement.}
\label{wolfire_test}
\end{figure}

The phase diagram we obtained is plotted in Figure~\ref{wolfire_test}. We also plot the 
corresponding  values from \citet{w03}, taken from their Figure~10. 
Although our results do not match those of 
\citeauthor{w03} precisely, the differences are small and almost certainly result from 
the fact that we do not include as wide a range of physical processes as \citeauthor{w03};
for instance, they model the chemistry of the gas in far more detail than we are able 
to, and so their values for the free electron abundance, and hence for the photoelectric
heating rate, will be more accurate than ours. The largest difference occurs at 
$n \sim 1 \: {\rm cm^{-3}}$, near the middle of the thermally unstable regime, and even
here our computed temperatures differ from those of \citeauthor{w03} by no more than a 
factor of two. Given the approximations made in our treatment of the chemistry of the ISM, 
we consider this degree of agreement to be acceptable.

\section{Initial conditions}
\label{init_conds}
\subsection{Box size and initial number density}
\label{box_size}
Since our aim in this paper is to model the transition from atomic to 
molecular gas, we have chosen to consider relatively small volumes of gas,
which we visualize as being small sub-regions within larger, gravitationally 
collapsing regions, such as those found in the simulations of \citet{krav03}
or \citet{lmk05}.
For simplicity, and also to allow us to use an FFT-based gravity solver, we take
our simulation volumes to be cubical, with periodic boundary conditions on all 
sides. For most of our simulations, we used boxes of size $20 \: {\rm pc}$ or 
$40 \: {\rm pc}$, although in a small number of cases, we used other sizes.
The particular size used for each individual simulation is summarized in 
Table~\ref{static_runs}. This table also lists the values adopted 
for the other main adjustable parameters, discussed in more detail below. Our 
choice of box size was guided by two major considerations. The first of these was 
the fact that as we reduce the size of the box, we increase the stability of the gas 
with respect to gravitational collapse. Indeed, if we choose to make the box too 
small, it will contain less than a Jeans mass of gas and so the gas will simply not 
collapse. The larger that we make the box, the more gas it will contain, and the
more representative it will be of a real gravitationally unstable region. On the other
hand, by reducing the size of the box, we increase our physical resolution at any
given numerical resolution. This is important since the physical resolution of our 
simulations determines the extent to which we can follow the gravitational collapse
of the gas (see \S~\ref{truelove} below). Obviously, these two 
considerations conflict, but in choosing our box sizes we have done our best to find 
an appropriate balance between them.

Within the box, we assume that the gas is initially distributed uniformly, with a
number density $n_{\rm i}$. In the majority of our simulations, we take $n_{\rm i} = 100 
\: {\rm cm^{-3}}$, as the inferred mean densities of many observed molecular 
clouds lie near to this value \citep{mk04}. However, we also explore the effects of
reducing $n_{\rm i}$, examining values as small as $10 \: {\rm cm^{-3}}$ (see 
\S~\ref{static_n0}).  At a density of $100 \: {\rm cm^{-3}}$, the 
atomic gas lies well within the CNM 
regime \citep{w95,w03},  and so the initial temperature of the gas has essentially 
no impact on the results of the simulations, as the gas rapidly cools, reaching
thermal equilibrium within 0.05~Myr. In most 
of our simulations, we therefore use a rather arbitrary initial temperature 
$T_{\rm i} = 1000 \: {\rm K}$, but as we demonstrate in \S~\ref{static-temp}, 
simulations with $T_{\rm i} = 100 \: {\rm K}$ produce essentially  identical 
results for times $t \simgreat  0.05 \: {\rm Myr}$.

Finally, we note that in all of the simulations described in \S~\ref{static_results}, 
we start with the gas at rest. Therefore, in order to trigger gravitational
collapse in runs performed using the local shielding approximation, it is necessary
to introduce small density perturbations to disrupt the symmetry of the initial
conditions (as this would otherwise artificially prevent gravitational collapse
from occurring). These perturbations are not required in runs performed using
our six-ray approximation, as the pressure-driven flows in these runs serve
the same purpose (see the discussion in \S~\ref{role_of_grav} below), but we 
include them nevertheless. To produce these perturbations,
we select for each grid zone a random number $N$ in the interval $[-0.5, 0.5]$, and 
then perturb the density of that zone by a factor of $(1 + 2\delta N)$, where $\delta$ 
is an adjustable parameter. In most of our simulations, we set $\delta = 0.05$, although 
in \S~\ref{static-box} we explore the effects of using larger values of $\delta$.

\subsection{Initial magnetic field}
\label{init_mag}
Since there is considerable observational evidence for the presence of 
dynamically significant magnetic fields in interstellar gas, much of 
which is discussed in recent reviews by \citet{beck01} and \citet{hc05},
it is clearly important to take the effects of magnetic fields into account in our 
simulations. For this reason, most of our simulations involve magnetized gas. 

For simplicity, in all of our MHD simulations we assumed that the 
field was initially uniform and oriented parallel to the $z$-axis of
the simulation. The strength of the field was a free parameter, and
the values used in our various simulations are summarized in  
Table~\ref{static_runs}. Observational 
determinations of the local magnetic field strength give a typical value 
of $6 \pm 2 \mu$G, and so we ensured that our fiducial value for the initial 
magnetic field strength, $B_{\rm i, fid} = 5.85 \mu{\rm G}$ was consistent 
with this value.

The reason for our slightly odd choice of fiducial value is that we
wanted to ensure that in each of our simulations the mass-to-flux ratio 
of the gas would be some simple multiple of the critical value \citep{nn78}
\begin{equation}
 \left( \frac{M}{\Phi} \right)_{\rm crit} = \frac{1}{2\pi \sqrt{G}}
\end{equation}
at which magnetic pressure balances gravity in an isothermal slab.
For a gas cloud in which all of the hydrogen is already fully molecular, 
\citet{cnwk04} show that the mass-to-flux ratio of the cloud can be written in 
units of this critical value as
\begin{equation}
 \lambda \equiv \frac{\left(M/\Phi\right)}{\left(M/\Phi\right)_{\rm crit}}
  = 7.6 \times 10^{-21} \frac{N_{\mHt}}{B},
\end{equation}
where $N_{\mHt}$ is the $\mHt$ column density of the cloud in units
of ${\rm cm^{-2}}$ and $B$ is the strength of the magnetic field,
in units of $\mu{\rm G}$. For a fully atomic cloud, the equivalent expression is
\begin{equation}
 \lambda = 3.8 \times 10^{-21} \frac{N_{\mH}}{B}, \label{m2f-atom}
\end{equation}
where $N_{\mH}$ is column density of atomic hydrogen. For a simulation with
an initial atomic hydrogen number density of $100 \: {\rm cm^{-3}}$ and a 
box size of $40 \: {\rm pc}$, we have $N_{\mH} \simeq 1.23 \times 10^{22} \: 
{\rm cm^{-2}}$, and hence $\lambda = 46.82 / B_{\rm i}$, where $B_{\rm i}$ is the 
initial magnetic field strength. Therefore, in our fiducial example in which
$B_{\rm i} = B_{\rm i, fid} = 5.85 \: \mu{\rm G}$, we have $\lambda_{\rm fid} = 8$.
Observations of magnetic field strengths in molecular cloud cores, summarized in 
\citet{cru99} and \citet{cnwk04}, find a smaller mean value, $\bar{\lambda} 
\simeq 2$, although there is significant scatter around this value. There are also
many sight-lines for which there are only upper limits on $B$ and hence lower
limits on $\lambda$.

In view of the uncertainty in the appropriate value of $\lambda$, we ran a number 
of simulations using larger values of the initial field strength, as described in
\S~\ref{field-init}, in order to explore the extent to which the $\mHt$ formation rate 
and morphology depend on the field strength. For completeness, we also ran
a few simulations in which we set ${\mathbf B} = 0$. Although unrealistic in the
sense that we expect all interstellar gas to be magnetized to some extent, setting
${\mathbf B} = 0$ does serve as a guide to the behaviour of the gas in the 
more realistic case that ${\mathbf B}$ is non-zero but is too small to significantly affect 
the gas dynamics on the scales of interest to us.

\subsection{Resolution issues}
\subsubsection{The Truelove criterion}
\label{truelove}
\begin{figure}  
\centering
\epsfig{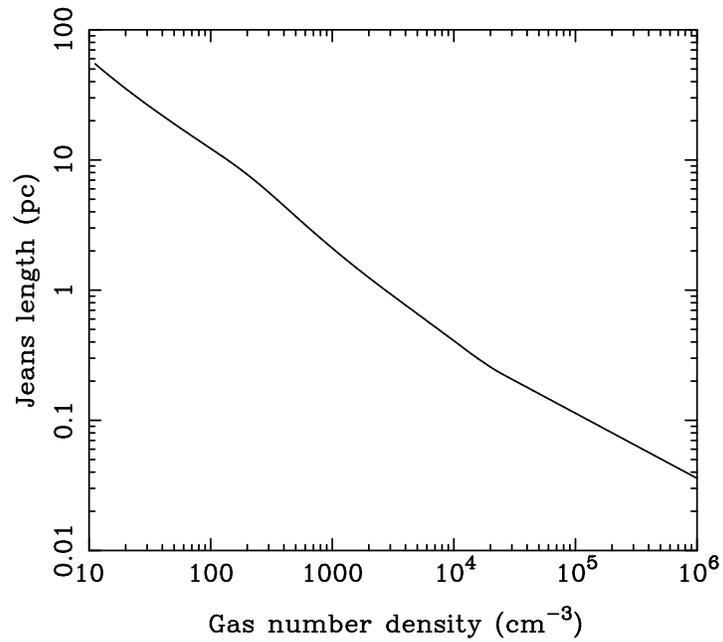}
\caption{Jeans length, $L_{\rm J}$, plotted as a function of density, 
for gas which is in thermal and chemical equilibrium. This plot was
produced using the same treatment of gas chemistry, heating and 
cooling as was used in our simulations, and using our local
approximation to treat $\mHt$ self-shielding and dust shielding.}
\label{jeans_length}
\end{figure}

As discussed in \S~\ref{box_size}, one of the important considerations
influencing our choice of box size was the desire to be able to accurately 
follow the gravitational collapse of gas over as large a range of densities as
possible. \citet{true97} showed that in order to
properly resolve collapse and avoid artificial fragmentation, it is necessary
to resolve the local Jeans length of the gas by at least four grid zones. In 
other words, collapse is resolved only while
\begin{equation}
 \Delta x \leq \frac{1}{4} L_{\rm J}(\rho, T),
\end{equation}
where $\Delta x$ is the width of a single grid zone, and $L_{\rm J}$ is
the Jeans length, given by
\begin{equation}
 L_{\rm J} = \frac{\pi^{1/2}c_{\rm s}}{\sqrt{G\rho}},
\end{equation}
where $c_{\rm s}$ is the adiabatic sound speed. In Figure~\ref{jeans_length},
we plot $L_{\rm J}$ as a function of density for gas which is in
thermal and chemical equilibrium, with the effects of self-shielding and dust 
shielding calculated using our local shielding approximation and assuming 
a zone size of $0.1 \: {\rm pc}$. Although somewhat 
simplified, this model gives a fair representation of how we expect the Jeans 
length to evolve with density in a real simulation.

It is clear from the figure that it becomes increasingly difficult to resolve
$L_{\rm J}$ as we move to higher densities, and so the smaller that we
can make $\Delta x$, the further we will be able to follow the collapse.
Now, for a uniform cubical grid, $\Delta x = L / N^{1/3}$, where $L$ is the box
size and $N$ is the total number of grid zones, so we can make $\Delta x$ small
either by increasing $N$ or by decreasing $L$. Since there are practical
limits on how large we can make $N$, dictated by the computational resources 
required to run the simulation, the easiest way to increase our resolution is
by decreasing $L$. However, as noted previously, if we make $L$ too small, then 
the total mass of gas in the box will fall far below the initial Jeans mass, and so the 
gas will be gravitationally stable and will not collapse.

For any given combination of $N$ and $L$, we can use Figure~\ref{jeans_length} to
determine the maximum number density, $n_{\rm max}$, at which the Truelove criterion 
is still satisfied. We have computed  $n_{\rm max}$ for a range of different numerical 
resolutions between $64^3$ and $512^3$ for boxes of size 20, 40 and 60 pc, and summarize 
the results in Table~\ref{resn_table}. The values of $n_{\rm max}$ we obtain vary from 
$490 \: {\rm cm^{-3}}$ for a $64^{3}$ simulation in a 60 pc box to $5.6 \times 10^{4} \: 
{\rm cm^{-3}}$ for a $512^{3}$ simulation in a 20 pc box. Note, however, that these 
figures are only exact for gas which is in chemical equilibrium; if the $\mHt$ fraction 
is out of equilibrium, then this will affect the value of $c_{\rm s}$, through its dependence
on $\gamma$ and on the mean molecular mass, and so these numbers will change slightly.

\subsubsection{The cooling length and the Field length}
\begin{figure}
\centering
\epsfig{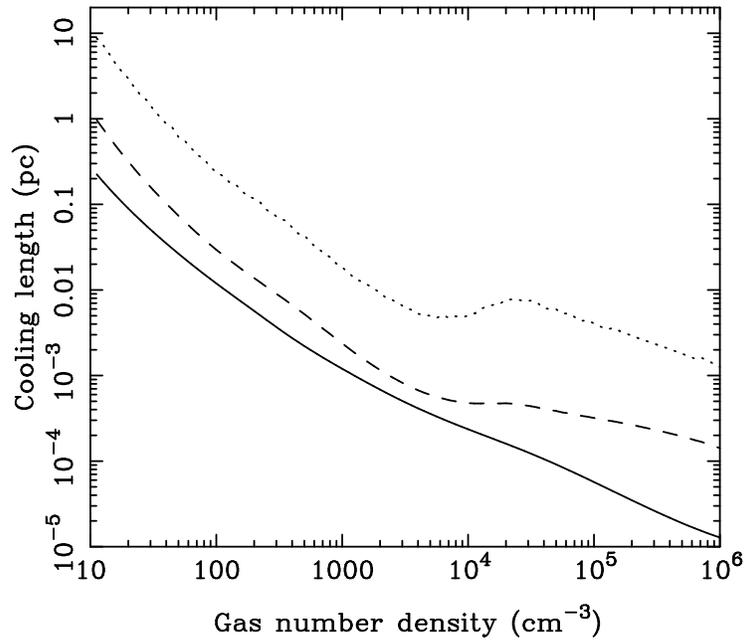}
\caption{Cooling length, $L_{\rm cool}$, plotted as a function of density, 
for gas which is $1 \: {\rm K}$ (dotted line), $10 \: {\rm K}$ (dashed line) or 
$100 \: {\rm K}$ (solid line) warmer than the thermal equilibrium temperature. Note 
that as $L_{\rm cool} \propto \Lambda^{-1}$, the cooling length becomes infinite 
for gas with a temperature which is precisely equal to the equilibrium value.}
\label{cool_length}
\end{figure}

Aside from $L_{\rm J}$, there are two other important length scales which
we would like to resolve: the cooling length, $L_{\rm cool}$, and the 
Field length, $L_{\rm F}$. A convenient definition of the cooling length is the 
product of the cooling time with the sound speed of the gas:
\begin{equation}
L_{\rm cool} = c_{\rm s} t_{\rm cool}.
\end{equation}
To properly resolve the thermal state of the flow, particularly features
involving strong thermal gradients (e.g.\  shock fronts), we must be able to
resolve $L_{\rm cool}$. In Figure~\ref{cool_length}, we plot 
$L_{\rm cool}$ as a function of density, calculated assuming temperatures
which are $1 \: {\rm K}$ (dotted line), $10 \: {\rm K}$ (dashed line) and
$100 \: {\rm K}$ (solid line) greater than the thermal equilibrium 
temperature.\footnote{Note that for gas in thermal equilibrium, the cooling rate
is zero, by definition, and so $t_{\rm cool}$ and $L_{\rm cool}$ are both
formally infinite.}

It is clear from the figure that the resolution required to resolve the cooling 
length of the flow is much greater than that required to resolve the Jeans 
length. This is not unexpected, since $L_{\rm J} \sim c_{\rm s} t_{\rm ff}$,
and $t_{\rm cool} \ll t_{\rm ff}$ for the conditions of interest to us. However,
it does mean that we cannot reasonably expect to use ZEUS-MP to resolve 
the cooling length in gas with a number density $n \simgreat 100 \: {\rm cm^{-3}}$, 
unless the temperature of the gas is extremely close to its equilibrium value, 
given the range of values that we use for the box size $L$. Nevertheless, we argue
that this does not represent a major problem for the simulations presented in this 
paper, provided that we are careful about the inferences that we draw from them. 

The main consequence of failing to resolve $L_{\rm cool}$ is that we
overproduce warm gas, particularly in post-shock regions, since the width of any
such region is clearly constrained to be at least one grid zone wide. Therefore,
any conclusions that we draw about the temperature distribution of the gas should
be treated with caution, as we know at the outset that we will have too much
warm gas, and therefore too little cool, dense gas. The effects on the dynamics of 
the flow are difficult to assess, but it is reasonable to suppose that gravitational 
collapse is inhibited to some degree. Similarly, it is likely that less $\mHt$ 
is produced than would be produced in a higher resolution simulation, since
the $\mHt$ formation rate decreases rapidly with temperature for $T > 170 \: {\rm K}$,
and also since the artificially warm gas is less dense than it would be if it were able to 
cool. The net effect of this is that our simulations produce less dense gas and 
less $\mHt$ within a given timeframe than would be the case if we could resolve 
$L_{\rm cool}$. However, the mass of warm, shocked gas in our simulations is never
large and so we do not expect our failure to resolve $L_{\rm cool}$ to significantly
affect our conclusions regarding the timescale of $\mHt$ formation.

\begin{figure}
\centering
\epsfig{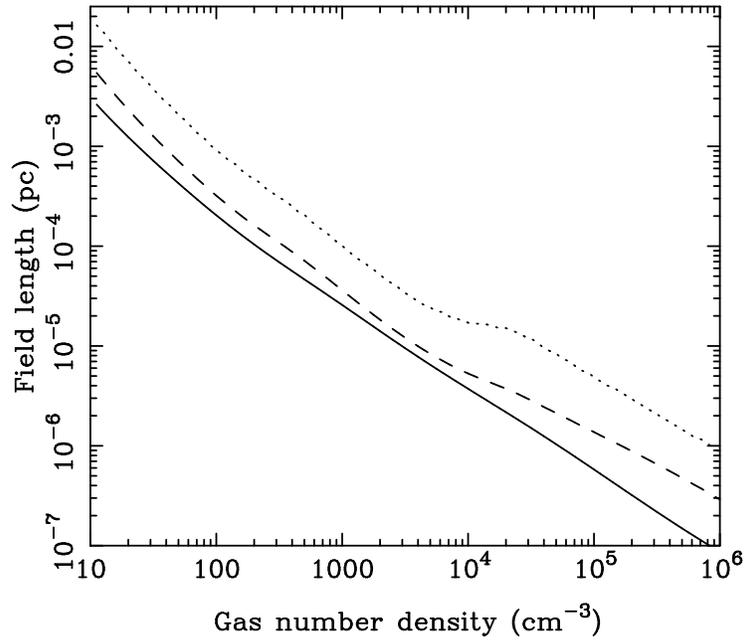}
\caption{Thermal conduction length, or Field length, $L_{\rm F}$,
calculated assuming temperatures which are $1 \: {\rm K}$ (dotted line), 
$10 \: {\rm K}$ (dashed line) and $100 \: {\rm K}$ (solid line) greater 
than the thermal equilibrium temperature.}
\label{field_length}
\end{figure}

The other length scale of interest is the Field length, which is the length scale on 
which thermal conduction stabilizes the growth of thermally unstable density perturbations, 
and  which is given by \citep{field65}
\begin{equation}
L_{\rm F} = \left(\frac{\kappa T}{\Lambda}\right)^{1/2},
\end{equation}
where $\kappa$ is the coefficient of thermal conductivity. In Figure~\ref{field_length}, 
we plot the value of $L_{\rm F}$ as a function of density for temperatures that are
1, 10 and $100 \: {\rm K}$ greater than the thermal equilibrium temperature.
To compute these values, we used a value of $\kappa_{\mH} = 2.5 
\times 10^{3} T^{1/2} \: {\rm ergs} \: {\rm cm^{-3}} \: {\rm K}^{-1} \: {\rm s^{-1}}$ 
for the coefficient of thermal conductivity of neutral atomic hydrogen, taken 
from \citet{park53}, and assumed that $\kappa$ would be of a similar order of
magnitude in fully molecular hydrogen, and in a mixed gas of atoms and molecules. At the
temperatures and densities of interest here, thermal conduction by electrons is 
unimportant and can be neglected.

In a recent paper, \citet{ki04} argue that in order to properly resolve the dynamics
of gas in a thermally bistable medium, it is necessary to resolve $L_{\rm F}$
with at least three grid zones in order to obtain converged results. Now, it is
clear from Figure~\ref{field_length} that it is even harder to resolve $L_{\rm F}$ 
than it is to resolve $L_{\rm cool}$ and so even if we were to include the
effects of thermal conduction in our simulations, we could not realistically hope
to resolve its effects. However, we argue that this does not represent a major problem
given the particular scenario that we are investigating. The reason is that while gas
at the densities studied in our simulations may be thermally {\em unstable} to isobaric
perturbations if its temperature is greater than the equilibrium temperature, it is not 
thermally {\em bistable}: it will all cool rapidly to the cold phase, rather than 
forming a two phase medium. Moreover, although isobaric thermal instability may operate
during this initial cooling phase, amplifying density enhancements on scales 
$L_{\rm F} < L < L_{\rm crit}$, where $L_{\rm crit} = c_{\rm s} 
t_{\rm cool}$ is the maximum length scale for an isobaric perturbation, the final density
will be at most a factor $(T_{\rm i} / T_{\rm eq})$ larger than the initial density, 
where $T_{\rm i}$ is the initial temperature and $T_{\rm eq}$ the equilibrium temperature
of the gas, and the resulting density perturbations will not be gravitationally bound.

To see why, consider an isobaric density perturbation with an initial size 
$L_{\rm i} < L_{\rm crit}$. Before its growth, this perturbation will be
unstable to gravitational collapse only if $L_{\rm i} > L_{\rm J}$,
which will be possible only if $L_{\rm crit} > L_{\rm J}$, or in other
words if $t_{\rm cool} > t_{\rm ff}$. However, we know that in fact 
$t_{\rm cool} \ll t_{\rm ff}$ for the temperatures and densities of interest to us, 
and so the initial perturbation must be gravitationally stable. Now consider the
situation once the gas in the perturbation has cooled from $T_{\rm i}$ to 
$T_{\rm eq}$: its density will have increased by a factor of $(T_{\rm i} / T_{\rm eq})$,
while its local Jeans length will have decreased by the same factor, since
$L_{\rm J} \propto T^{1/2} \rho^{-1/2}$. At the same time, its linear size will
only have decreased by a factor of $(T_{\rm i} / T_{\rm eq})^{1/3}$, assuming an
approximately spherical perturbation. Clearly, therefore, the perturbation will be
less stable against gravitational collapse than it was initially. However, for it to
become unstable, its initial size must satisfy the following inequality
\begin{equation}
 L_{\rm i} > \left(\frac{T_{\rm eq}}{T_{\rm i}}\right)^{2/3} L_{\rm J},
\end{equation}
and since  $L_{\rm i} < L_{\rm crit}$, this requires that 
\begin{equation}
 L_{\rm crit} > \left(\frac{T_{\rm eq}}{T_{\rm i}}\right)^{2/3} L_{\rm J},
\end{equation}
which in turn will only be satisfied if 
\begin{equation}
 \frac{t_{\rm cool}}{t_{\rm ff}}  > \left(\frac{T_{\rm eq}}{T_{\rm i}}\right)^{2/3}.
\end{equation}
Now, for $T_{\rm i} = 1000 \: {\rm K}$ and $T_{\rm eq} \simeq 65 \: {\rm K}$
-- the values appropriate for fiducial run MS256, described in 
\S~\ref{h2_timescale} below -- we have
$(T_{\rm eq} / T_{\rm i})^{2/3} \simeq  0.16$. However, if the initial number 
density is $100 \: {\rm cm^{-3}}$, as it is in most of our simulations, then
$t_{\rm cool}/t_{\rm ff} \simeq 0.007$. Therefore, isobaric perturbations starting
with this temperature and density remain gravitationally stable throughout their
evolution.  A similar result can be derived for the other combinations of
initial temperature and density examined in this paper. 

If the overdensities created by the isobaric thermal instability are not 
gravitationally unstable and therefore do not collapse further once they have 
reached $T_{\rm eq}$, then what actually happens to them? As the surrounding
gas cools towards $T_{\rm eq}$, the overdense regions will expand in order to
stay in pressure equilibrium with the lower density gas, and so their density will 
fall until ultimately they become indistinguishable from the surrounding gas.
The net effect will therefore simply be the injection of some additional kinetic
energy into the gas (as in \citealt{kn02a}). The magnitude of the kinetic energy
input depends on the fraction of the gas processed through overdense regions,
but it will never be larger than the initial thermal energy content of the gas. 
Since the main aim of the simulations presented in this paper is to allow us
to determine how quickly $\mHt$ forms in the {\em absence} of turbulence, 
neglecting this energy input is probably justified. 

\section{Results}
\label{static_results}
Before we began the study of $\mHt$ formation in turbulent, self-gravitating gas that is
described in paper II,  we first spent 
some time examining the much simpler case of gas that was initially at rest. The 
main aim of these runs was to determine how quickly $\mHt$ would form in the absence 
of turbulence (\S~\ref{h2_timescale}), the role that gravity plays in driving this
process (\S~\ref{role_of_grav}) and the morphology of the resulting 
$\mHt$ (\S~\ref{static_morph}), primarily so that we could later compare these 
results with the results of the turbulent runs discussed in paper II. These runs also act
as more comprehensive, albeit less quantifiable, tests of the code than the tests 
discussed in \S~\ref{tests}; thanks to the relative simplicity of the dynamics, 
unphysical behaviour is much easier to spot here than in the turbulent simulations of
paper II.  We also examined the temperature distribution of the gas and
how its maximum and minimum temperatures evolve with time (\S~\ref{static-temp}), 
and determined how sensitive our results are to variations in input parameters such 
as the box size (\S~\ref{static-box}), the amplitude of the initial density perturbations 
(\S~\ref{pert-init}), the magnetic field strength (\S~\ref{field-init}) and the mean
density of the gas in the box (\S~\ref{static_n0}). Parameters for the runs discussed 
here are listed in Table~\ref{static_runs}. 

\subsection{The $\mHt$ formation timescale}
\label{h2_timescale}
To quantify the rate at which $\mHt$ forms in our simulations, there are
various quantities that we might choose to examine. One of the simplest 
is the volume averaged molecular fraction, $\Htvol$, which we can 
calculate simply by summing up the molecular fraction in every grid zone 
and dividing by the total number of zones, i.e.\
\begin{equation}
\Htvol = \frac{1}{N}\sum_{i,j,k} x_{\mHt}(i,j,k),
\end{equation}
where $x_{\mHt}(i,j,k)$ represents the molecular fraction
in the zone with coordinates $(i,j,k)$ and $N$ is the total number of 
zones. However, although $\Htvol$ is very easy to calculate, it is 
of limited use to us, as in an inhomogeneous gas, the volume average
will tend to be dominated by low density regions, while much of the 
actual gas resides in high density regions. Consequently, a more useful
quantity to compute is the mass-weighted mean molecular fraction,
$\Htmass$, which is given by
\begin{equation}
\Htmass = \frac{\sum_{i,j,k} \rho_{\mHt}(i,j,k) \Delta V(i,j,k)}{M_{\rm H}},
\end{equation}
where $\rho_{\mHt}(i,j,k)$ is the mass density of $\mHt$ in zone $(i,j,k)$,
$\Delta V(i,j,k)$ is the volume of zone $(i,j,k)$,
$M_{\rm H}$ is the total mass of hydrogen present in the simulation, and
where we sum over all grid zones. Finally, we might also look at 
the total mass of $\mHt$ in the simulation, $M_{\mHt}$. However, since this 
can be written in terms of $\Htmass$ as
\begin{equation}
 M_{\mHt} = M_{\rm H} \Htmass,
\end{equation}
there is no real benefit to be gained from studying $M_{\mHt}$ rather than 
$\Htmass$. 

To begin our investigation, we first selected a set of parameters that
we could treat as a fiducial example of collapse from static initial
conditions. Our aim here is to discuss the results of this example in 
some detail, and then to explore the effects of varying each of the main 
input parameters by focusing on how the results vary compared to this 
fiducial case. Given the large number of simulations that we have run,
this strategy seems to us to be more efficient than discussing in
detail the results of each individual simulation.

As mentioned previously, the initial density and 
temperature used for our fiducial runs were $n_{\rm i, fid} = 100 \: 
{\rm cm^{-3}}$ and $T_{\rm i, fid} = 1000 \: {\rm K}$ respectively. A 
simple calculation shows that with these parameters, the initial Jeans 
length of the gas is $L_{\rm J} = 47.8 \: {\rm pc}$, which means 
that if we want the gas in our simulation to be gravitationally unstable
at $t = 0.0$, we would need to use a simulation volume of size 
$L \simgreat 50 \: {\rm pc}$. However, since the gas in all of our 
simulations very rapidly cools from $T_{\rm i, fid}$ to a value close
to the equilibrium temperature of the gas,  it is actually sufficient
to choose a value for $L$ that is greater than the value of $L_{\rm J}$ 
in the cool gas. In the simulations which we performed using our
local approximation to treat photodissociation and photoheating,
the equilibrium temperature is initially the same throughout the
simulation volume, and for $n_{\rm i} = 100 \: {\rm cm^{-3}}$ has
a value $T_{\rm eq} \sim 65 \: {\rm K}$. At $n = 100 \: {\rm cm^{-3}}$ and 
$T \sim 65 \: {\rm K}$, we have $L_{\rm J} \sim 12 \: {\rm pc}$ 
for fully atomic gas, and so in these simulations we require 
$L \simgreat 12 \: {\rm pc}$.  In practice, we would like to
ensure that we have multiple Jeans masses of material in 
our simulation volume, and so for our fiducial runs we chose to 
set $L = 40 \: {\rm pc}$.  In the simulations performed using 
the six-ray approximation, the equilibrium temperature is
no longer the same everywhere, as the amount of dust 
shielding, and hence the value of the photoelectric heating 
rate, varies with position within the simulation volume. 
However, even the largest initial value of $T_{\rm eq}$ in
these simulations is smaller than the initial value of 
$T_{\rm eq}$ in the runs performed using the local shielding 
approximation, and so setting $L = 40 \: {\rm pc}$ again 
ensures that multiple Jeans masses of gas will be present.
The effect of varying $L$ is explored in \S~\ref{static-box}. 

The other main parameter that we had to specify for our fiducial runs was
the strength of the initial magnetic field. We chose to set $B_{\rm i, fid}
 = 5.85 \: \mu{\rm G}$, for the reasons discussed in \S~\ref{init_mag}. 
For this combination of $B$, $n$ and $L$, our initial mass-to-flux ratio 
(in units of the critical value) is $\lambda = 8$. 

\begin{figure}[Htb]
\centering
\epsfig{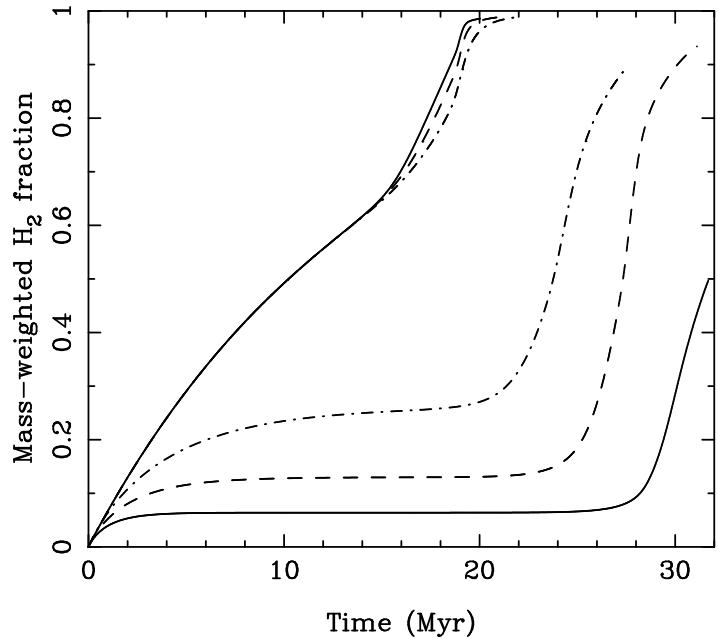}
\caption{Time evolution of $\Htmass$ in several sets of runs.
All of the simulations were performed using our fiducial set of
initial conditions. The three lines in the bottom right of the plot
show the results from runs MS64 (dot-dashed line),  MS128 
(dashed line) and MS256 (solid line), performed using the
local shielding approximation. The three lines in the upper
left indicate the results of runs MS64-RT (dot-dashed line),
MS128-RT (dashed line) and MS256-RT (solid line), 
performed using the six-ray shielding approximation.
The digits in the run name indicate the numerical resolution; 
i.e.\ runs MS64 and MS64-RT were both performed with 
$64^{3}$ zones resolution etc.}
\label{H2-MS}
\end{figure}

In Figure~\ref{H2-MS}, we plot the evolution with time of $\Htmass$ for
our fiducial parameters for six different runs. Runs MS64, MS128 and
MS256 were performed using the local shielding approximation 
and differ solely in the numerical resolution used in the runs. 
Runs MS64-RT, MS128-RT and MS256-RT were performed using
the six-ray approximation and again differ from each other only 
in terms of numerical resolution.

It is immediately apparent from Figure~\ref{H2-MS} that the choice of shielding
approximation makes a significant difference in the outcome of these uniform
density simulations.
Far more $\mHt$ is produced at early times in runs which make use of the 
six-ray  approximation than in those using the local shielding approximation.
In the six-ray runs, the $\mHt$ fraction grows steadily with time, with the 
gas becoming approximately 50\% molecular after $10 \: {\rm Myr}$. The $\mHt$ 
formation rate is independent of the numerical resolution of the simulation until 
$t \sim 15 \: {\rm Myr}$, following which the fastest rate is found in the highest
resolution simulation (although the difference between the three runs is never
great). In the runs performed using the local shielding approximation, on the 
other hand, the growth of the $\mHt$ fraction quickly stalls, and the value of 
$\Htmass$ remains small for at least $20 \: {\rm Myr}$, particularly in the 
higher resolution simulations.
At $t \simgreat 20 \: {\rm Myr}$, however, $\Htmass$ suddenly begins to increase 
rapidly in these runs, and by the end of the simulations, the values of $\Htmass$
obtained are starting to become comparable with those found in the six-ray runs.
Finally, it is clear from Figure~\ref{H2-MS} that the results 
that we obtain when using the local shielding approximation are resolution
dependent: the higher the resolution of the simulation, the less $\mHt$ is produced.
This resolution dependence is a consequence of the fact that in the local
shielding approximation, the self-shielding factor $f_{\rm shield}$ and dust 
extinction $e^{-\tau_{\rm d, 1000}}$ depend explicitly on the physical size of
the grid zone, $\Delta x$, with the result that the equilibrium $\mHt$ fraction, 
$\xhteq$, also explicitly depends on the zone size.

\begin{figure}
\centering
\epsfig{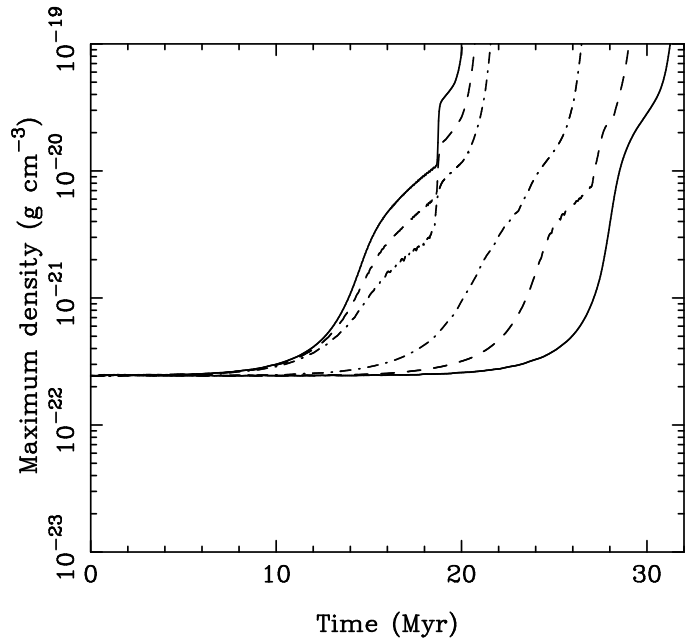}
\caption{Time evolution of the maximum gas density, $\rho_{\rm max}$.
The three lines on the left-hand side of the plot correspond to runs
MS64-RT (dot-dashed line), MS128-RT (dashed line) and MS256-RT
(solid line); the lines on the right-hand side correspond to runs
MS64 (dot-dashed line), MS128 (dashed line) and MS256 (solid line).}
\label{dmax-MS-fid}
\end{figure}

In Figure~\ref{dmax-MS-fid}, we plot the evolution with time of the maximum
gas density, $\rho_{\rm max}$, found in each of these six simulations. We 
again see that the results depend on the choice of shielding approximation:
gas in runs performed using the six-ray approximation collapses 
roughly 5--10~Myr earlier than gas in runs performed using the local 
shielding approximation. Moreover, the two sets of runs display a different
sensitivity to the numerical resolution used: in the six-ray runs, increasing
the resolution causes the collapse to occur slightly earlier, while in the
other set of runs, increasing the resolution {\em delays} the collapse.

\begin{figure}
\centering
\epsfig{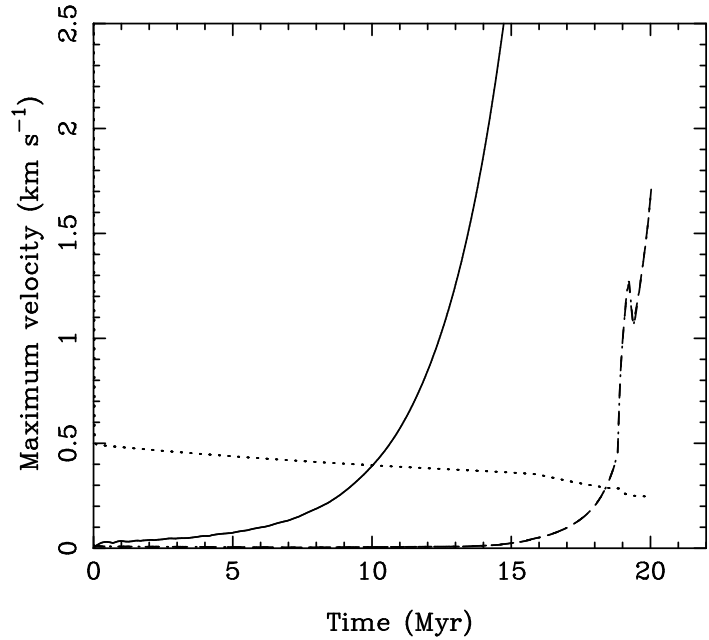}
\caption{Time evolution of the maximum velocity in $256^{3}$ runs
MS256-RT (solid line) and MS256 (dot-dashed line). The evolution
of the sound speed in the densest gas in run MS256 is indicated by
the dotted line; comparable results are found for the sound speed
in run MS256-RT, but we omit them here for clarity. The larger
maximum velocities found in the six-ray run at early times are
caused by pressure waves driven in from the boundaries by 
photoelectric heating there. \label{vmax-fid}}
\end{figure}

One obvious question to ask is why collapse occurs so much earlier in the
six-ray runs than in the runs performed using the local shielding 
approximation. The gas temperatures in the former runs are smaller than
in the latter (see \S~\ref{static-temp} below), and so the gas is more 
gravitationally unstable, but the temperature difference is not great and it would be
surprising if this were to be responsible. In fact, the true culprit is something
quite different. When we use the local shielding approximation, the 
amount of dust shielding, and hence the magnitude of the photoelectric
heating rate, is purely a function of the local gas density. Since the gas 
starts with a nearly uniform density distribution, this means that the heating
rate is the same throughout, meaning that the equilibrium gas temperature
is also almost the same throughout. In our six-ray runs, on the other hand,
this is not the case. Gas near the edges of the simulation volume is shielded
less than gas at the centre and so is heated more. This means that at early
times the equilibrium temperature of the gas near the edges is higher than
that of the gas in the centre. Since the density is still almost the same 
throughout, this results in the development of a pressure gradient, acting
towards the center of the box. Although the resulting flow velocities are 
small, as Figure~\ref{vmax-fid} makes clear, they are large enough to create
non-linear overdensities within about $10 \: {\rm Myr}$, which subsequently
merge and collapse on the free-fall timescale of $~ 5 \: {\rm Myr}$. In comparison, the
density inhomogeneities present in runs MS64, MS128 and MS256 at
$t=10 \: {\rm Myr}$ are very much smaller and so gravitational collapse 
is delayed for a much longer period. 

\begin{figure}
\centering
\epsfig{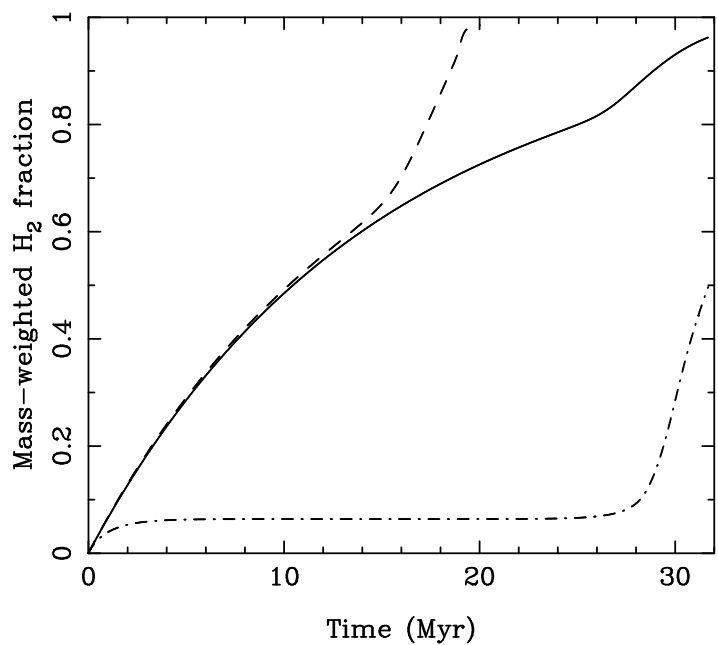}
\caption{Time evolution of $\Htmass$ in $256^{3}$ runs MS256-nr (solid line),
MS256-RT (dashed line) and MS256 (dot-dashed line). Run MS256-nr
was performed with the strength of the radiation field set to zero. The fact 
that {\em more} $\mHt$ forms in run MS256-RT than in run MS256-nr is
attributable to the compression of the cloud that occurs in the former case
due to the temperature and  pressure gradients set up by the non-uniform
heating of the gas.}
\label{H2uv}
\end{figure}

An interesting consequence of this is that $\mHt$ formation can actually
occur {\em faster} when an ultraviolet field is present then when one is
absent, as we show in Figure~\ref{H2uv}. Of course, this is not a new 
discovery: what we are seeing is essentially a form of radiation-driven 
implosion, which is a phenomena that has been discussed by a number
of previous authors, albeit primarily in the context of driving by 
photoionization, rather than by photoelectric heating
\citep[see e.g.][]{swk82,swk84,b89,kdb03}. 

Given the large disparity between the two sets of results, 
which set of results should we believe? In other words, which of the
two approximations that we have used does better at capturing the true
behaviour of the gas? From our discussion in \S~\ref{h2_phdis}  we know that the 
main disadvantages of the six-ray approximation are its lack of angular
resolution and its insensitivity to the effects of velocity differences along the
line of sight. However, at early times in these simulations, the gas distribution
is nearly uniform and the velocities are small, and so this approximation
should give very accurate results. On the other hand, we know that the
local approximation underestimates the true amount of shielding, and from
our results here it is apparent that we underestimate the shielding by quite
a large factor, enough to significantly alter the outcome of the simulations. 
We can therefore conclude that this approximation does not work well for 
treating $\mHt$ photodissociation in this case. Of course, this does
not come as a great surprise, as the physical conditions which best motivate 
the use of a local approximation -- large variations in the gas velocity from 
zone to zone, and significant density inhomogeneities -- are not present at 
early times in these runs. At late times, once a substantial density 
inhomogeneity has developed, the local shielding approximation does a 
much better job of modelling the photodissociation. Moreover, as we
will see in paper II, the local shielding approximation works far better for
treating $\mHt$ photodissociation in supersonically turbulent flow than it 
does here.

As larger density inhomogeneities develop, and in particular as large flow
velocities develop, the accuracy of our six-ray approximation degrades.
However, Figures~\ref{dmax-MS-fid} and \ref{vmax-fid} demonstrate that 
the flow remains subsonic and the density inhomogeneities remain small 
at $t < 10 \: {\rm Myr}$, and so the six-ray approximation should remain
accurate throughout this period. Moreover, at later times enough $\mHt$
is present to provide effective self-shielding for almost all of the gas, and
so we would not expect our results at late times to be particularly sensitive
to the growing inaccuracy of our approximation.

It is also clear that when we use the six-ray approximation, we do a much
better job of modelling the dynamics of the flow, as we recover the large
scale pressure gradient that is missed by the local approximation. It 
should be noted, however, that one reason that we find so much difference
between the dynamics of the two sets of simulations is our choice of 
initial conditions. Because the gas is initially at rest, any small 
velocities, such as those produced by the large-scale pressure gradient,
are significant and can have a large effect on the outcome. In simulations
where the gas does not start at rest (such as the turbulent models presented
in paper II), we would expect to see far less difference in the dynamics.

In view of the higher accuracy of the results from our six-ray runs, we
will focus on these in particular in the remainder of this paper. However,
for the purposes of comparison we include some discussion of results 
from runs performed using the local shielding approximation 

Finally, it should be noted that even in the six-ray runs, it takes 
roughly $10 \: {\rm Myr}$ to form a significant amount of $\mHt$.
This is the most important result of the simulations reported on in
this paper, as this timescale is greater than a gravitational free-fall 
time and is significantly longer than the timescales derived for the 
turbulent cloud models studied in paper II.

\subsection{The role of gravity}
\label{role_of_grav}
From the comparison of Figures~\ref{H2-MS} and \ref{dmax-MS-fid}, it is clear
that the growth in $\Htmass$ at late times in runs MS64, MS128 and MS256
is being driven by the gravitational collapse of the gas: the increase in
density caused by the collapse increases both the $\mHt$ formation rate
and the amount of shielding, allowing $\Htmass$ to rise rapidly. Similarly,
the growth of $\Htmass$ at late times in the six-ray runs appears to be
accelerated by the collapse of the gas.

\begin{figure}
\centering
\epsfig{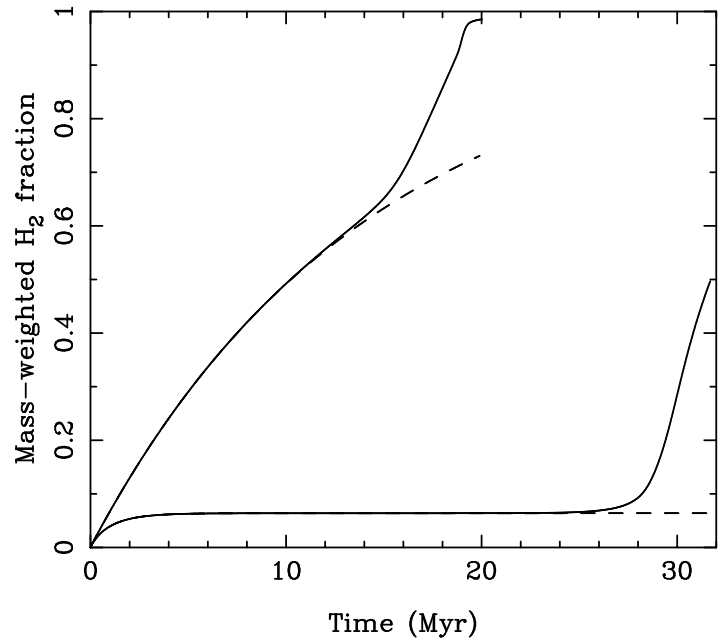}
\caption{Time evolution of $\Htmass$ in $256^{3}$ runs performed
with and without self-gravity. The two lines on the left-hand side of 
the plot indicate the results of runs MS256-RT and MS256-RT-ng
(solid and dashed lines respectively). The two lines on the right-hand
side indicate the results of runs MS256 (solid line) and MS256-ng 
(dashed line). Runs MS256-RT and MS256 include the effects of 
self-gravity, while runs MS256-RT-ng and MS256-ng do not.}
\label{H2ng}
\end{figure}

To confirm this, we performed a set of runs in which we did not
include the effects of self-gravity.  These runs, designated MS64-ng, MS128-ng,
MS256-ng and MS256-RT-ng, used the same input parameters as runs
MS64, MS128, MS256 and MS256-RT respectively. The evolution of $\Htmass$ 
in runs MS256, MS256-ng, MS256-RT and MS256-RT-ng is shown in
Figure~\ref{H2ng}. At $t < 15 \: {\rm Myr}$ in runs MS256-RT and MS256-RT-ng,
and at $t < 27 \: {\rm Myr}$ in runs MS256 and MS256-ng, the results of the
simulations agree, demonstrating that at early times the effects of self-gravity
are unimportant. However, it is also clear that at later times, the simulation 
results diverge, with gravitational collapse providing a distinct boost to the 
$\mHt$ formation rate in the self-gravitating simulations. 

Since gravity plays an important role in the outcome of our simulations,
it is also natural to investigate how well we are resolving the gravitational
collapse of the gas. Specifically, we would like to know at what point the
Truelove criterion is first violated and how much of the $\mHt$ formation 
that we see in our runs occurs before this point.

Using the results of our simulations, it is relatively easy to determine when
the Truelove criterion is first violated for any given model, and what the value 
of $\Htmass$ is at that time. The values we obtain for each of our runs are
summarized in Table~\ref{xh2_at_end_stat}. In runs performed using the
local shielding approximation, most $\mHt$ formation occurs {\em after} 
the run has violated the Truelove criterion. On the other hand, in runs that
use the six-ray approximation, the majority of the $\mHt$ forms before
the Truelove criterion is violated. 

\begin{figure}
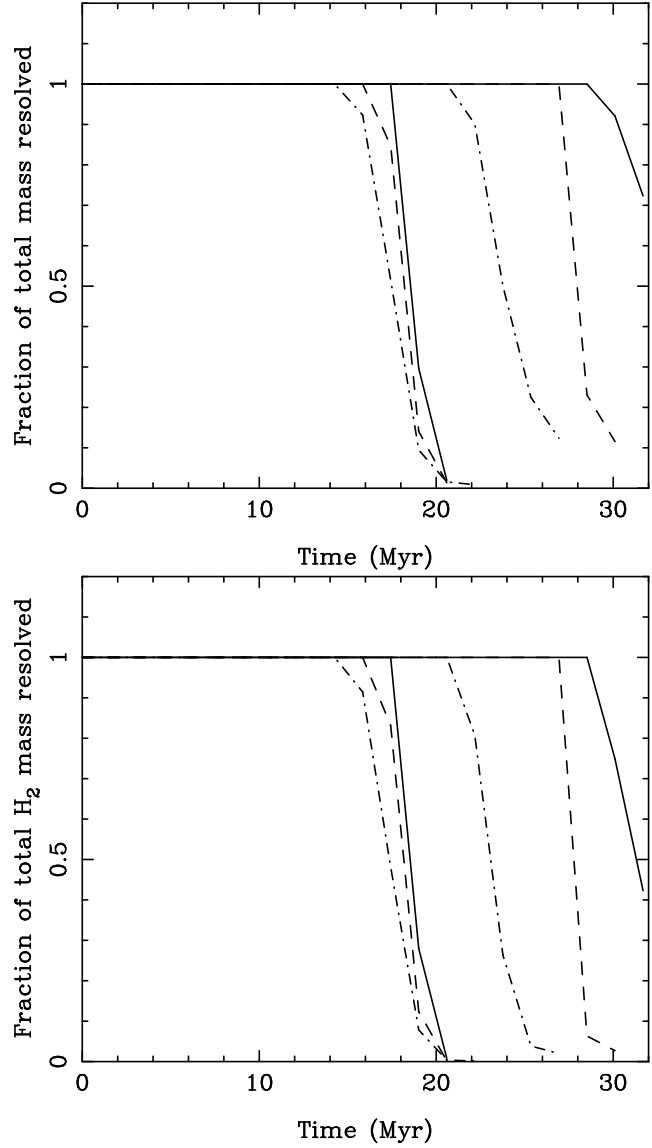

\centering
\epsfig{figure=fig10a.eps,width=18pc,angle=270,clip=}
\epsfig{figure=fig10b.eps,width=18pc,angle=270,clip=}
\caption{(a) Fraction of the total gas mass situated in resolved regions (i.e.\ in
zones which satisfy the Truelove criterion) plotted as a function of time for 
runs MS64-RT (left-hand dot-dashed line), MS128-RT (left-hand dashed line),
MS256-RT (left-hand solid line),  MS64 (right-hand dot-dashed line), MS128 
(right-hand dashed line) and MS256 (right-hand solid line). 
(b) As (a), but for the fraction of the total mass of molecular gas.
\label{fres-MS-fid}}
\end{figure}

Of course, the fact that we no longer properly resolve gravitational collapse
in dense gas once the Truelove criterion is violated does not necessarily invalidate 
all of the subsequent results from our simulations. For instance, if only a few percent of
the total $\mHt$ were found in unresolved regions, then the fact that we are no 
longer able to resolve these regions properly would be likely to have only a minor 
effect on $\Htmass$. Conversely, if most of the $\mHt$ were in unresolved regions,
then our results for $\Htmass$ would be trustworthy only up to the point at which
the Truelove criterion was violated. In order to quantify how much gas and how much 
$\mHt$ ends up in unresolved regions in our simulations, we examined
intermediate output dumps of density, internal energy and $\mHt$ fraction 
from each of our fiducial runs, and for each dump determined which zones, 
if any, were unresolved. We then computed $f_{\rm res}$, the fraction of the total
gas mass in resolved regions, and $f_{\rm res, \mHt}$ the fraction of the 
total $\mHt$ mass in the same resolved regions, for each output time. The 
resulting values of $f_{\rm res}$ and $f_{\rm res, \mHt}$ are plotted in 
figures~\ref{fres-MS-fid}a and \ref{fres-MS-fid}b respectively.
It is clear from these figures that once the Truelove criterion is violated,
the majority of the gas and of the $\mHt$ soon ends up being located 
in unresolved regions, suggesting that our simulation results are only 
trustworthy up to this point.

We obtain similar results if we perform the same analysis for any of the
other runs in Table~\ref{static_runs} which violate the Truelove criterion. 
We should therefore not place too much weight on results coming from very
late times in any of these simulations. Nevertheless, it should be clear
that these concerns do not affect the basic timescale for $\mHt$ formation 
that we find from our simulations and so do not undermine our main results.

\clearpage

\subsection{${\mathbf H_{2}}$ morphology}
\label{static_morph}
As noted above, even in our highest resolution fiducial simulations, 
most of the gas is located in unresolved regions at the end of the run. It 
is therefore not particularly enlightening to examine the morphology of either 
the gas or the $\mHt$ at the end of the run, as both will be inaccurate in
ways not easily quantified. Moreover, in a real molecular cloud,
we would expect star formation to occur rapidly once gas has collapsed
to high densities, and since we do not include any feedback in these 
simulations from effects such as protostellar outflows, our simulations are
missing some of the physics necessary to accurately model the 
morphology at late times. 

\begin{figure}
\centering
\epsfig{figure=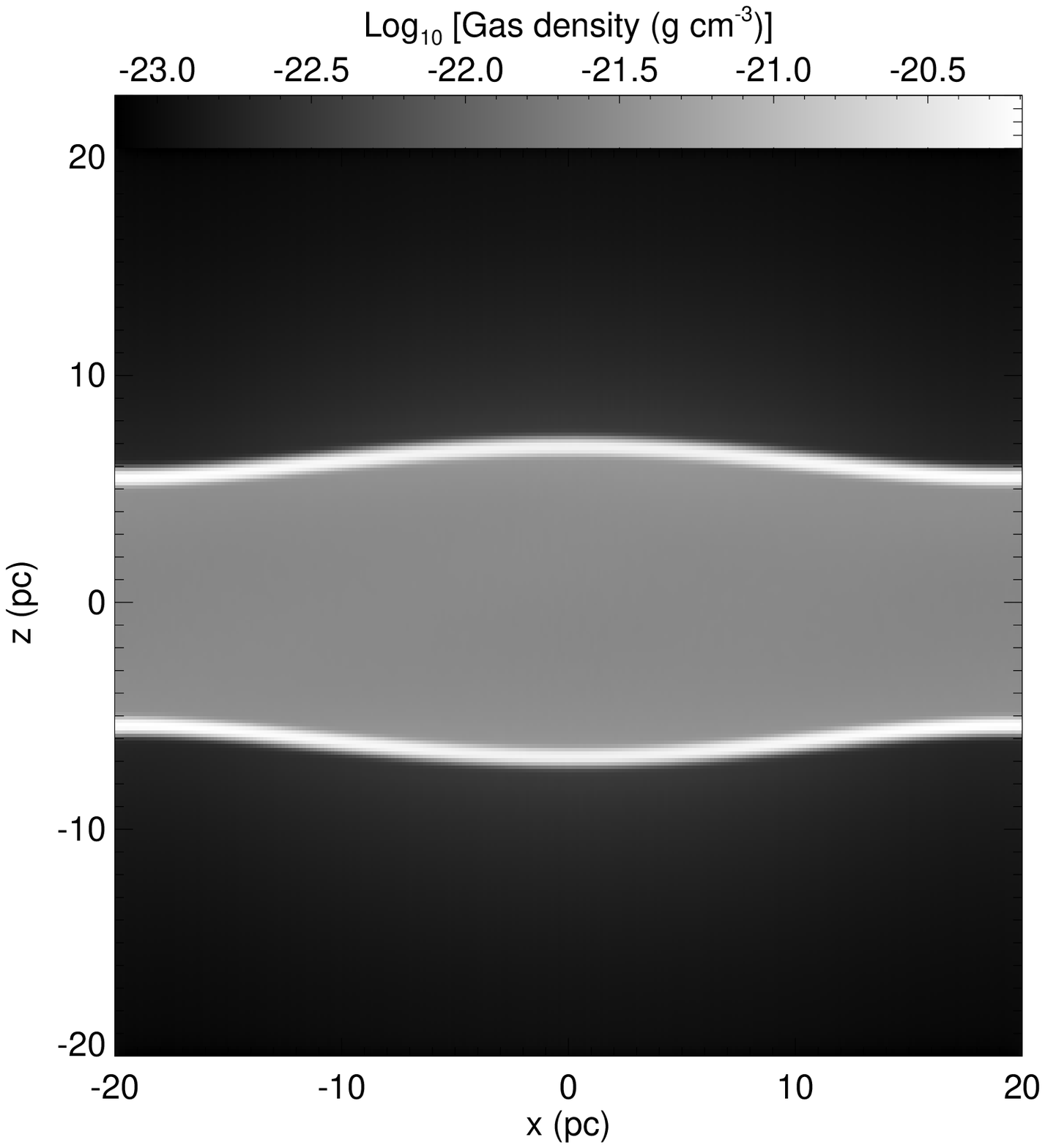,width=20pc,angle=0,clip=}
\epsfig{figure=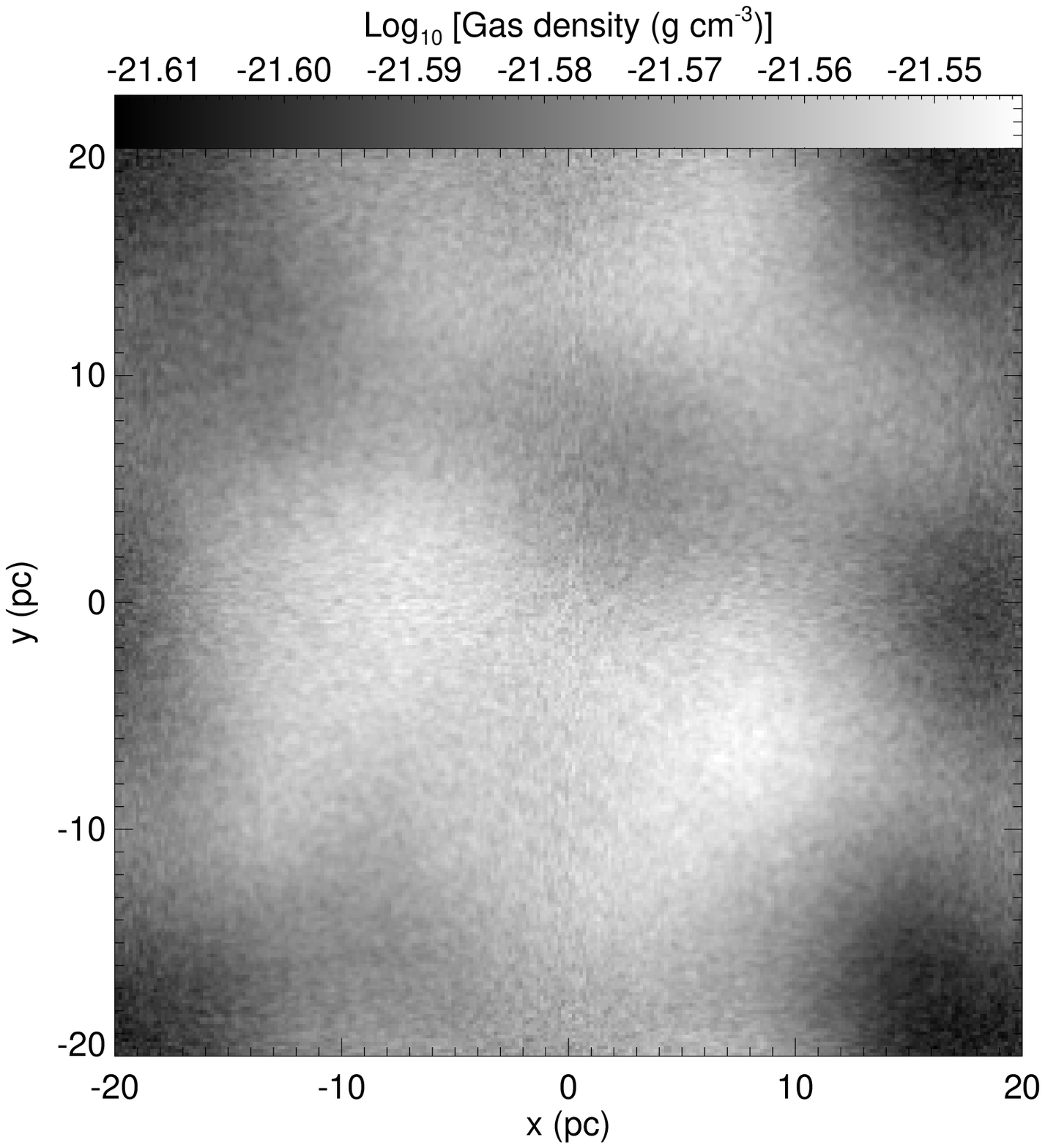,width=20pc,angle=0,clip=}
\figcaption{(a) Slice in the $x$-$z$ plane through the density field in 
$256^{3}$ zone run MS256-RT at a time $t = 17.4 \: {\rm Myr}$. 
The slice is centered on the midpoint of the simulation volume.
(b) As (a), but for a slice in the $x$-$y$ plane. Note the change
in density scale compared to figure (a). \label{den-MS-RT}}
\end{figure}

\begin{figure}
\centering
\epsfig{figure=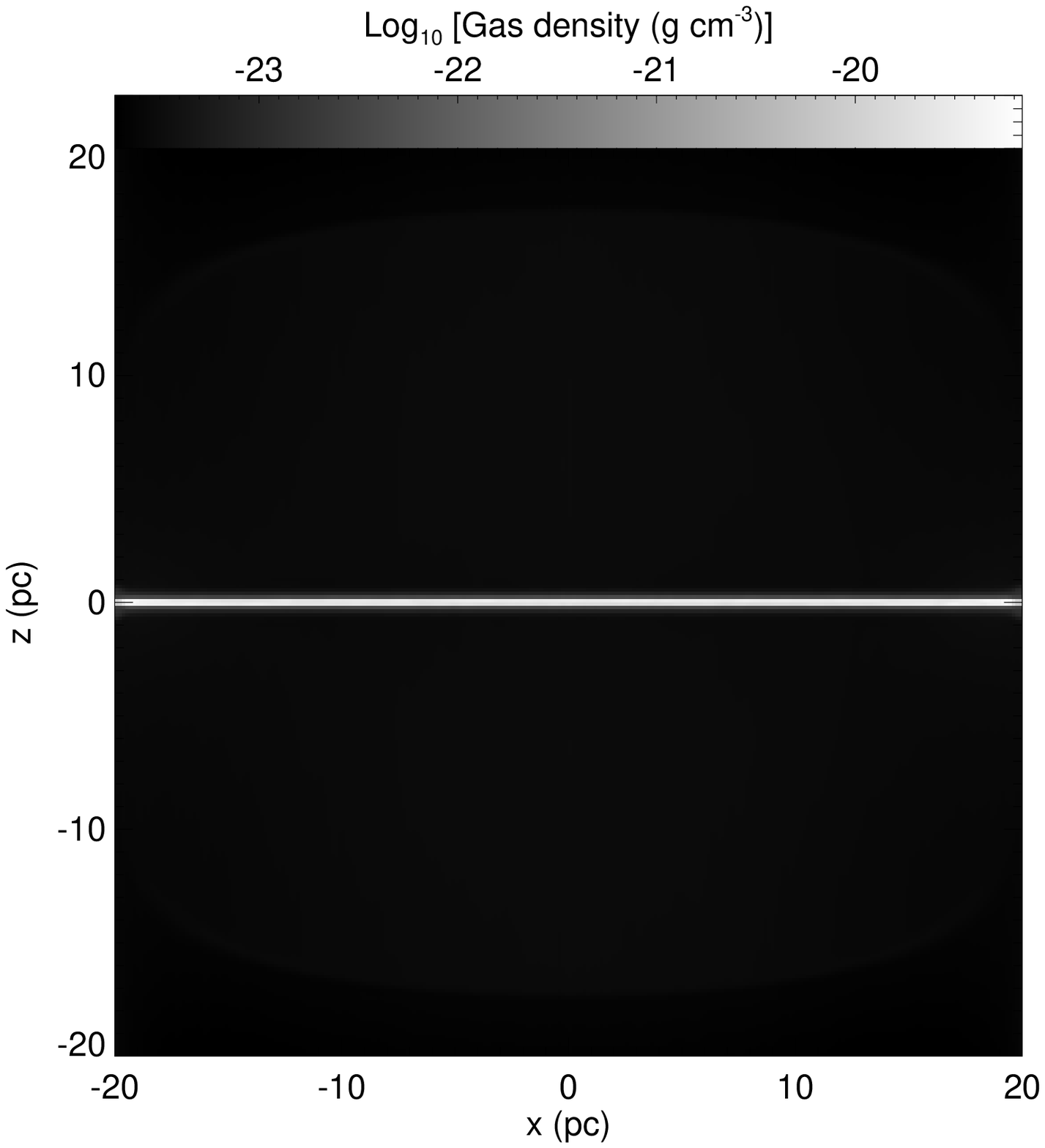,width=20pc,angle=0,clip=}
\figcaption{Slice in the $x$-$z$ plane through the density field in 
$256^{3}$ zone run MS256-RT at a time $t = 20.6 \: {\rm Myr}$.
The slice is centered on the midpoint of the simulation volume.
\label{den-post-collapse}}
\end{figure}

On the other hand, what we can do with some degree of confidence is to 
examine the morphology of the gas at a time shortly before the Truelove
criterion is first violated, when star formation has presumably 
yet to occur. The morphology of the cloud at this time depends upon 
our choice of shielding approximation. In the runs performed with the
six-ray approximation, a thick slab of gas has formed, oriented
perpendicularly to the magnetic field. The density within most of this slab 
is only slightly elevated over the initial density of the gas, but the 
thin layer of gas bounding the slab shows a large overdensity. Plots of
the gas density in the x-z and x-y planes in run MS256-RT at time 
$t = 17.4 \: {\rm Myr}$ are shown in Figure~\ref{den-MS-RT}. The gas 
in this simulation is gravitationally collapsing in a direction parallel 
to the magnetic field lines, with the result that shortly after the time of 
this output dump, the two large overdensities merge as the gas forms 
a thin dense sheet located at $z = 0$ (see Figure~\ref{den-post-collapse}). 
It is reasonable to expect that this sheet would then  fragment into a 
number of filaments and cores \citep{lar85}, but by this point in our 
simulation, the sheet itself was unresolved and so we were unable 
to follow the further dynamical evolution of the simulated cloud. 

\begin{figure}
\centering
\epsfig{figure=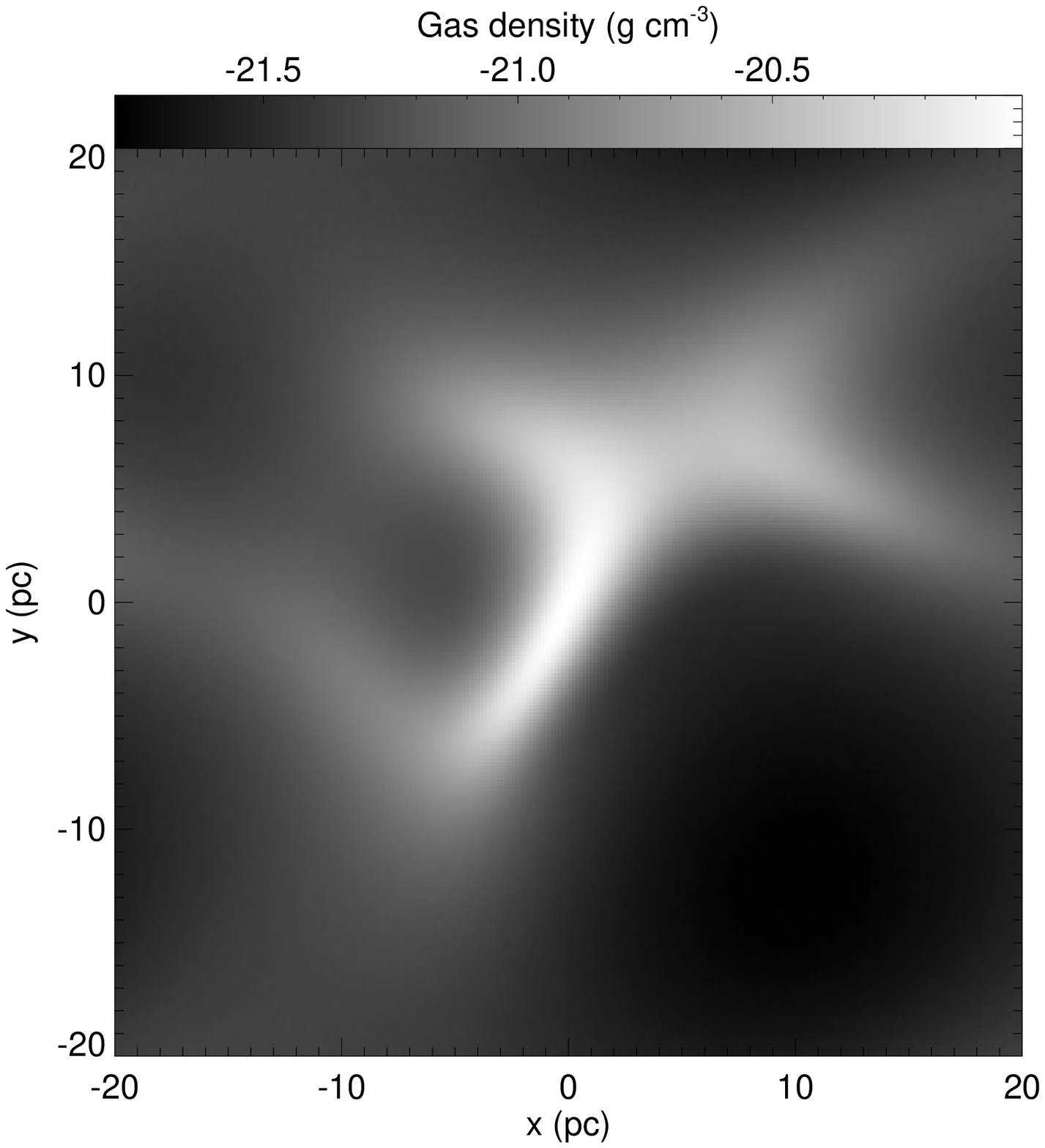,width=20pc,angle=0,clip=}
\epsfig{figure=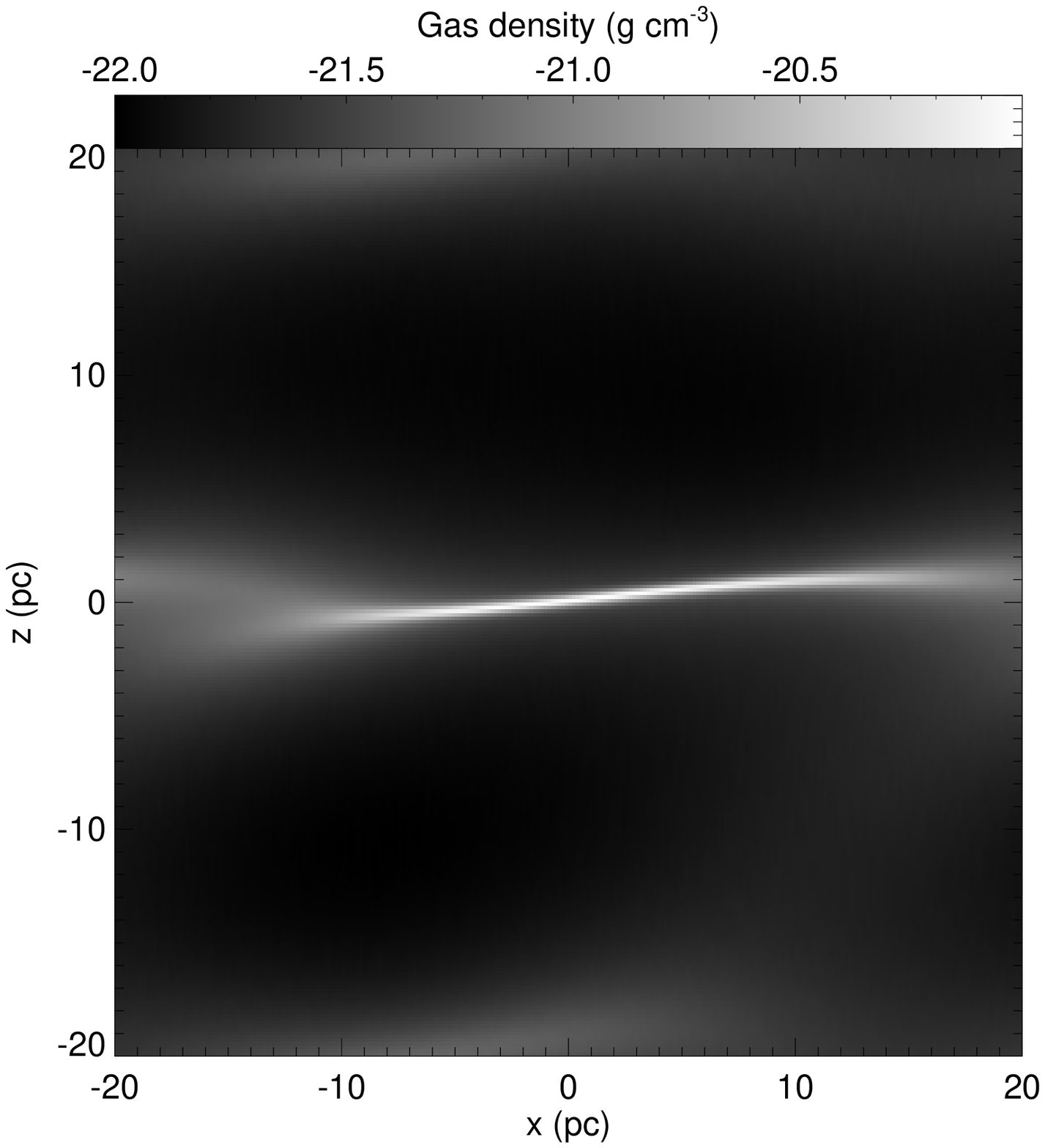,width=20pc,angle=0,clip=}
\figcaption{(a) Slice in the $x$-$y$ plane through the density field in 
$256^{3}$ zone run MS256 at a time $t = 28.5 \: {\rm Myr}$. The periodic 
boundary conditions employed in the simulation have allowed us to shift 
the image so that the densest region lies at the centre of the figure.  
(b) As (a), but for a slice in the $x$-$z$ plane. \label{den-MS-fid}}
\end{figure}

The morphology of the gas in run MS256 at an output time shortly 
before the Truelove criterion is violated is also sheet-like, as we 
can see from Figure~\ref{den-MS-fid}. However, the width of the
sheet in run MS256 is much smaller, and the density cross-section
is rather different, with the maximum density being found at the
midplane, rather than at the top and bottom edges. As we have
already discussed, these differences stem from the differing
dynamics of the flow: in this run, photoelectric heating by the 
ultraviolet background is initially uniform, and so there is no 
large scale temperature or pressure gradient. The fact that in 
both runs gravitational collapse produces a sheet-like structure
is a consequence of the presence of the magnetic field,
which strongly suppresses gas flow perpendicular to the 
field, but which does not affect flow in the $z$ direction, 
parallel to the field.

\begin{figure}
\centering
\epsfig{figure=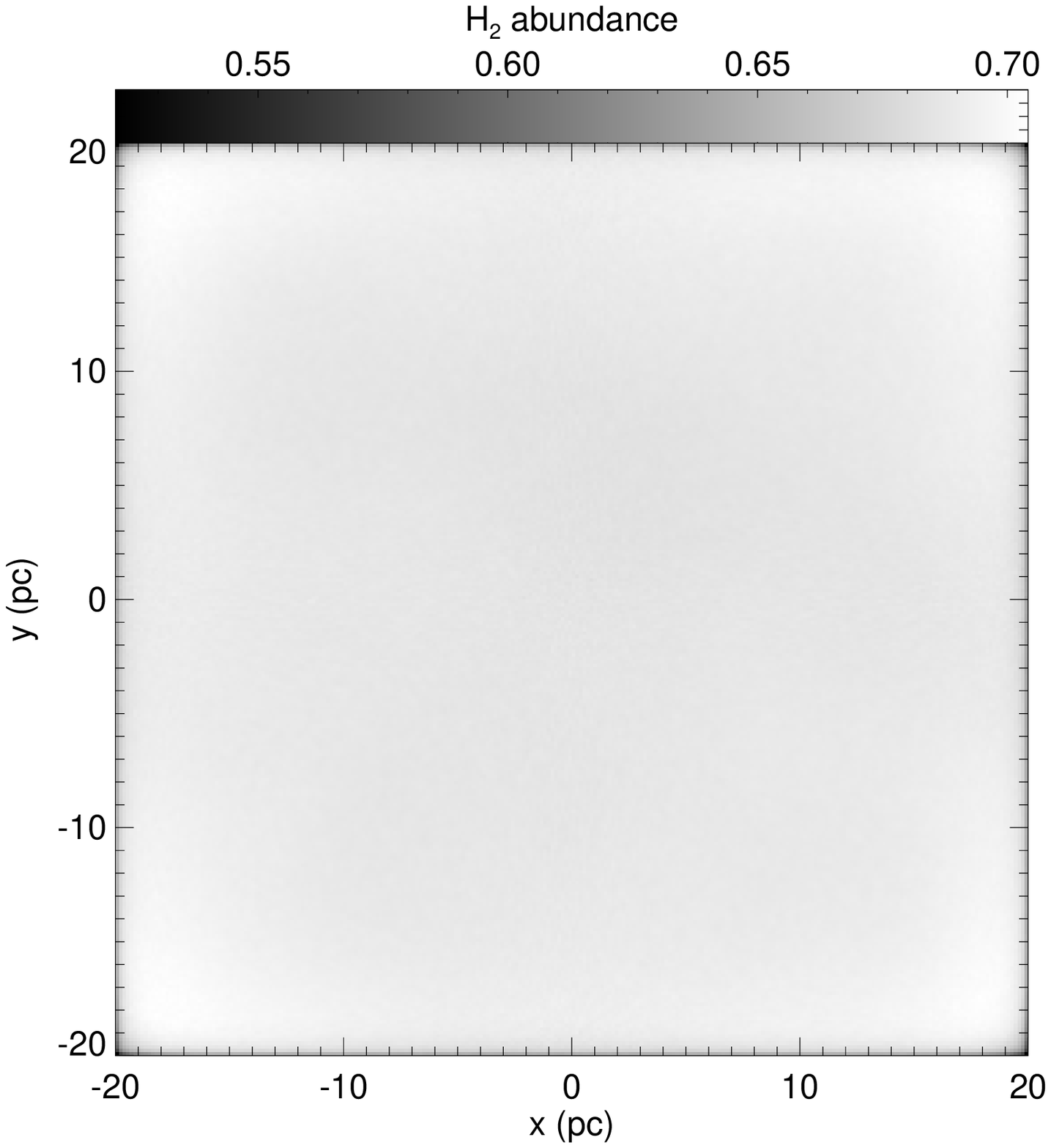,width=20pc,angle=0,clip=}
\epsfig{figure=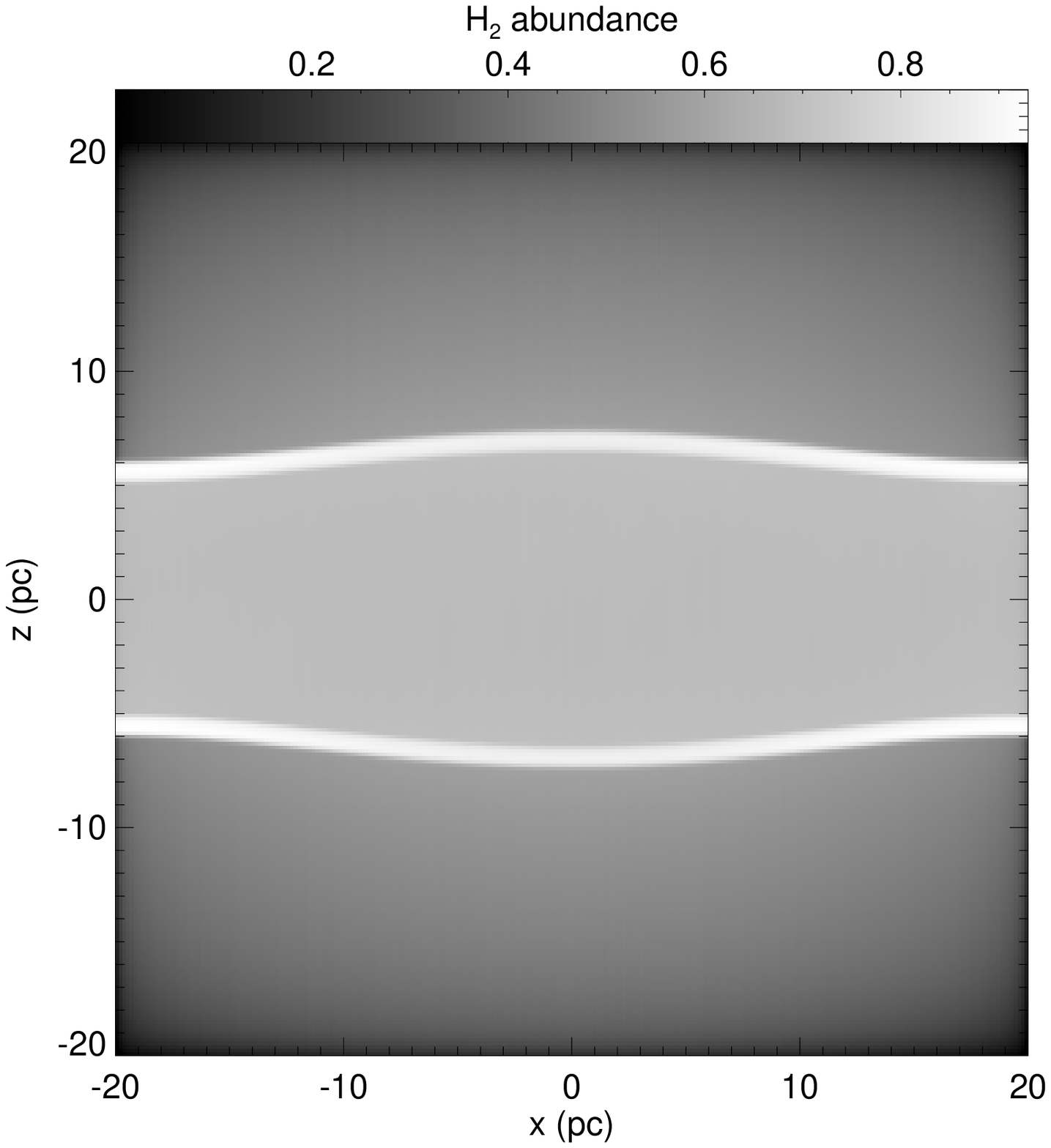,width=20pc,angle=0,clip=}
\caption{(a) Slice in the $x$-$y$ plane through $256^{3}$ zone run 
MS256-RT at time $t = 17.4 \: {\rm Myr}$ showing the spatial variation 
of the $\mHt$ fraction. The slice is centered on the midpoint of the
simulation volume.
(b) As (a), but for a slice in the $x$-$z$ plane. \label{H2-slice-RT}}
\end{figure}

\begin{figure}
\centering
\epsfig{figure=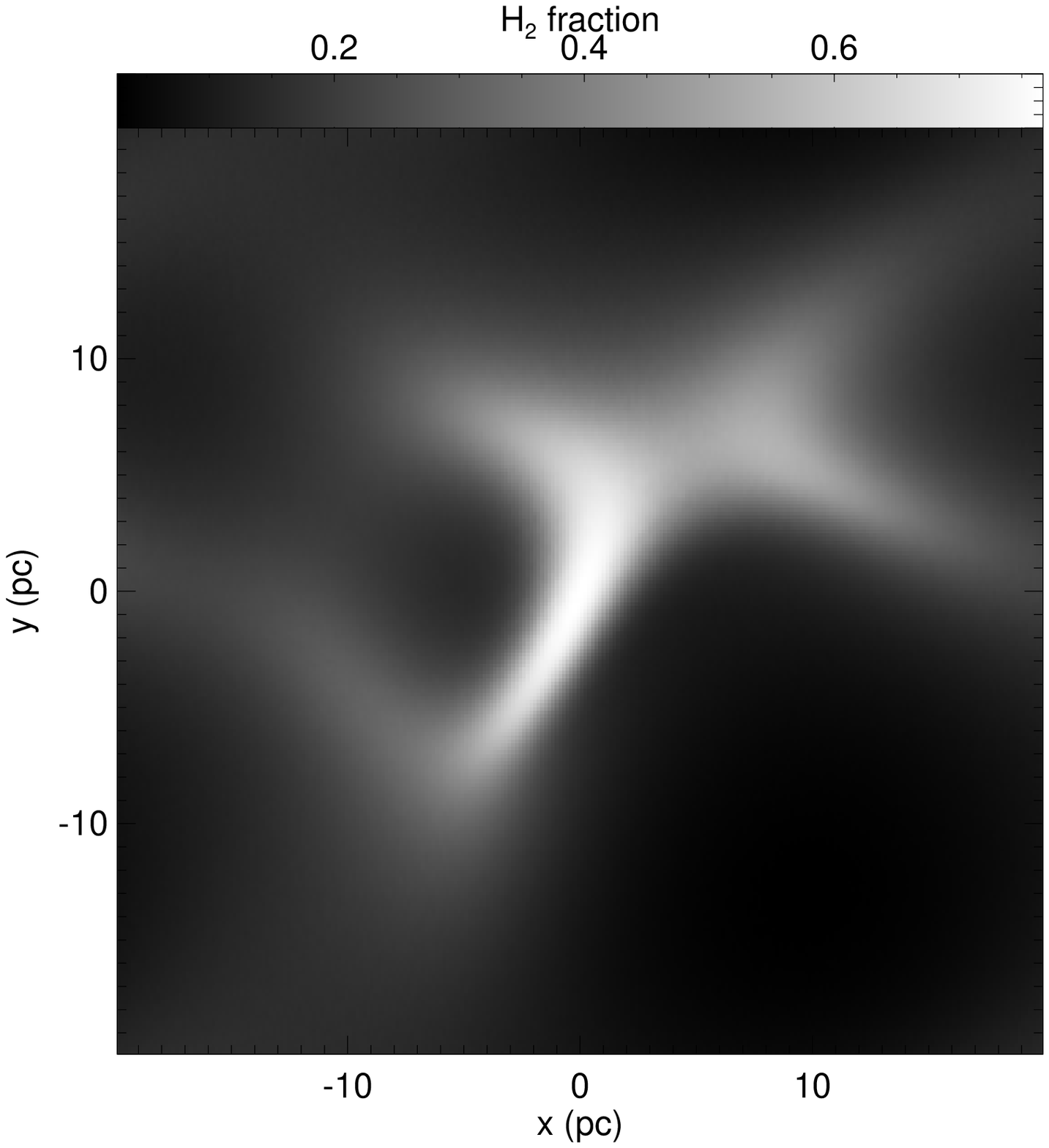,width=20pc,angle=0,clip=}
\epsfig{figure=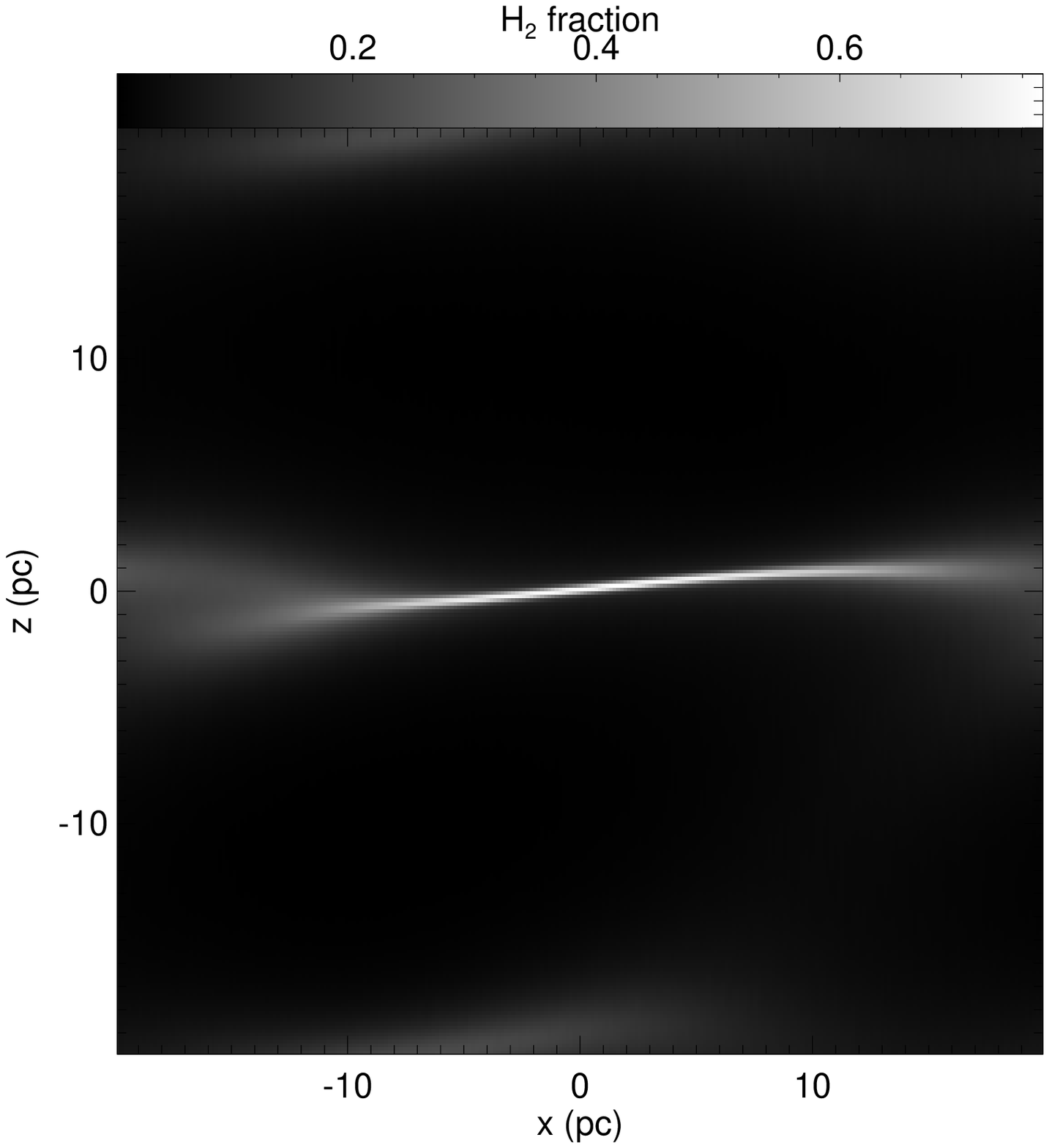,width=20pc,angle=0,clip=}
\caption{(a) Slice in the $x$-$y$ plane through $256^{3}$ zone run 
MS256 at time $t = 28.5 \: {\rm Myr}$ showing the spatial variation 
of the $\mHt$ fraction. The peak value in this slice is 0.77. 
(b) As (a), but for a slice in the $x$-$z$ plane. \label{H2-slice}}
\end{figure}

The spatial distribution of $\mHt$ in these simulations is also of interest.
In Figure~\ref{H2-slice-RT} we show how the $\mHt$ fraction varies with
position in run MS256-RT. Figure~\ref{H2-slice} gives the corresponding
results for run MS256. In both runs, there is an obvious correlation 
between the gas density and the $\mHt$ fraction, with the largest 
molecular fractions being found in the densest gas. However, this
correlation is much stronger in run MS256 than in run MS256-RT, 
because in the former run, the amount of shielding is purely a function
of the local gas density, so $\mHt$ both forms faster and is photodissociated
more slowly in denser gas. In run MS256-RT, the correlation between
local gas density and photodissociation rate is much weaker (although
not completely absent).

\begin{figure}
\centering
\epsfig{figure=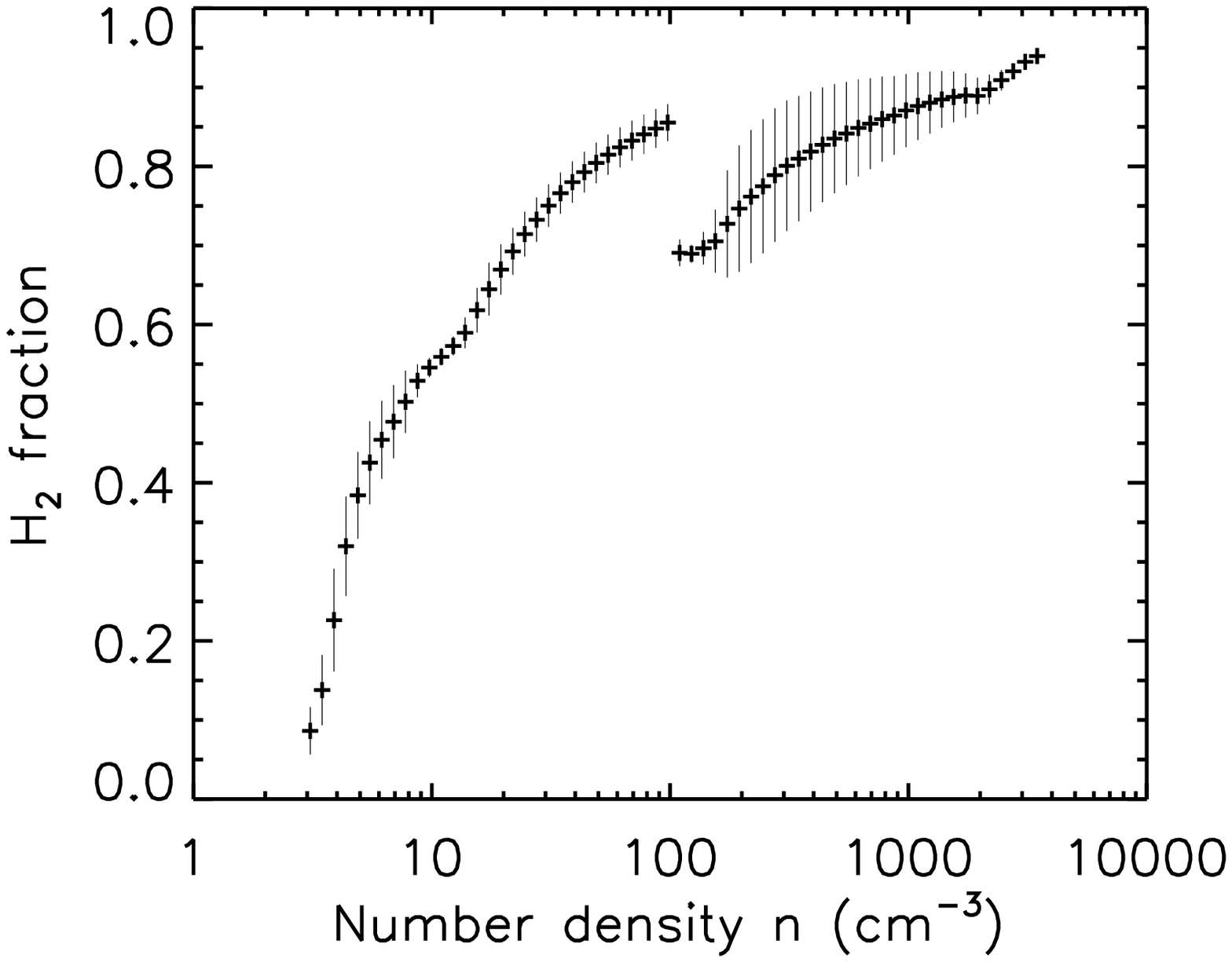,width=25pc,angle=0,clip=}
\epsfig{figure=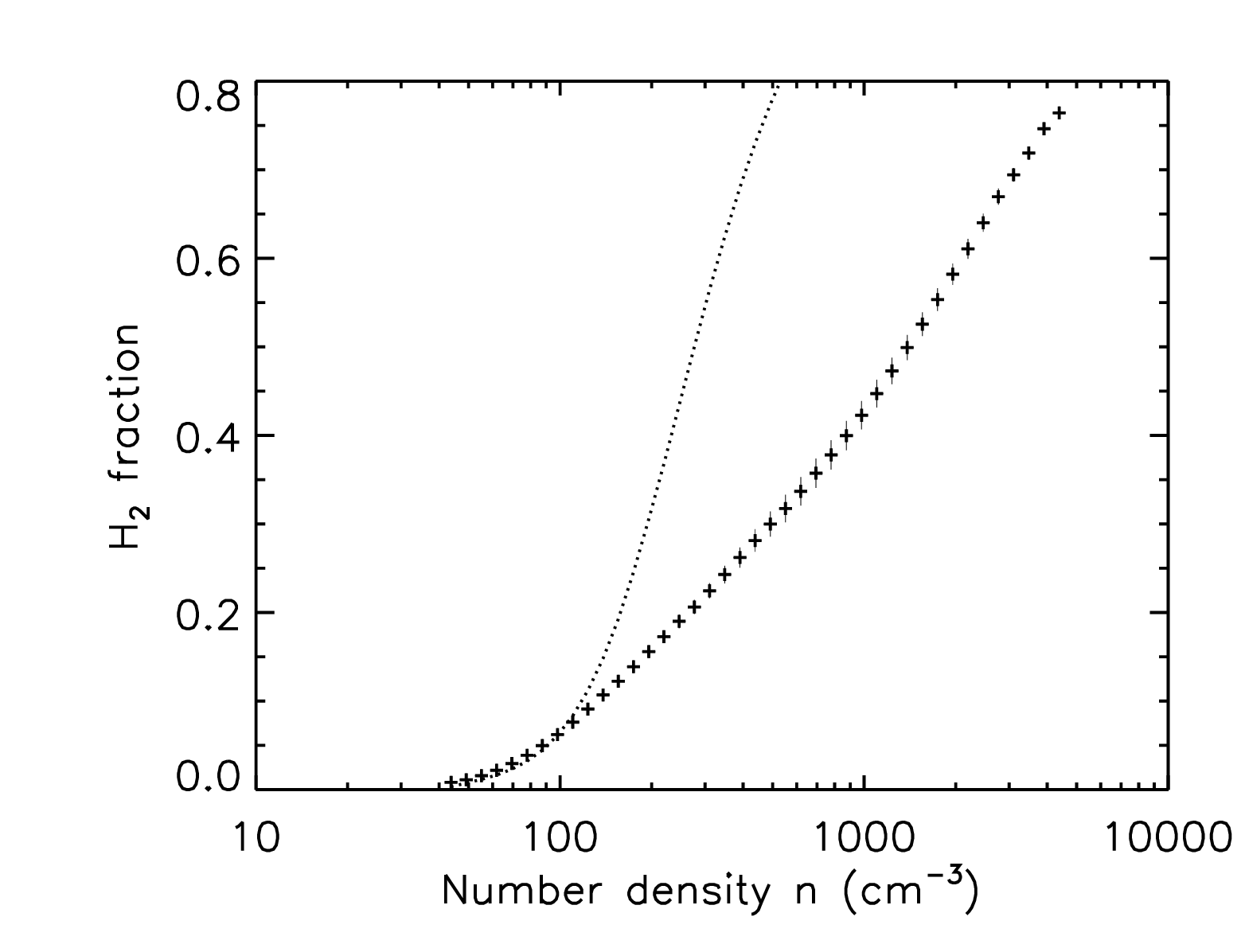,width=25pc,angle=0,clip=}
\caption{(a) $\mHt$ fraction as a function of the number density of the
gas (crosses) for $256^{3}$ zone run MS256-RT at time $t = 17.4 \: {\rm Myr}$. 
To compute these values, we binned the data by number density, using bins of 
width 0.05 dex, and computed the mean value of $x_{\mHt}$ for each bin. The 
standard deviation in the value of $x_{\mHt}$ in each bin is also indicated 
(where it exceeds the size of the symbol). 
(b) As (a), but for run MS256 at time $t = 28.5 \: {\rm Myr}$. In this figure, 
we also indicate the mean value of $\xhteq$ in each bin (dotted line).
\label{xh2vn-RT}}
\end{figure}

To better quantify the relationship between gas density and $\mHt$
fraction, we examined how the mean $\mHt$ fraction varied as a
function of $n$ in each simulation. To do this, we computed 
$x_{\mHt}$ and $n$ for each grid zone, and then binned the data
by number density, using bins of width 0.05 dex. We then computed
the mean and standard deviation for $x_{\mHt}$ for each bin.
The resulting values are plotted is Figure~\ref{xh2vn-RT}a for run 
MS256-RT and Figure~\ref{xh2vn-RT}b for run MS256. Although 
the mean values that we compute here are volume weighted, we 
would not expect the mass weighted values to differ greatly, since 
the narrow width of our density bins means that there is little 
variation in the gas mass from zone to zone within a given bin.

An immediately obvious feature of Figure~\ref{xh2vn-RT}a is the discontinuity
at $n = 100 \: {\rm cm^{-3}}$. This feature is a result of the fact that some of 
the $\mHt$ that forms in the overdense regions bounding the collapsing slab
is left behind by the collapse, and so finds itself ultimately in a region with
$n < 100 \: {\rm cm^{-3}}$. In other words, most of the gas at 
$n < 100 \: {\rm cm^{-3}}$ did not form in situ, but instead formed at higher 
densities and has been transported to lower densities. On the other hand, 
gas with $n > 100 \: {\rm cm^{-3}}$ resides in the dense slab, and much of
this gas has yet to be greatly affected by the collapse (as is apparent from
Figure~\ref{H2-slice-RT}). 

We do not see a comparable feature in Figure~\ref{xh2vn-RT}b. This is
not unexpected, as the dynamics of the flow are quite different in this case,
with collapse happening far more slowly. Moreover, the local shielding
approximation used in run MS256 leads naturally to a tight relationship 
between density and $\mHt$ fraction at low densities, as the gas is close
to photodissociation equilibrium. This can be seen clearly in 
Figure~\ref{xh2vn-RT}b if one compare the simulation results with the
curve indicating how $\xhteq$ varies with density. The latter curve was
constructed by first computing $\xhteq$ for each grid zone (assuming
shielding to be described by our local shielding approximation)
and then binning and averaging these values using the same procedure 
as for $x_{\mHt}$. Note that in run MS256-RT we expect $\xhteq$ to vary
with both density {\em and} with position within the simulation volume, 
and so we have not constructed a similar curve for this run.

\begin{figure}
\centering
\epsfig{figure=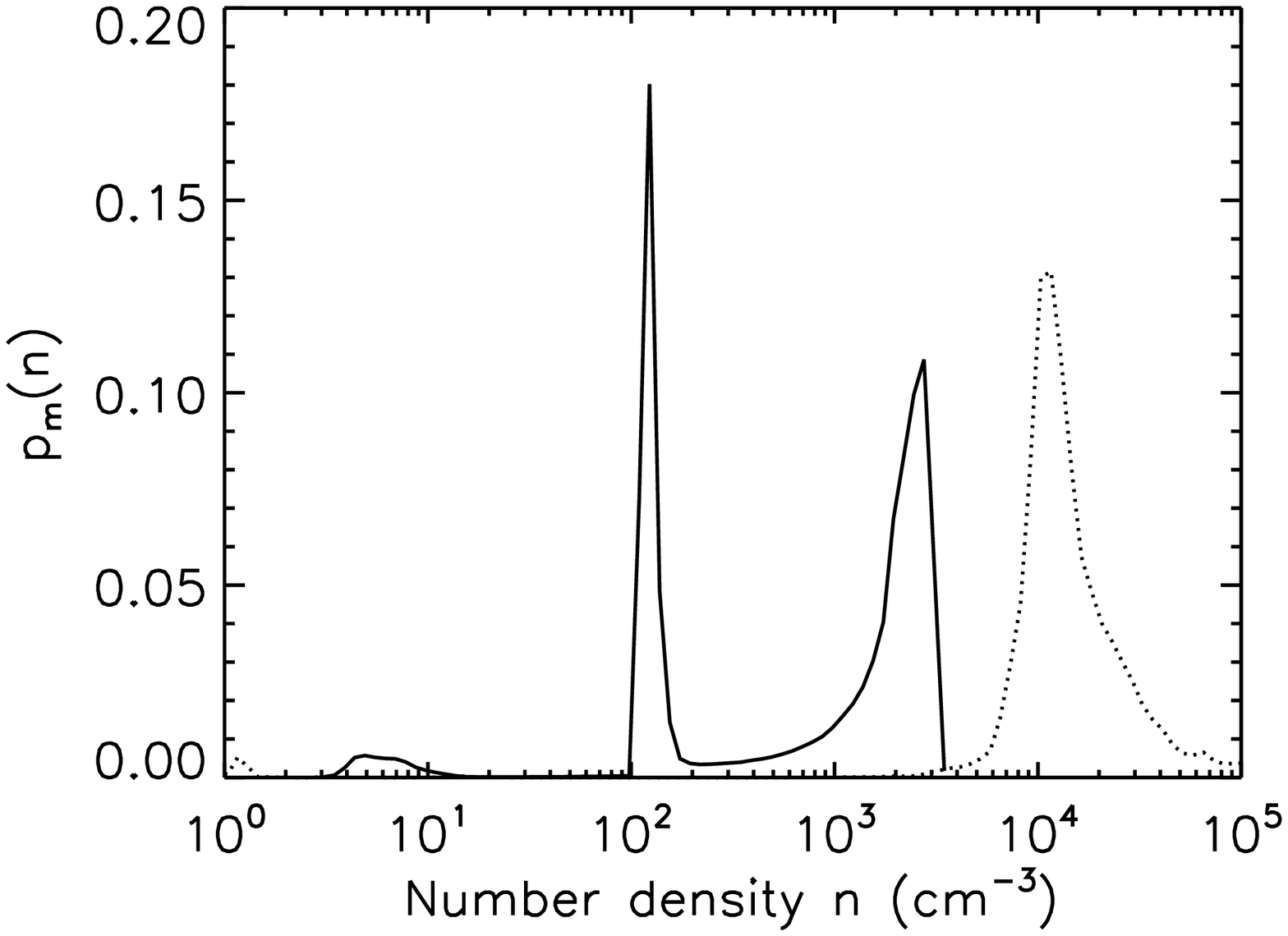,width=25pc,angle=0,clip=}
\epsfig{figure=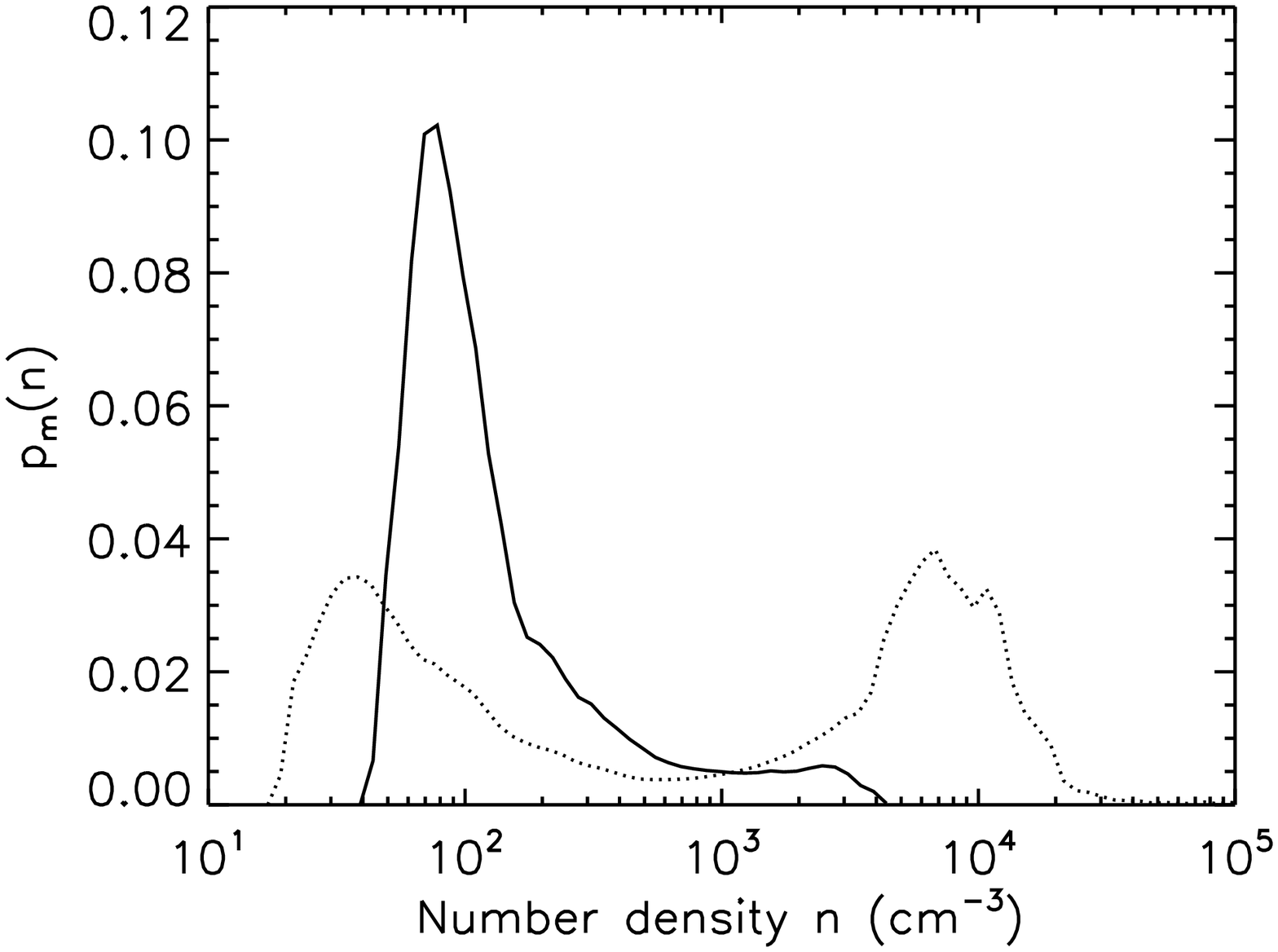,width=25pc,angle=0,clip=}
\caption{(a) Mass-weighted density PDF, $p_{m}(n)$, in 
$256^{3}$ zone run MS256-RT at times  $t = 17.4 \: {\rm Myr}$ 
(solid line) and $t = 20.6 \: {\rm Myr}$ (dotted line). Note that
gas at $n >  5500 \: {\rm cm^{-3}}$ is not properly resolved by 
the code, and so the gas distribution at $t = 20.6 \: {\rm Myr}$
may not be quantitatively accurate (although it should be
qualitatively correct). 
(b) As (a), but for run MS256 at output times $t = 28.5 \: {\rm Myr}$ 
(solid line) and $t = 31.7 \: {\rm Myr}$ (dotted line).
\label{den-PDF}}
\end{figure}

It is also of interest to ask how the $\mHt$ {\em mass} is 
distributed as a function of density.  The plots given above show
how the $\mHt$ {\em fraction} varies with density, but contain no 
information on how much gas is to be found at any given density. 
This information is given by the mass-weighted density probability
distribution function (PDF), which we have plotted for runs MS256-RT 
and MS256 in Figures~\ref{den-PDF}a and \ref{den-PDF}b respectively. 

\begin{figure}
\centering
\epsfig{figure=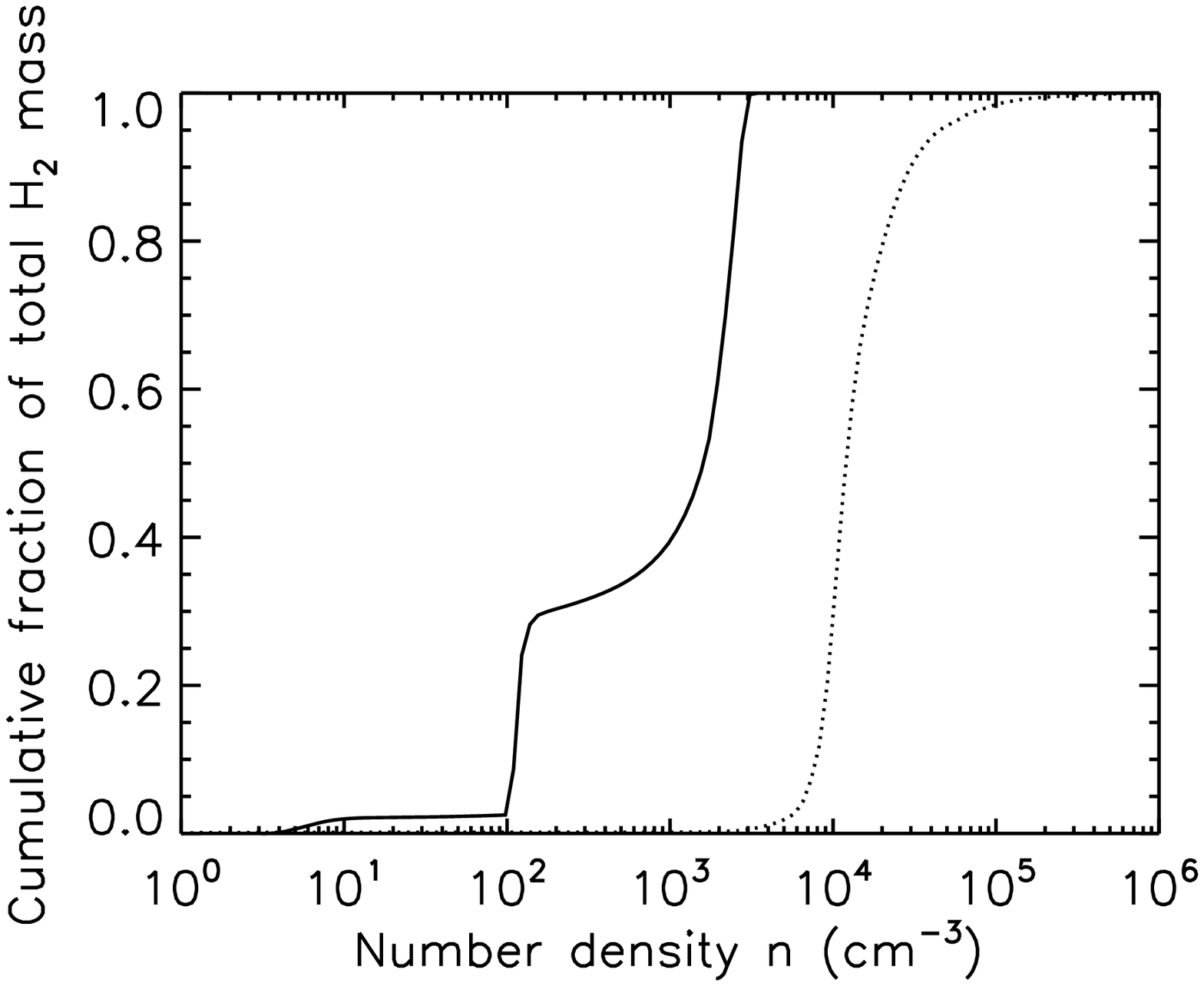,width=25pc,angle=0,clip=}
\epsfig{figure=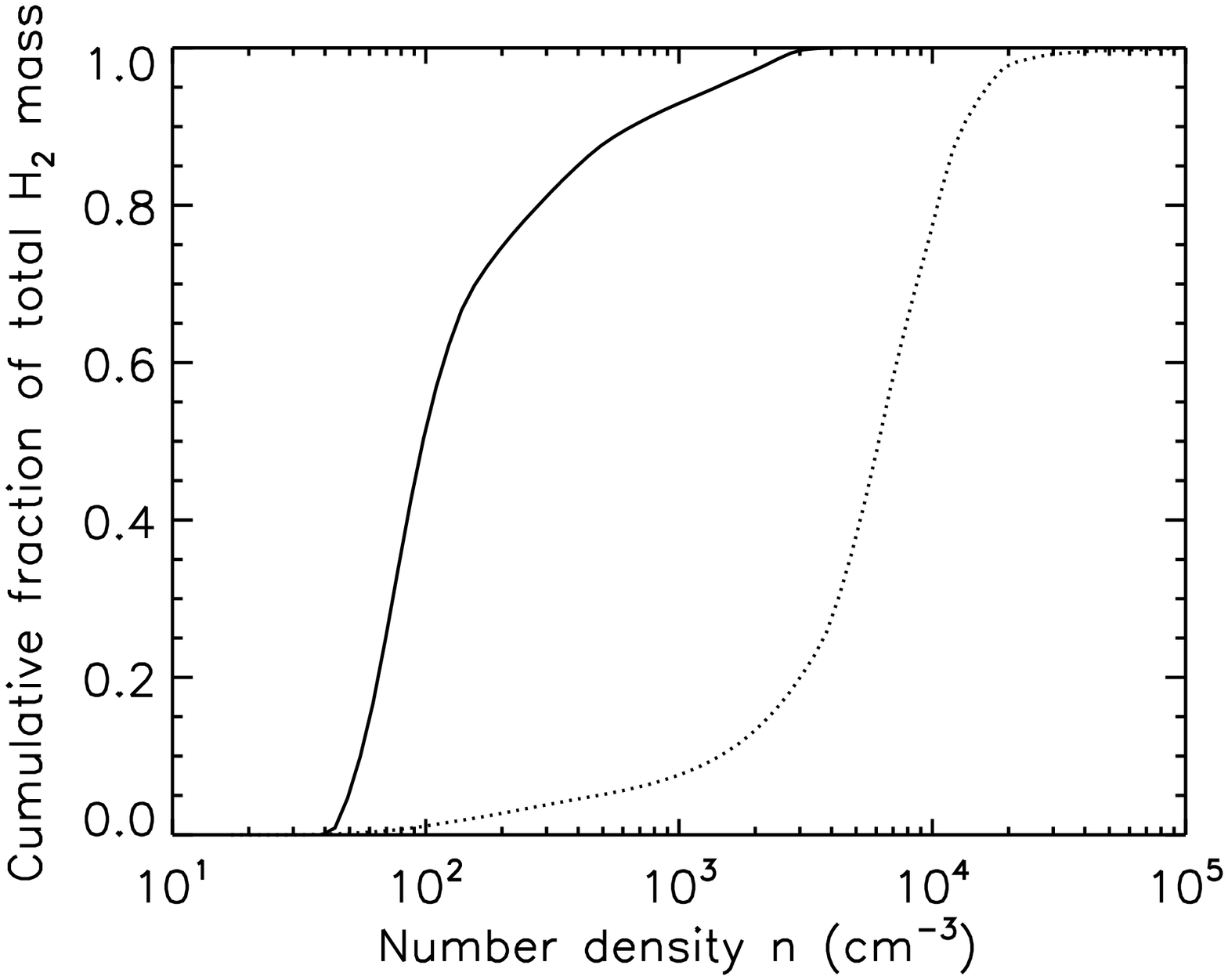,width=25pc,angle=0,clip=} 
\caption{(a) Cumulative mass distribution of $\mHt$ with $n$ in 
$256^{3}$ zone run MS256-RT at times $t = 17.4 \: {\rm Myr}$
(solid line) and $t = 20.6 \: {\rm Myr}$ (dotted line).
(b) As (a), but for run MS256 at output times 
$t = 28.5 \: {\rm Myr}$ (solid line) and $t = 31.7 \: {\rm Myr}$ (dotted line). 
\label{cuml-H2}}
\end{figure}

Figure~\ref{den-PDF} demonstrates that in both runs, much of the gas
remains close to the initial density. In run MS256-RT there is also a
substantial amount of gas at $n > 1000 \: {\rm cm^{-3}}$, corresponding
to gas associated with the overdensities bounding the collapsing slab.
In run MS256, this feature is absent, but there is nevertheless a high 
density tail, extending up to $n \simeq 4000 \: {\rm cm^{-3}}$. In run
MS256-RT, about 30\% of the total $\mHt$ mass is associated with 
the relatively unperturbed gas within the slab, with the other 70\% 
being associated with the overdense, collapsing gas (see 
Figure~\ref{cuml-H2}). On the other hand, in run MS256, most of the 
$\mHt$ is located at densities close to the initial gas density; indeed, 
only half of the total amount of $\mHt$ is found at $n > 100 \: {\rm cm^{-3}}$.

However, as the gas continues to gravitationally collapse, we expect 
both the PDF and the cumulative mass distribution of $\mHt$ to alter 
greatly, as we know already that most of the mass and most of the 
$\mHt$ will ultimately be located in dense, unresolved gas. This is 
borne out by the results from the end of the simulations which are
also plotted in Figures~\ref{den-PDF} and \ref{cuml-H2} for runs MS256-RT 
and MS256.

\subsection{Gas temperature: evolution and distribution}
\label{static-temp}
\begin{figure}
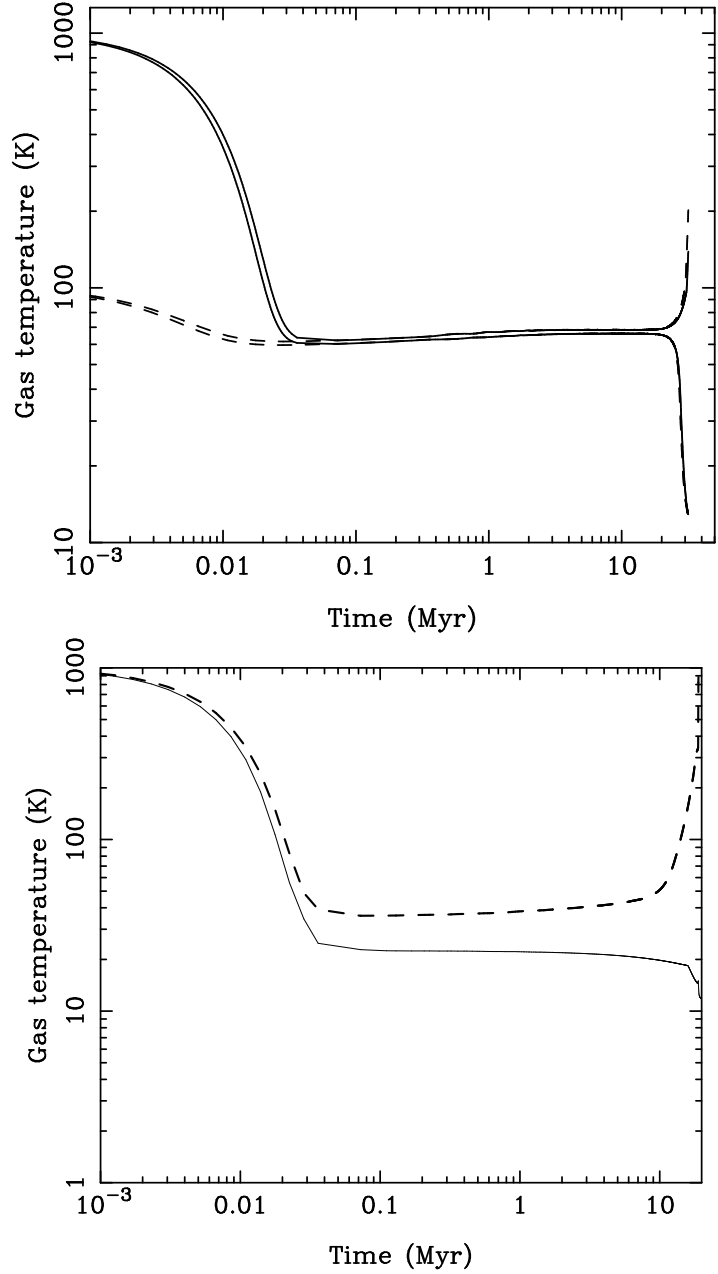

\centering
\epsfig{figure=fig19a.eps,width=20pc,angle=270,clip=}
\epsfig{figure=fig19b.eps,width=20pc,angle=270,clip=}
\caption{(a) Evolution with time of the minimum and maximum gas 
temperatures, $T_{\rm min}$ and $T_{\rm max}$, in $256^{3}$ zone runs 
MS256 (solid lines) and MS256-T100 (dashed lines), which were performed 
with initial temperatures of 1000~K and 100~K respectively.
(b) Evolution with time of $T_{\rm min}$ and $T_{\rm max}$ in $256^{3}$ zone 
run MS256-RT. \label{T-MS-fid}}
\end{figure}

As we have previously noted, the cooling time of the gas in our
simulations is much shorter than the dynamical timescale.
Therefore, the gas very quickly cools to the thermal equilibrium 
temperature (which for a gas density of $100 \: {\rm cm^{-3}}$ is 
approximately $65 \: {\rm K}$, if we use our local shielding 
approximation) regardless of the initial temperature of the cloud.
This is clearly illustrated in Figure~\ref{T-MS-fid}, where we plot the 
evolution with time of the minimum and maximum gas temperatures
$T_{\rm min}$ and $T_{\rm max}$ in run MS256 (which has an 
initial gas temperature $T_{\rm i} = 1000 \: {\rm K}$) and run 
MS256-T100 (which has $T_{\rm i} = 100 {\rm K}$ but is otherwise 
identical to run MS256). Following an initial period of cooling that lasts 
for approximately $0.05 \: {\rm Myr}$, the behaviour of the two runs 
becomes essentially identical. We see a similarly rapid initial phase
of cooling in runs performed using the six-ray shielding 
approximation.

In run MS256, the temperature distribution following the initial cooling
phase is almost uniform, and remains so until the runaway gravitational 
collapse of the gas begins at $t \sim 20 \: {\rm Myr}$. On the other hand,
in run MS256-RT we see that even at early times there is a temperature
difference of approximately 20~K between $T_{\rm min}$ and 
$T_{\rm max}$. This temperature differential is a result of the fact that 
gas at the center of the box is more shielded from photoelectric heating
than gas at the edge, and, as we have already discussed, this gives rise
to a pressure gradient that drives the subsequent gas flow.

\begin{figure}
\centering
\epsfig{figure=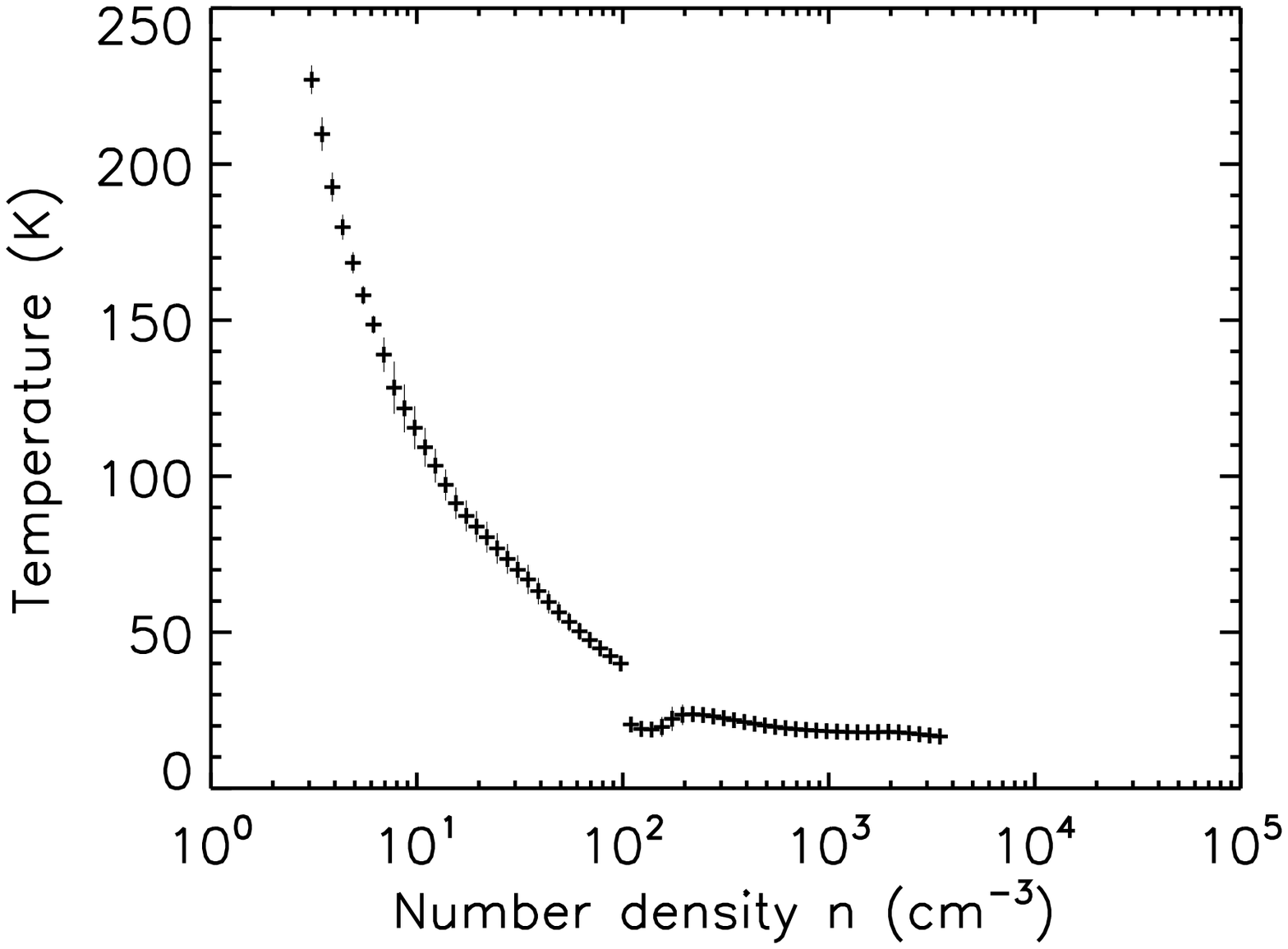,width=25pc,angle=0,clip=}
\epsfig{figure=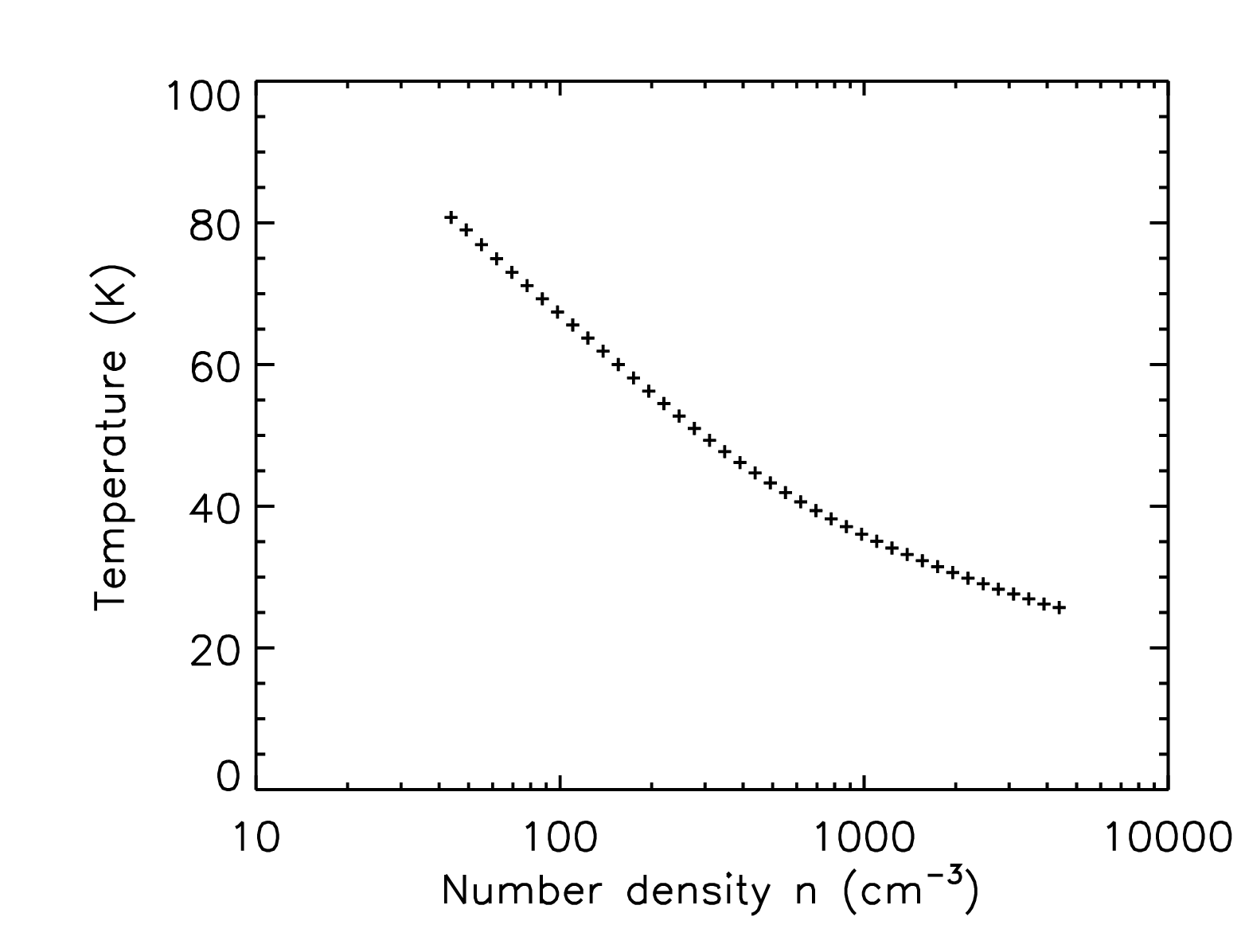,width=25pc,angle=0,clip=}
\caption{(a) Mean gas temperature $T$ plotted as a function of the number density
$n$ in $256^{3}$ zone run MS256-RT at time $t = 17.4 \: {\rm Myr}$. The data were 
binned as in Figure~\ref{xh2vn-RT} above. The standard deviation in each bin is
also indicated whenever it is larger than the size of the symbols used in the plot.
(b) As (a), but for run MS256 at time $t = 28.5 \: {\rm Myr}$. \label{Tvn-MS-fid}}
\end{figure}

In both runs, we find tight correlations between the gas density and the
temperature, as we demonstrate in Figure~\ref{Tvn-MS-fid}. In run 
MS256-RT, we see that there is a discontinuity at $n=100 \: {\rm cm^{-3}}$.
Gas at densities $n > 100 \: {\rm cm^{-3}}$ is located within the collapsing
slab, and so is shielded from the ultraviolet background by the gas in the
overdense regions that bound the slab. On the other hand, gas at 
$n < 100 {\rm cm^{-3}}$ is located either above or below the slab and so 
only receives the benefit of this shielding in one direction. Since the 
contribution from the lightly shielded direction dominates, the result is 
that the photoelectric heating rate is considerably higher in this gas 
than in the dense gas in the slab, and so consequently $T_{\rm eq}$ is
also higher. No comparable effect is seen in run MS256, as this is a
consequence of the non-local nature of the dust shielding and so is 
not captured by the local shielding approximation.

We have also examined whether one can usefully describe the behaviour 
of gas in these simulations using a polytropic equation of state. In
other words, if we write the gas pressure as a polynomial function of
density:
\begin{equation}
 p = K \rho^{\gamma_{\rm eff}},
\end{equation}
then what functional form must $\gamma_{\rm eff}$ have if this relation
is to accurately describe the gas? If $\gamma_{\rm eff}$ is a constant,
independent of the density, then the gas is a simple polytrope. Even if
$\gamma_{\rm eff}$ is not constant, however, this description can
be useful provided that $\gamma_{\rm eff}$ varies smoothly and simply
with $\rho$.

\begin{figure}
\centering
\epsfig{figure=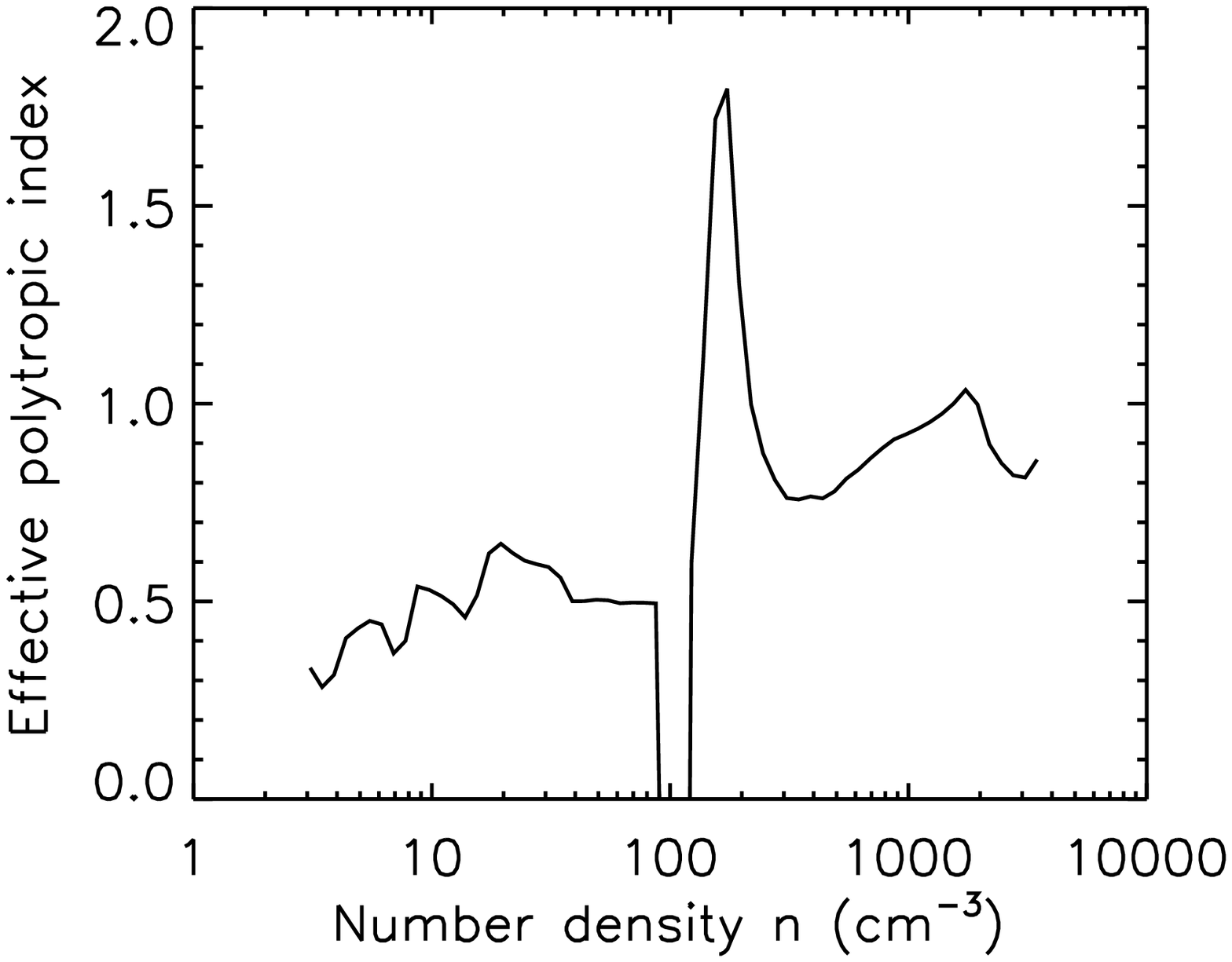,width=25pc,angle=0,clip=}
\epsfig{figure=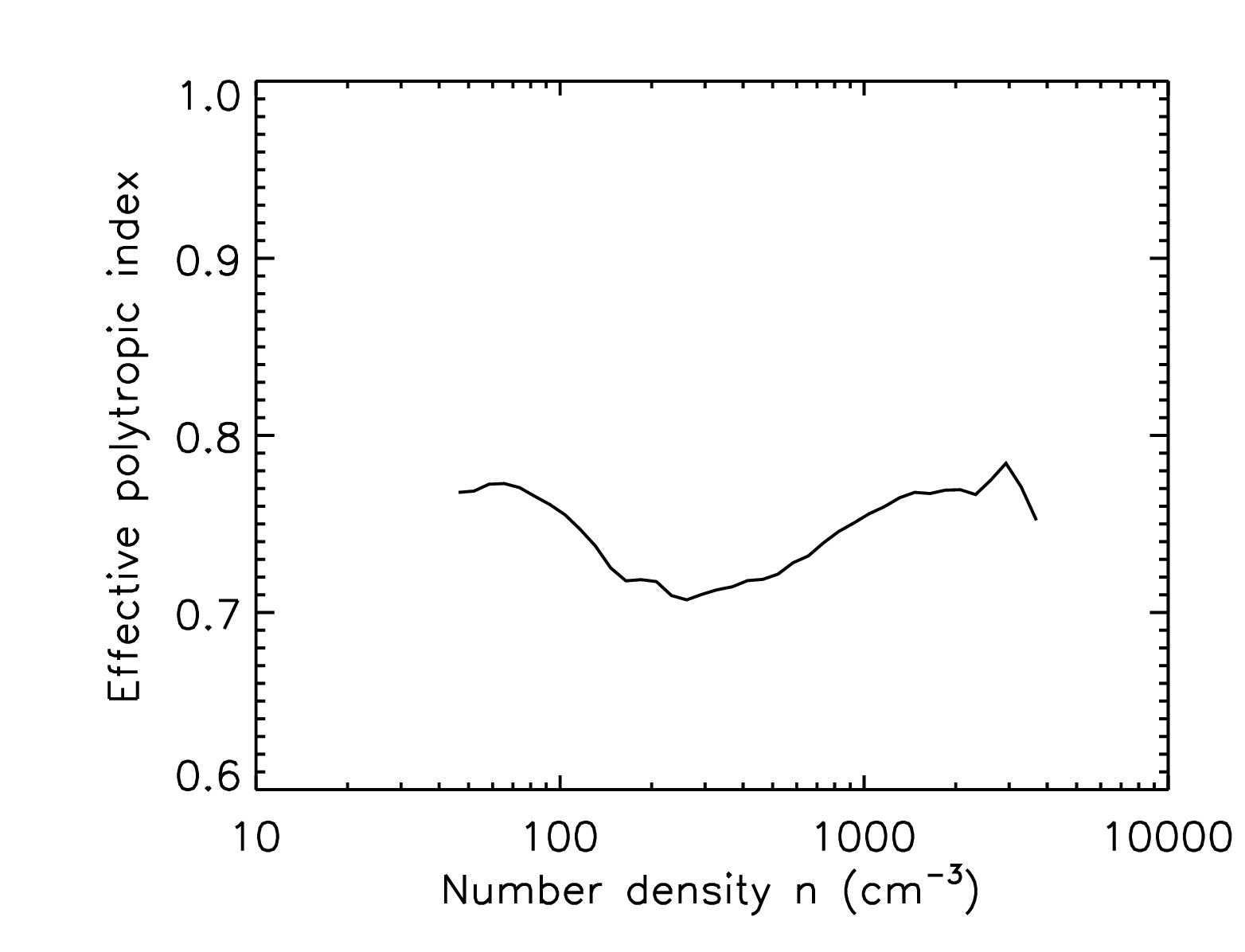,width=25pc,angle=0,clip=}
\caption{(a) Value of $\gamma_{\rm eff}$ as a function of $n$ in $256^{3}$ zone 
run MS256-RT at time $t = 17.4 \: {\rm Myr}$. The data were binned as indicated
in the text.
(b) As (a), but for run MS256 at time $t = 28.5 \: {\rm Myr}$. Note the
difference in the vertical scale compared to Figure 21a.
\label{gamma-MS}}
\end{figure}

To investigate how $\gamma_{\rm eff}$ varies as a function of density at
late times in each run, we computed the thermal pressure $p$ for each
grid zone and then binned the data with density just as we did for the
temperature above. We then used the resulting curve, together with the
fact that
\begin{equation}
 \gamma_{\rm eff} = \frac{{\rm d} \ln p}{{\rm d} \ln \rho}
\end{equation}
to compute $\gamma_{\rm eff}$ as a function of density. The values we
obtained for runs MS256-RT and MS256 are plotted in 
Figures~\ref{gamma-MS}a and \ref{gamma-MS}b respectively. 

Figure~\ref{gamma-MS}a demonstrates clearly that the gas in run 
MS256-RT is not well described by a polytropic equation of state. 
The prominent feature at $n \sim 100 \: {\rm cm^{-3}}$ is a 
consequence of the discontinuity in the temperature curve at 
this density and should perhaps be disregarded (since one might
still hope to use a polytropic description for the gas within and 
without the slab, even if the polytropic index for one differs from
that for the other). However, from the figure it is clear that even at 
densities $n \gg 100 \: {\rm cm^{-3}}$ and $n \ll 100 \: {\rm cm^{-3}}$
a polytropic equation of state is not really appropriate, as there is
considerable variation in $\gamma_{\rm eff}$ with $n$. Nevertheless,
it is clear that the equation of state of the gas is softer
than isothermal, considerably so at low densities.

In run MS256, the polytropic approximation fares much better.
Although the gas is not a polytrope, since $\gamma_{\rm eff}$ varies 
with density, the rate of change of $\gamma_{\rm eff}$ is typically small. 
For the range of gas densities covered by our simulations, 
$\gamma_{\rm eff} \sim 0.7$--$0.8$, and so for many applications, treating 
the gas as a simple polytrope with a constant polytropic exponent that is 
within this range of values may be a reasonable approximation.
We hypothesize that the polytropic description does much better in 
this case because when we use the local shielding approximation, 
all of the heating and cooling terms in the energy equation become
smooth functions of density. This is not the case in the six-ray runs,
as the non-local shielding disrupts the simple relationship between
density and photoelectric heating rate.

It is also interesting to examine how the values of $\gamma_{\rm eff}$ derived
here compare with previous suggestions in the literature. Our values are
significantly smaller than those derived by \citet{ss00} from their numerical
models of solar metallicity gas with this range of densities. However, the
values found in run MS256 are only slightly smaller than the value of 
$\gamma_{\rm eff} = 0.725$ derived by \citet{japp05} from a synthesis of 
a variety of observational and theoretical sources. In run MS256-RT, the
value of $\gamma_{\rm eff}$ at low densities is considerably smaller still,
but at higher densities there is better agreement. Nevertheless, 
it is important to bear in mind that our treatment of the gas chemistry 
remains highly approximate, and that this, plus the crude nature of our
shielding approximations,  will limit the accuracy with which we can 
model the gas temperature. It is therefore unclear how seriously we 
should take the differences between our results and those of 
\citet{ss00} or \citet{japp05}.  It would be interesting to revisit 
this point in the future with simulations that include far more of the 
relevant carbon and oxygen chemistry and a better treatment of the 
photoelectric heating.

\begin{figure}
\centering
\epsfig{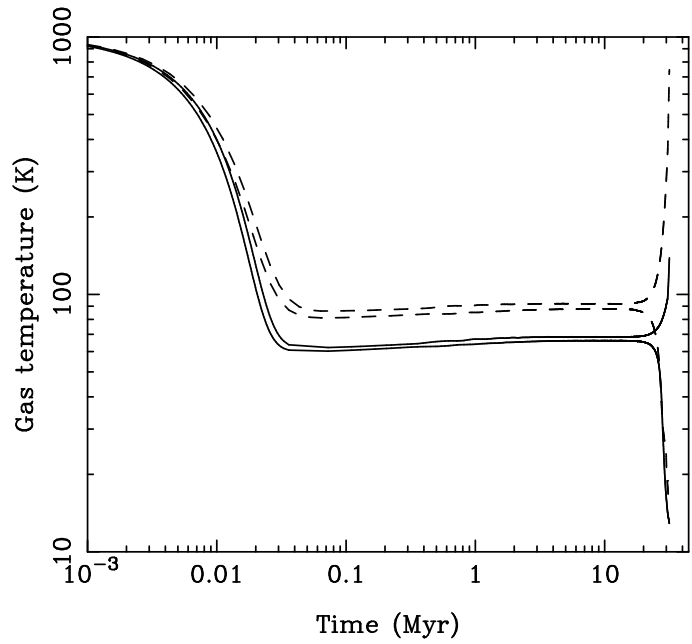}
\caption{Evolution with time of the minimum and maximum gas 
temperatures, $T_{\rm min}$ and $T_{\rm max}$, in $256^{3}$ zone runs 
MS256 (solid lines) and MS256-CR (dashed lines), which were
performed with cosmic ray ionization rates $\zeta = 10^{-17} \: {\rm s^{-1}}$
and $\zeta = 10^{-15} \: {\rm s^{-1}}$ respectively.}
\label{T-MS-CR}
\end{figure}

Finally, we have investigated how the current uncertainty in the value
of the cosmic ray ionization rate, discussed previously in \S~\ref{chem_sec},
affects the temperature evolution of the gas in our simulations. 
To do this, we performed a run, designated MS256-CR, with the same
input parameters as run MS256, but taking a large value of
$\zeta = 10^{-15} \: {\rm s^{-1}}$ for the cosmic ray ionization rate. 
This value lies at the high end of current determinations 
\citep[see e.g.][]{mac03} and so this run and run MS256 should bracket
the true behaviour. We compare the evolution of $T_{\rm min}$ and 
$T_{\rm max}$ in these two runs in Figure~\ref{T-MS-CR}. We see from 
the figure that prior to the onset of runaway gravitational collapse,
the effect of the higher cosmic ray ionization rate is to increase both
$T_{\rm min}$ and $T_{\rm max}$ by about 20~K. However, once collapse 
begins, the difference between the runs is quickly erased.

The higher temperatures resulting from the higher ionization rate 
have some influence on the $\mHt$ formation rate. At early times,
the net effect is small: the higher gas temperature leads to a 
slightly higher rate coefficient for $\mHt$ formation, and so to 
a higher $\mHt$ abundance, but the difference is at the 
level of about 10\%. At late times, once gravitational collapse
is well underway, the numbers reported in Table~\ref{xh2_at_end_stat}
suggest a rather more substantial difference between the runs. 
However, most of this difference is illusory, and is a consequence 
of the fact that the higher temperature of the gas in run MS256-CR 
is matched by a higher Jeans mass, allowing us to resolve the 
collapse for about 0.5~Myr longer. We therefore resolve more of 
the $\mHt$ formation that occurs during this run.

We also briefly investigated the effects of using a higher value of
$\zeta$ in combination with our six-ray approximation. However,
low resolution test runs suggested that the differences were no 
larger than the differences between runs MS256 and MS256-CR, so we
did not pursue this line of investigation further.

\subsection{Sensitivity to variations of the input parameters}
\label{sense}
Having explored in some detail the formation and distribution
of $\mHt$ in our standard runs, and the thermal behaviour of the 
gas, it is now time to turn our attention to examining what happens
when our input parameters are varied from their standard values.
Therefore, in \S~\ref{static-box}--\ref{static_n0} below, we 
examine the effects of varying the box size $L$, the initial perturbation
amplitude $\delta$, the initial magnetic field strength $B_{\rm i}$ 
and the initial density $n_{\rm i}$, while holding the other parameters
constant.

\subsubsection{Box size}
\label{static-box}
\begin{figure}
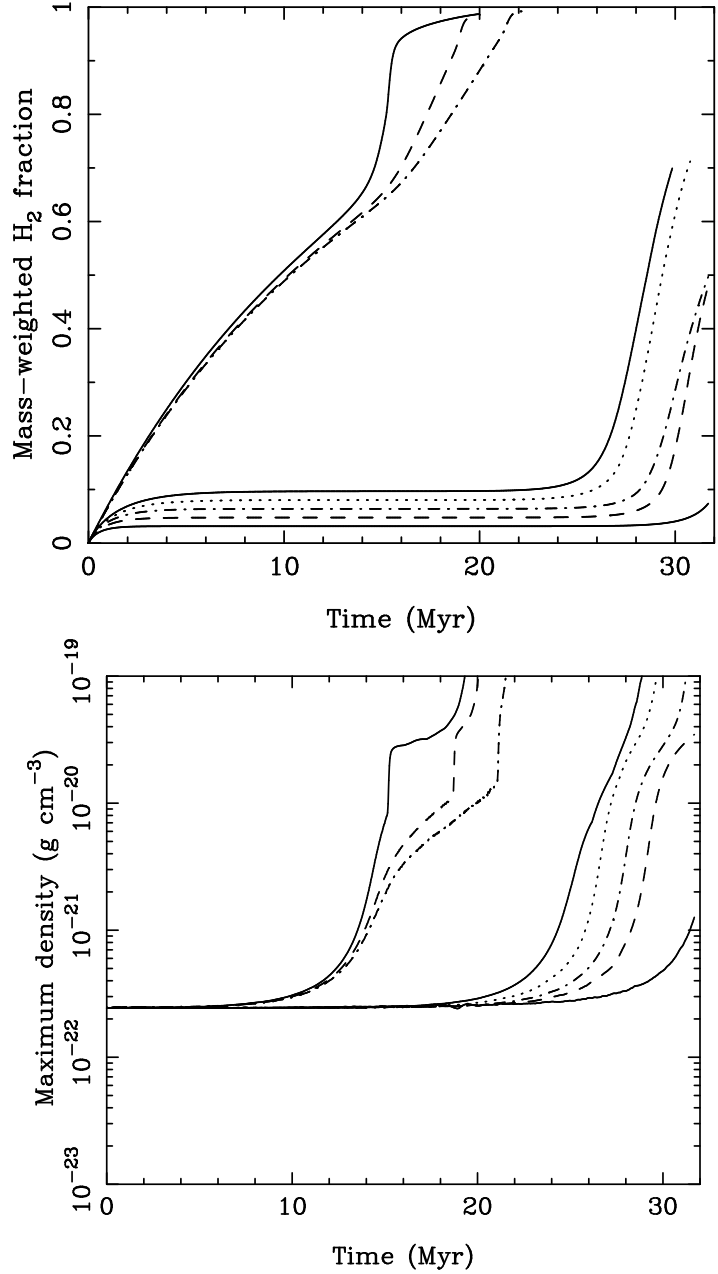

\centering
\epsfig{figure=fig23a.eps,width=20pc,angle=270,clip=}
\epsfig{figure=fig23b.eps,width=20pc,angle=270,clip=}
\caption{(a) Evolution of $\Htmass$ with time in a set of 
$256^{3}$ zone runs in which $L$ was varied from 20~pc to 60~pc. 
The three lines in the upper left represent runs MS256-RT-L20
(solid line), MS256-RT (dashed line) and MS256-RT-L60
(dot-dashed line). The five lines in the bottom right represent
runs MS256-L20 (lower solid line), MS256-L30 (dashed line),
MS256 (dot-dashed line), MS256-L50 (dotted line) and MS256-L60
(upper solid line).
(b) As (a), but for the maximum gas density, $\rho_{\rm max}$.
\label{MS-box}}
\end{figure}

By varying the size of the box while keeping the density of gas
within it fixed, we can alter the number of Jeans masses of gas which
the box contains and so make the gas either more stable against
gravitational collapse (if we decrease $L$) or less stable against
collapse (if we increase $L$). If the results of our fiducial simulations
are to be considered to be properly representative of the behaviour
of real gas, then we should ensure that they are insensitive to small
changes in $L$. We therefore performed several additional simulations
with our standard set of input parameters, but using different values
for $L$. Specifically, we performed four additional $256^{3}$ runs 
using the local shielding approximation, with $L = 20$, 30, 50 and
$60 \: {\rm pc}$ respectively, and two additional runs using the
six-ray approximation, with $L = 20$ and $60 \: {\rm pc}$
respectively. The results of these runs are plotted in 
Figure~\ref{MS-box}, along with the results from runs MS256 and
MS256-RT for comparison.

We see that in the runs using the local shielding approximation,
increasing $L$ at fixed numerical resolution increases the value
of $\Htmass$ at all times. This is easy to understand as in 
these runs, the amount of shielding is proportional to the grid
zone size $\Delta x$, and if we increase $L$ while keeping the
numerical resolution fixed, then we increase $\Delta x$. It is
also apparent that in these runs the rapid increase in $\Htmass$ 
that is caused by the runaway collapse of the gas occurs earlier 
in runs with a larger value of $L$. Figure~\ref{MS-box}b demonstrates 
that this is because runaway gravitational collapse occurs at 
progressively earlier times as $L$ is increased and at later 
times as $L$ is decreased. This is a straightforward consequence
of the presence of an increased number of Jeans masses in the 
larger simulations compared to the smaller ones. 
However, it should be noted that the changes to $L$ change
the collapse time of the gas by no more than 20\%. Therefore, 
while magnetic and thermal pressure are clearly playing some 
role in retarding the collapse of the gas in our fiducial simulations, 
the effect is not so large as to render the results of these runs 
unrepresentative of the general case, particularly given the level 
of approximation to which we are working.

In the runs using the six-ray approximation, we see
much less sensitivity to the box size, which again is not 
unexpected, given that the shielding in these runs is not directly 
dependent on $\Delta x$. The main effect that is apparent is that 
$\mHt$ forms slightly more efficiently at early times and 
considerably more efficiently at late times in runs with a smaller
$L$. This appears to be a consequence of the pressure-driven 
dynamics of the flow, which we have already discussed in some 
detail in previous sections. In runs with a smaller $L$ it takes less 
time for the overdense gas layers to propagate to the center of the
box, and so therefore it also takes less time for the gas to reach
a state of runaway gravitational collapse, as can be seen 
clearly in Figure~\ref{MS-box}b. The more rapid density evolution
in runs with smaller $L$ leads to a more rapid growth of $\mHt$. 
Nevertheless, the difference between the runs remains relatively
small, suggesting again that our fiducial simulation is adequately
representative of the general case.

\subsubsection{Initial perturbation amplitude}
\label{pert-init}
Since we use periodic boundary conditions in our simulations, a
perfectly uniform distribution of gas will be in dynamical
equilibrium (albeit an unstable equilibrium on scales larger than
the Jeans length) and so will not collapse. To provoke collapse,
it is necessary to perturb the distribution. In our simulations
of turbulent gas, described in paper II, large perturbations are rapidly 
created by the motion of the gas.  Similarly, in the runs described 
here which used the six-ray shielding approximation, the 
pressure gradient caused by the non-uniform photoelectric heating 
drives a flow that creates density perturbations large enough to
trigger gravitational collapse. However, in the runs we performed 
that used the local shielding approximation, it was necessary to add
some form of initial perturbation by hand. Our technique for doing so 
has already been described, but it is important to understand to what 
extent our results depend on our choice of $\delta$, the maximum 
amplitude of the initial random perturbations. Specifically, we would 
like to know whether the long delay before runaway gravitational 
collapse begins in earnest in our simulations is a consequence of 
our choice of a small value for $\delta$. Therefore, we have run 
several simulations in which we have varied $\delta$ while keeping 
the other parameters fixed at their fiducial values.

\begin{figure}
\centering
\epsfig{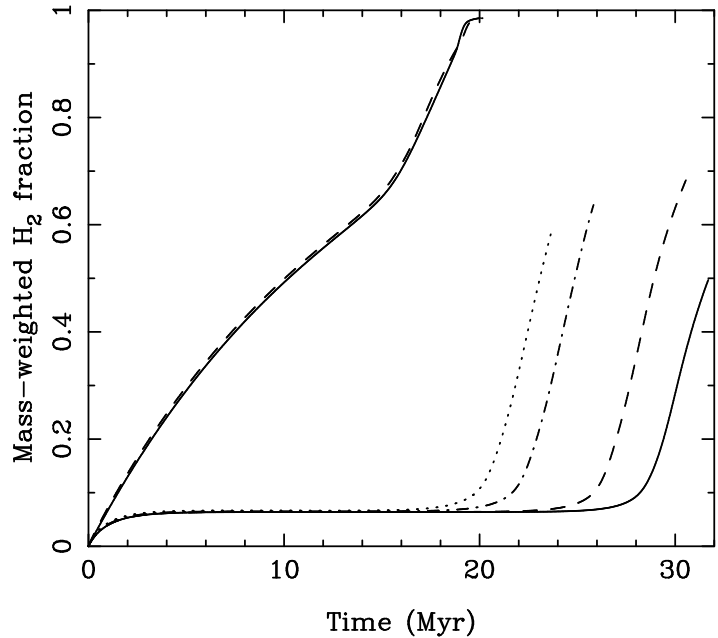}
\caption{Evolution of $\Htmass$ with time in a set of $256^{3}$
zone runs in which the maximum amplitude of the initial 
density perturbations, $\delta$, was varied. Results are plotted for 
runs MS256-RT and MS256-RT-d100 (solid and dashed lines in 
the upper left) as well as for runs MS256 (solid line), 
MS256-d10 (dashed line), MS256-d50 (dot-dashed line)
and MS256-d100 (dotted line). The first pair of runs have
$\delta = 0.05$ and 1.0 respectively; the latter four have
$\delta = 0.05$, 0.1, 0.5 and 1.0 respectively.
\label{H2-MS-d}}
\end{figure}

The growth of $\Htmass$ with time in $256^{3}$ zone runs with 
$\delta = 0.1$, 0.5 and 1.00, as well as in run MS256 is shown in
Figure~\ref{H2-MS-d}. Clearly, increasing $\delta$ does decrease the 
time that elapses before runaway collapse occurs, as can also be seen 
from the results summarized in Table~\ref{xh2_at_end_stat}. 
Nevertheless, the time required to form significant quantities of 
$\mHt$ remains long even in the $\delta = 1.00$ case, and while we 
could reduce the timescale even further by considering even larger 
inhomogeneities, at this point we would essentially be considering a 
collection of smaller, denser clouds, rather than a single cloud with a 
well-defined initial density. Therefore, although the timescale for collapse 
in these runs does depend on the initial density structure of our cloud, 
we can nevertheless be certain that it is of the order of 20 Myr or 
more for an approximately uniform cloud initially at rest.

We also investigated the sensitivity of runs that use our six-ray
approximation to variations in $\delta$, since these runs are also
seeded with small perturbations to the initially uniform density.
As expected, these runs show essentially no sensitivity to 
$\delta$: the aforementioned pressure-driven effects dominate.

\subsubsection{Initial magnetic field strength}
\label{field-init}
\begin{figure}
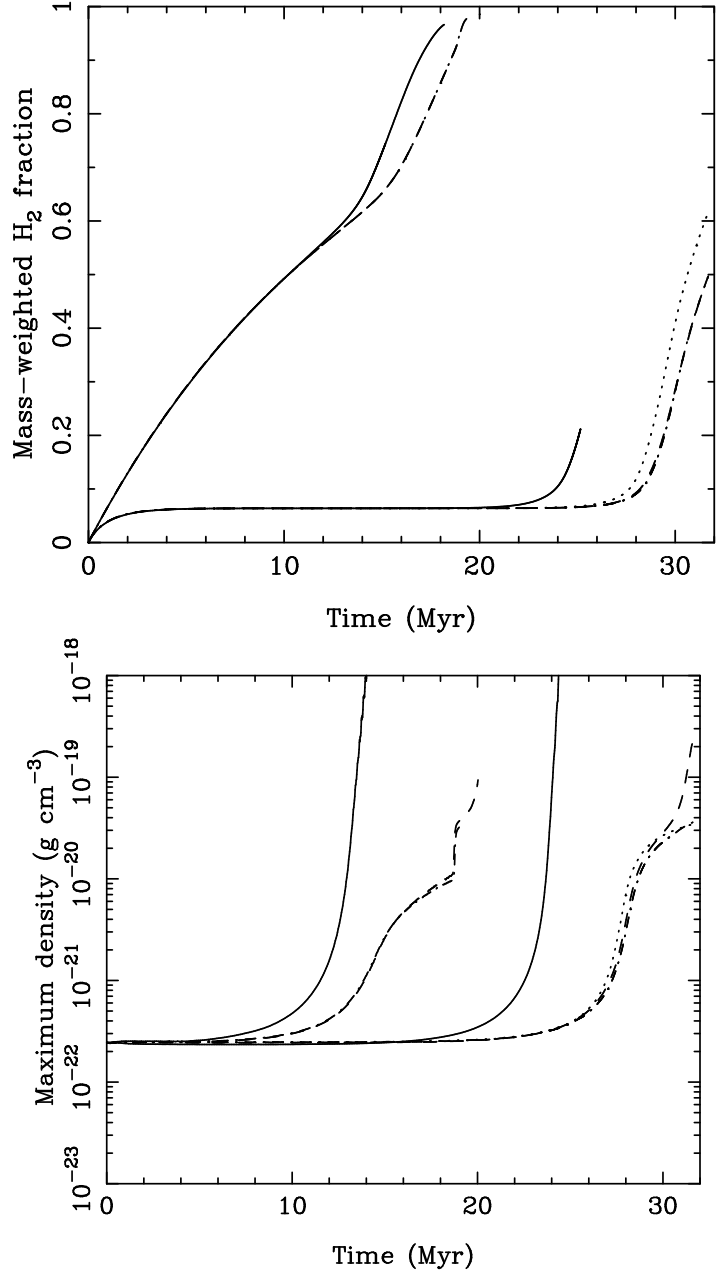

\centering
\epsfig{figure=fig25a.eps,width=20pc,angle=270,clip=}
\epsfig{figure=fig25b.eps,width=20pc,angle=270,clip=}
\caption{(a) Evolution of $\Htmass$ with time in a set of $256^{3}$ zone
runs in which the strength of the initial magnetic field was varied.
Results are plotted for seven runs. Three of these runs used the 
six-ray shielding approximation: runs HS256-RT (upper solid line), 
MS256-RT (upper dashed line) and MS256-RT-Bx4 (upper dot-dashed line), 
which had initial magnetic field strengths $B_{\rm i} = 0.0$, 5.85 and 
$23.4 \: \mu{\rm G}$ respectively. The other four runs used the 
local shielding approximation: runs HS256 (lower solid line), 
MS256 (lower dashed line), MS256-Bx2 (lower dot-dashed line) and 
MS256-Bx4 (lower dotted line), which had initial magnetic field 
strengths $B_{\rm i} = 0.0$, 5.85, 11.7 and $23.4 \: \mu{\rm G}$ 
respectively. In the magnetized runs, there is so little sensitivity
to $B_{\rm i}$ that the lines are hard to distinguish from each other 
in the plot. On the other hand, in the $B_{\rm i} = 0$ runs we see
a significant difference in behaviour.
(b) As (a), but for the maximum gas density $\rho_{\rm max}$.
\label{H2-MS-mag-comp}}
\end{figure}

In order to explore how our results depend on the strength of the initial
magnetic field, we performed several sets of runs in which we varied the
strength of the field but kept all of the other input parameters the same
as in our fiducial runs. Specifically, we performed two additional runs
using the six-ray shielding approximation: one with an initial field 
strength $B_{\rm i, fid} = 23.4 \: \mu{\rm G}$, corresponding to a 
mass-to-flux ratio of two, in units of the critical value, and one with a
field strength of zero, i.e.\ a purely hydrodynamical run. The evolution
with time of $\Htmass$ in these runs is plotted in 
Figure~\ref{H2-MS-mag-comp}a, along
with the corresponding values from run MS256-RT. We see that at
$t > 14 \: {\rm Myr}$,  the $\mHt$ fraction rises more rapidly in our 
hydrodynamical run than in our MHD runs, and also that the behaviour
of the latter runs is essentially indistinguishable. Figure~\ref{H2-MS-mag-comp}b, 
which shows the evolution with time of $\rho_{\rm max}$ in these three
runs, demonstrates that difference in behaviour between the run 
with $B = 0$ and the runs with $B \neq 0$ is caused by the fact that
gravitational collapse occurs more rapidly in the former run than in
the latter runs. This is a simple consequence of the fact that in the
absence of the magnetic field, gas can flow freely in every direction,
allowing the cloud to collapse along all three of its axes at once,
while when a magnetic field is present, the gas flow is channeled
primarily along the field lines, leading to collapse along only one 
axis. This also explains why our results are not particularly sensitive 
to the strength of the field, when one is present, as this will have no
effect on the velocity of the flow parallel to the field lines. 

We also examined the sensitivity of the results from the local shielding 
approximation simulations to the value of $B_{\rm i}$, using values
of 0.0, $11.7 \: \mu{\rm G}$ and $23.4 \: \mu{\rm G}$, and found very 
similar results: collapse occurs faster and $\mHt$ forms more
quickly in the complete absence of a magnetic field, but if a field is 
present then the timescales for either are not very sensitive to its 
strength. 

\subsubsection{Initial density}
\label{static_n0}
A final interesting topic to examine is the sensitivity of our results to the initial
density of the gas. For instance, if we reduce the initial number density $n_{\rm i}$
from its fiducial value of $100 \: {\rm cm^{-3}}$, which is on the high side for the 
CNM, to $30 \: {\rm cm^{-3}}$ or $10 \: {\rm cm^{-3}}$, what effect will this have 
on the timescale for $\mHt$ formation? Obviously, if we reduced $n_{\rm i}$ but
kept $L$ constant, then we would reduce the amount of gas in our simulation 
volume and thereby alter how well the gas could resist gravitational collapse, 
just as we did when we reduced $L$ while keeping $n_{\rm i}$  constant. Therefore,
to examine the effect of changing the density without significantly affecting the
gravitational stability of the gas, we must increase $L$ at the same time that
we decrease $n_{\rm i}$.  Similarly, if we decrease $n_{\rm i}$ and increase $L$,
we must also alter $B_{\rm i}$ if we wish to keep the mass-to-flux ratio constant,
and hence ensure that the field cannot completely prevent the gas from collapsing.

\begin{figure}
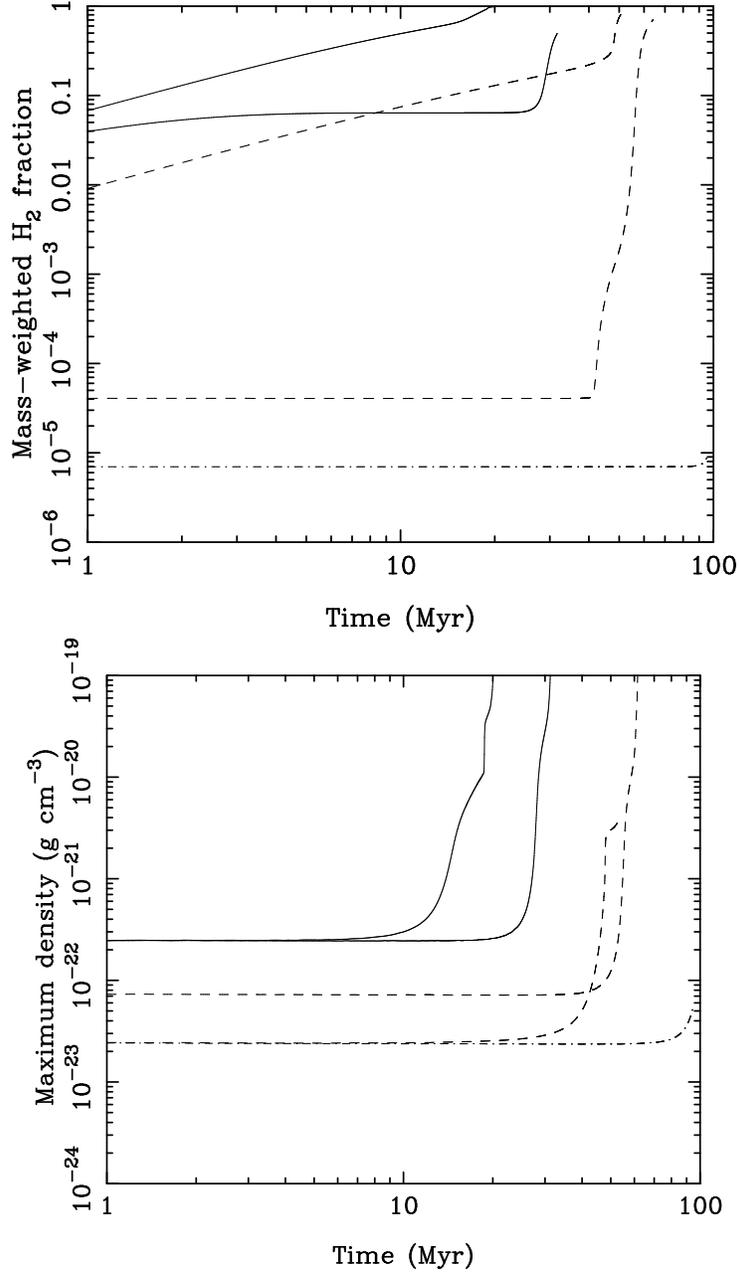

\centering
\epsfig{figure=fig26a.eps,width=20pc,angle=270,clip=}
\epsfig{figure=fig26b.eps,width=20pc,angle=270,clip=}
\caption{(a) Evolution of $\Htmass$ with time in a set of $256^{3}$ 
runs in which the initial gas density was varied. Results are plotted for runs
MS256-RT (lefthand solid line), MS256-RT-n10 (lefthand dashed line), MS256 
(righthand solid line), MS256-n30 (righthand dashed line) and MS256-n10 
(dot-dashed line). Note that $L$ and $B_{\rm i}$ were varied in these runs so as
to keep the number of Jeans masses in the simulation volume and the mass-to-flux
ratio approximately constant.
(b) As (a), but for $\rho_{\rm max}$.
\label{MS-n0-comp}}
\end{figure}

We therefore performed three simulations with lower initial densities than in our
fiducial simulations. In run MS256-RT-n10, we set $n_{\rm i} = 10 \: {\rm cm^{-3}}$
and used our six-ray shielding approximation. In runs MS256-n10 and MS256-n30,
we set $n_{\rm i} = 10$ and $30 \: {\rm cm^{-3}}$ respectively, and used our local
shielding approximation. In all three cases, the box size and magnetic field
strength were adjusted so as to keep the number of Jeans masses in the simulation
volume and the mass-to-flux ratio approximately the same as in our fiducial runs
(see Table~\ref{static_runs} for the values used). The evolution of $\Htmass$
in these simulations, as well as in runs MS256 and MS256-RT for comparison,
is plotted in Figure~\ref{MS-n0-comp}a. In Figure~\ref{MS-n0-comp}b, we show a 
similar plot for $\rho_{\rm max}$

Comparison of the evolution of $\Htmass$ in the various runs 
(see Figure~\ref{MS-n0-comp}a) demonstrates that the reduction in
$n_{\rm i}$ has a huge effect on the evolution of $\Htmass$ at early
times in runs performed using the local shielding approximation.
This is a consequence of the dramatic reduction in $\xhteq$ in these
runs, which decreases from $\sim 0.1$ in run MS256 to 
$\sim 4 \times 10^{-5}$ in run MS256-n30 and to $\sim 7 \times 10^{-6}$
in run MS256-n10. Substantial quantities of $\mHt$ are produced in 
run MS256-n30 only once the gravitational collapse of the gas is well
underway and the density in the collapsing gas has become high enough
to allow for effective self-shielding on small scales. Since
the reduction in $n_{\rm i}$  from $100 \: {\rm cm^{-3}}$ to 
$30 \: {\rm cm^{-3}}$ more than doubles the time required for the 
gas to collapse, this means that we do not see significant $\mHt$
formation before $t \sim 40 \: {\rm Myr}$. In run MS256-n10, substantial 
quantities of $\mHt$ are never produced, even after 100~Myr
(although we suspect that if we were to run the simulation for
considerably longer, we would see significant $\mHt$ formation, as 
the gas in this run has only just begun to collapse at the point 
when the simulation is ended). 

In run MS256-RT-n10, the effect of the reduced density is less
pronounced, but is significant nonetheless. The reduction of 
$n_{\rm i}$ by an order of magnitude leads to a reduction in 
$\Htmass$ by roughly a factor of seven at early times. The
reduction in $n_{\rm i}$ again leads to an increase in the time
required for gravitational collapse to occur, and the gas 
in run MT256-RT-n10 does not become dominated by $\mHt$ until 
collapse occurs at $t \sim 50 \: {\rm Myr}$.  

In practice, of course, we do not expect the gravitational collapse of gas clouds 
with mean densities of 10 or $30 \: {\rm cm^{-3}}$ to take quite as long as these 
simulations suggest, as the use of periodic boundary conditions to treat real
clouds with these densities, and with the sizes considered here, is unlikely to be
a good approximation --  there is simply too much inhomogeneity in the ISM on
these scales\footnote{For instance, the scale height of molecular gas in the 
disk of the Milky Way at the solar Galactrocentric radius is estimated to be only
65~pc \citep{sss84}.}. Nevertheless, even if we were to disable the
periodic boundary conditions and to treat the gas as an isolated cloud in free-fall
collapse (i.e.\ if we could ignore its thermal and magnetic energy content), then
we would still expect gravitational collapse, together with the associated $\mHt$
formation, to occur on a timescale of the order of 10~Myr or more. We show in
paper II that the results of turbulent simulations with the same mean
density are dramatically different.

\section{Summary}
\label{summary}
In this paper, we have discussed in detail how we have modified the ZEUS-MP
hydrodynamical code to allow us to simulate the formation and destruction of 
$\mHt$ in the ISM. We have used two different approaches for modeling the
shielding of $\mHt$ against photodissociation. The first of these is a local 
shielding approximation, novel to this work, which is computationally efficient
and which has the advantage of being biased in a known way: it will always 
underestimate the amount of shielding, and so we will overestimate the 
photodissociation rate of $\mHt$. The second approach that we have used is
our so-called six-ray approximation, in which we compute $\mHt$ and dust
column densities only along the three coordinate axes of the simulation 
volume. This approach is similar to one used previously by \citet{nl97,nl99}
and by \citet{yahs03}, but we believe that this is the first time that it has
been applied in grid-based simulations of this scale. This approximation 
does a much better job of capturing the shielding due to dust, but will tend
to overestimate the amount of $\mHt$ self-shielding once the flow velocities
become supersonic. For the simulations presented in this paper, it proves to 
be the better choice, but for the simulations of turbulent gas presented in 
paper II, the local shielding approximation proves to be more useful.

Our modifications to ZEUS-MP also include a detailed treatment of the thermal
behaviour of the gas. By computing the effects of the most important heating 
and cooling processes active in the ISM, we are able to follow the thermal 
evolution of the gas with far more accuracy than if we relied upon a description
in terms of a simple equation of state, whether isothermal or polytropic. In dense
gas, our accuracy is limited by our inability to fully resolve the cooling length of 
the gas, with the result that we will tend to overproduce warm gas. However, the
impact of this on the simulations presented in this paper appears to be small.

In \S~\ref{tests}, we discussed how the modified code was tested, and 
presented the results of a test designed to reproduce the \citet{w03} analysis
of the thermal equilibrium state of the neutral atomic ISM. Although we do not
reproduce their results exactly -- most likely due to the fact that the chemical
network used in our simulations is very much simpler than that adopted in their
analysis -- we do find qualitatively similar behaviour. Moreover, a quantitative
comparison of the two sets of results suggests that at most densities, our value
for the equilibrium gas temperature is accurate to within 50\% or better; only for 
gas at densities $n \sim 1 \: {\rm cm^{-3}}$, which lies in the middle of the thermally 
unstable regime, do we find larger errors, and even here we are accurate to 
within a factor of two.

Although our main goal in making these modifications was to study $\mHt$ formation
in turbulent gas (with results that we report on in paper II), we also applied our
modified hydrodynamical code to the problem of $\mHt$ formation in gravitationally 
collapsing gas without turbulence, starting from initial conditions in which the 
gas was smoothly distributed and initially at rest. We showed that with these 
initial conditions and an assumed mean density $n_{\rm i} = 100 \: {\rm cm^{-3}}$, 
gravitational collapse occurs on a timescale $t_{\rm coll} > t_{\rm ff}$. The precise
value of $t_{\rm coll}$ depends on our choice of shielding approximation (as this
significantly affects the pressure balance of the gas), and is also sensitive to 
the size of the simulation volume, the strength of the magnetic field, and in some
circumstances to the size of the initial density inhomogeneities. Our use of 
periodic boundary conditions probably artificially stabilizes the gas to some 
extent, but even if these were not used, we would still expect collapse to take 
{\em at least} one free-fall time (which for our standard initial density $n_{\rm i}$ 
is approximately $5 \: {\rm Myr}$), and in fact probably significantly longer, since the
thermal pressure of the gas is never negligible. 

In the simulations that we ran that used  the local shielding approximation, we 
found that $\mHt$ formation occurred in two phases. During the first
phase, which had a timescale of 5--$10 \: {\rm Myr}$, $\Htmass$ grew slowly 
until it reached a small limiting value, set by the fact that the gas had reached 
photodissociation equilibrium. Owing to our approximate treatment of $\mHt$ 
shielding, this limiting value was resolution dependent (and was smaller in higher
resolution simulations), but even in our lowest resolution, $64^{3}$ zone 
simulations, it corresponded to no more than about 25\% of the gas being molecular. 
The second, rapid phase of $\mHt$ formation was triggered by the runaway gravitational
collapse of the gas, as the increased density boosted the $\mHt$ formation rate 
while also increasing the amount of shielding against UV photodissociation.
By the end of the simulations, between 50 and 100\% of the gas had become molecular,
with most of the $\mHt$ having being formed during this collapse phase. 
Therefore, in these simulations, gravitational collapse drives $\mHt$ formation.

In the simulations that we ran that used the six-ray approximation, the 
situation was somewhat different. In these runs, the amount of shielding in
most of the box was considerably higher, and so the gas generally did not reach 
photodissociation equilibrium until very late times. Therefore, in these runs
$\mHt$ formation was reasonably efficient even in the absence of dynamical
effects. However, we have seen that in these runs the fact that the gas in
the center is shielded more than the gas at the edges leads to the development
of a significant temperature gradient and associated pressure gradient. This
drives a flow of gas towards the center of the box, and the presence of a 
dynamically significant magnetic field causes this flow to be oriented in a
direction parallel to the magnetic field lines. The resulting
pressure-driven compression
is ultimately responsible for triggering runaway gravitational collapse. Once 
this runaway collapse begins, we see a distinct acceleration in the rate of 
$\mHt$ formation, and so even in these simulations it is true to say that 
gravitational collapse plays an important role in driving $\mHt$ formation.
These results should be compared with the results from simulations of
supersonically turbulent gas that we present in paper II. They demonstrate 
that once a sufficiently high mean density is reached, whether through 
gravitational instability or other means, the turbulent compressions alone 
lead to substantial $\mHt$ formation, with gravitational collapse subsequently 
playing only a minor role.

We have also examined the spatial distribution of the $\mHt$ produced in our
simulations. We have shown that by the end of the simulation, most of the
$\mHt$ is to be found in dense gas, with at least 50\% in gas denser than
$5000 \: {\rm cm^{-3}}$. Moreover, since the initial distributions of the gas and
the magnetic field are nearly uniform, and since the subsequent dynamical 
evolution proceeds smoothly, the final spatial distribution is highly ordered.
Most of the $\mHt$, and, indeed, most of the gas, are found in a thin, dense
sheet oriented perpendicularly to the direction of the magnetic field. Visually, 
this distribution looks quite unlike that of the molecular gas in real molecular 
clouds, suggesting that more small-scale power is required in either the initial 
density distribution or velocity distribution (or both) in order to produce 
realistic-looking clouds.

Finally, we briefly examined the thermal behaviour of gas in our simulations, 
and showed that, as expected, most of the gas is in thermal equilibrium, owing
to the short cooling times at the simulated gas densities. However, the gas is
{\em not} isothermal, nor is it describable as a simple polytrope: while we can
describe the relationship between density and temperature with a function of
the form
\begin{equation}
 T \propto n^{\gamma-1},
\end{equation}
the polytropic exponent $\gamma$ is not constant, but varies with density as 
shown in Figure~\ref{gamma-MS}. In runs performed using the local shielding
approximation, $\gamma_{\rm eff} \sim 0.7$--$0.8$ over the range of densities
covered by our simulations, but in runs that used the six-ray approximation, 
we found a much wider range of $\gamma_{\rm eff}$. We conclude that while
there may be some applications for which treating the gas as a simple polytrope 
with a constant polytropic exponent is a reasonable approximation, this 
approach is probably not valid in the general case. 

Obviously, we do not claim that the results of the study presented in this paper 
are directly applicable to the real ISM: we have investigated a rather simplified
dynamical situation, which is missing one of the main physical ingredients 
present in the real ISM, namely supersonic turbulence. However, we believe 
that this study is interesting for the light it sheds on the rate at which molecular
clouds will form in the absence of turbulence: as we have seen, in this case 
substantial quantities of $\mHt$ are formed only on timescales $t > t_{\rm ff}$,
consistent with a cloud formation timescale of at least 5--$10 \: {\rm Myr}$.
As we shall see in paper II, the behaviour of supersonically  turbulent gas is 
very different.

\acknowledgments
The authors would like to thank the anonymous referee for suggesting the
use of the six-ray shielding approximation, and for a number of other 
comments that have helped us to improve the presentation of this paper.
They would also like to acknowledge valuable discussions on various aspects 
of this work with J.\ Black, R.\ Garrod, R.\ Klessen, C.\ Lintott, H.\ Liszt, J.\ Niemeyer, 
A.\ Rosen, W.\ Schmidt and M.\ Smith. Comments by F.\ Shu, R.\ Klein, and others 
at a meeting of the Berkeley-Santa Cruz-Ames Star Formation Center first inspired this
work. The simulations discussed in this paper were performed on the Parallel Computing 
Facility of the AMNH and on an Ultrasparc III cluster generously donated to the AMNH 
by Sun Microsystems.  We would  like to thank T.\ Grant, D.\ Harris, S.\ Singh, and, in 
particular, J.\ Ouellette for their invaluable technical assistance at various points during 
the simulation runs. Financial support for this work was provided by NASA grant 
NAG5-13028 and NSF grant AST-0307793.

\clearpage

\begin{deluxetable}{lc}
\tablecaption{The set of chemical reactions that make up our model of 
non-equilibrium hydrogen chemistry. \label{chem_model}}
\tablewidth{0pt}
\tablehead{
\colhead{Reaction}  & \colhead{Reference} }
\startdata
1. $\mH + \mH + {\rm grain} \rightarrow  \mHt + {\rm grain}$ & \citet{hm79} \\
 & \\
 2. $\mHt + \mH \rightarrow 3\mH$ & \citet{ms86} (low density), \\
 & \citet{ls83} (high density)  \\
 & \\
 3. $\mHt + \mHt \rightarrow 2\mH + \mHt$ & \citet{mkm98} (low density), \\
 & \citet{sk87} (high density) \\
 & \\
 4. $\mHt + \gamma \rightarrow 2\mH$ & {\rm See \S~\ref{h2_phdis}} \\
 & \\
 5. $\mH + {\rm c.r.} \rightarrow \Hp + \me$ & {\rm See \S~\ref{chem_sec}} \\
 & \\
 6. $\mH + \me \rightarrow \Hp + 2\me$ & \citet{a97} \\
 & \\
 7. $\Hp + \me \rightarrow \mH + \gamma$ & \citet{f92} \\
 & \\
 8. $\Hp + \me + {\rm grain} \rightarrow \mH + {\rm grain}$ & \citet{wd01} \\
& \\
 \enddata
 \end{deluxetable}

\begin{deluxetable}{ll}
\tablecaption{Processes included in our thermal model. \label{cool_model}}
\tablewidth{0pt}
\tablehead{
\colhead{Process}  & \colhead{References} }
\startdata
{\bf Cooling:} & \\
\cii fine structure lines &  Atomic data -- \citet{sv02} \\
&  Collisional rates ($\mHt$) -- \citet{fl77}  \\
&  Collisional rates ($\mH$, $T < 2000 \: {\rm K}$) -- \citet{hm89} \\
&  Collisional rates ($\mH$, $T > 2000 \: {\rm K}$) -- \citet{k86} \\
&  Collisional rates (${\rm e^{-}}$) -- \citet{wb02} \\
\oi fine structure lines &  Atomic data -- \citet{sv02} \\
& Collisional rates ($\mH$, $\mHt$) -- Flower, priv.\ comm.\ \\ 
& Collisional rates (${\rm e^{-}}$) -- \citet{bbt98} \\
& Collisional rates ($\Hp$) -- \citet{p90,p96} \\
\sii fine structure lines &  Atomic data -- \citet{sv02} \\
& Collisional rates ($\mH$) -- \citet{r90} \\
& Collisional rates (${\rm e^{-}}$) -- \citet{dk91} \\
$\mHt$ rovibrational lines & \citet{lpf99} \\
Gas-grain energy transfer$^1$ & \citet{hm89} \\
Recombination on grains & \citet{w03} \\
Atomic resonance lines & \citet{sd93} \\
$\mH$ collisional ionization& \citet{a97} \\
$\mHt$ collisional dissociation & See Table~\ref{chem_model} \\
& \\
{\bf Heating:} & \\
Photoelectric effect & \citet{bt94,w03} \\
$\mHt$ photodissociation & \citet{bd77} \\ 
UV pumping of $\mHt$ & \citet{bht90}  \\
$\mHt$ formation on dust grains & \citet{hm89} \\
Cosmic ray ionization & \citet{gl78}  \\
\enddata
\tablecomments{1: If $T_{\rm gas} < T_{\rm grain}$, the net flow of 
energy is from the grains to the gas, leading to heating instead of
cooling.}
\end{deluxetable}

\begin{deluxetable}{cccc}
\tablecaption{The number density at which the Truelove criterion is violated,
$n_{\rm max}$, computed for gas in thermal and chemical equilibrium for
various different box sizes and numerical resolutions. \label{resn_table}}
\tablewidth{0pt}
\tablehead{\colhead{Resolution}  & \colhead{Box size (pc)}
& \colhead{Zone size (pc)}  
& \colhead{$n_{\rm max} \: \left({\rm cm}^{-3} \right)$}}
\startdata
          & 20 & 0.31  & $2.0 \times 10^{3}$ \\
$64^{3}$  & 40 & 0.62  & $8.0 \times 10^{2}$ \\
          & 60 & 0.94  & $4.9 \times 10^{2}$ \\
\hline \hline
          & 20 & 0.16  & $5.4 \times 10^{3}$ \\
$128^{3}$ & 40 & 0.31  & $2.0 \times 10^{3}$ \\
          & 60 & 0.47  & $1.2 \times 10^{3}$ \\
\hline \hline
          & 20 & 0.078 & $1.5 \times 10^{4}$ \\
$256^{3}$ & 40 & 0.16  & $5.4 \times 10^{3}$ \\
          & 60 & 0.23  & $3.0 \times 10^{3}$ \\
\hline \hline
          & 20 & 0.039 & $5.6 \times 10^{4}$ \\
$512^{3}$ & 40 & 0.078 & $1.5 \times 10^{4}$ \\
          & 60 & 0.12  & $8.2 \times 10^{3}$ \\
\enddata
\end{deluxetable}

\begin{deluxetable}{ccccccc}
\tablecaption{Input parameters used for our runs
\label{static_runs}}
\tablewidth{0pt}
\tablehead{\colhead{Run}  & \colhead{$L$ (pc)}
& \colhead{$\delta$} 
& \colhead{$n_{\rm i}$ ($\rm{cm^{-3}}$)}  
& \colhead{$T_{\rm i}$ (K)} 
& \colhead{$B_{\rm i}$ ($\mu$G)} 
& \colhead{Notes}}
\startdata
MS64       & 40 & 0.05 &  100 & 1000 &  5.85   \\
MS128       & 40 & 0.05 &  100 & 1000 &  5.85   \\
MS256       & 40 & 0.05 &  100 & 1000 &  5.85   \\
MS64-RT   & 40 & 0.05 & 100 & 1000 & 5.85 \\
MS128-RT  & 40 & 0.05 & 100 & 1000 & 5.85 \\
MS256-RT  & 40 & 0.05 & 100 & 1000 & 5.85 \\
MS64-ng     & 40 & 0.05 &  100 & 1000 &  5.85 & 1 \\
MS128-ng    & 40 & 0.05 &  100 & 1000 &  5.85 & 1 \\
MS256-ng    & 40 & 0.05 &  100 & 1000 &  5.85 & 1 \\
MS256-RT-ng & 40 & 0.05 &  100 & 1000 &  5.85 & 1 \\
MS64-nr     & 40 & 0.05 &  100 & 1000 &  5.85 & 2 \\  
MS128-nr    & 40 & 0.05 &  100 & 1000 &  5.85 & 2 \\  
MS256-nr    & 40 & 0.05 &  100 & 1000 &  5.85 & 2 \\
MS256-T100  & 40 & 0.05 &  100 & 100  &  5.85 & \\
MS256-CR    & 40 & 0.05 &  100 & 1000 &  5.85 & 3 \\
MS256-L20   & 20 & 0.05 &  100 & 1000 &  5.85 & \\
MS256-RT-L20 & 20 & 0.05 &  100 & 1000 &  5.85 & \\
MS256-L30   & 30 & 0.05 &  100 & 1000 &  5.85 & \\
MS256-L50   & 50 & 0.05 &  100 & 1000 &  5.85 & \\
MS256-L60   & 60 & 0.05 &  100 & 1000 &  5.85 & \\
MS256-RT-L60 & 60 & 0.05 &  100 & 1000 &  5.85 & \\
MS256-Bx2   & 40 & 0.05 &  100 & 1000 &  11.7 & \\
MS256-Bx4   & 40 & 0.05 &  100 & 1000 &  23.4 & \\
MS256-RT-Bx4 & 40 & 0.05 &  100 & 1000 &  23.4 & \\
HS256       & 40 & 0.05 &  100 & 1000 &  0.0  & \\
HS256-RT    & 40 & 0.05 &  100 & 1000 &  0.0  & \\
MS256-d10   & 40 & 0.10 &  100 & 1000 &  5.85 & \\
MS256-d50   & 40 & 0.50 &  100 & 1000 &  5.85 & \\
MS256-d100  & 40 & 1.00 &  100 & 1000 &  5.85 & \\
MS256-RT-d100 & 40 & 1.00 &  100 & 1000 &  5.85 & \\
MS256-n10   & 85 & 0.05 &   10 & 1000 &  1.24 & \\
MS256-RT-n10 & 85 & 0.05 & 10 & 1000 & 1.24 & \\
MS256-n30   & 60 & 0.05 &   30 & 1000 &  2.63 & \\
\enddata
\tablecomments{1: runs with self-gravity disabled; 2: runs with $\chi = 0.0$;
3: run with a higher cosmic ray ionization rate, $\zeta = 10^{-15} \: {\rm s^{-1}}$.}
\end{deluxetable}

\begin{deluxetable}{ccccc}
\tablecaption{$t_{\rm res}$, $t_{\rm f}$, and associated values of $\Htmass$
for all runs in Table~\ref{static_runs}.
\label{xh2_at_end_stat}}
\tablewidth{0pt}
\tablehead{\colhead{Run}  & 
\colhead{$t_{\rm res}$ (Myr)} &
\colhead{$\Htmass (t_{\rm res})$} &
\colhead{$t_{\rm f}$ (Myr)} &
\colhead{$\Htmass (t_{\rm f})$}}
\startdata
MS64        & 21.3 & 0.30 & 27.3 & 0.89 \\
MS128       & 26.5 & 0.34 & 31.1 & 0.93 \\  
MS256       & 29.5 & 0.21 & 31.7 & 0.50 \\     
MS64-RT     & 15.0 & 0.65 & 22.1 & 0.99 \\ 
MS128-RT    & 16.9 & 0.74 & 20.9 & 0.99 \\
MS256-RT    & 18.4 & 0.89 & 20.6 & 0.99 \\
MS64-ng     & --- & --- & 31.7 & 0.26 \\  
MS128-ng    & --- & --- & 31.7 & 0.13 \\  
MS256-ng    & --- & --- & 31.7 & $6.4 \times 10^{-2}$ \\  
MS256-RT-ng & --- & --- & 19.9 & 0.73 \\
MS64-nr     & 18.3 & 0.70 & 29.6 & 0.99 \\  
MS128-nr    & 22.1 & 0.76 & 28.9 & 0.98 \\  
MS256-nr    & 26.6 & 0.83 & 31.7 & 0.96\\  
MS256-T100  & 28.9 & 0.23 & 31.7 & 0.62 \\ 
MS256-CR    & 30.0 & 0.42 & 31.7 & 0.76 \\ 
MS256-L20   & --- & --- & 31.7 & $7.4 \times 10^{-2}$ \\  
MS256-RT-L20 & 17.2 & 0.96 & 20.0 & 0.99 \\
MS256-L30   & 31.3 & 0.42 & 31.7 & 0.48 \\  
MS256-L50   & 27.7 & 0.20 & 30.8 & 0.71 \\  
MS256-L60   & 26.7 & 0.20 & 29.9 & 0.70 \\  
MS256-RT-L60 & 17.8 & 0.76 & 22.1 & 0.98 \\
MS256-Bx2   & 29.8 & 0.25 & 31.7 & 0.50 \\  
MS256-Bx4   & 29.2 & 0.27 & 31.7 & 0.62 \\  
MS256-RT-Bx4 & 18.7 & 0.92  & 19.0  & 0.96  \\
HS256       & 23.9 & 0.10 & 25.2 & 0.21 \\
HS256-RT    & 13.1 & 0.60 & 18.2 & 0.97 \\
MS256-d10   & 27.1 & 0.22 & 30.5 & 0.68 \\  
MS256-d50   & 22.9 & 0.20 & 25.8 & 0.64 \\  
MS256-d100  & 21.0 & 0.19 & 23.7 & 0.59 \\
MS256-RT-d100 & 17.1 & 0.80 & 20.2 & 0.99 \\
MS256-n10   & ---  & ---  & 95.1 & $8.7 \times 10^{-6}$ \\
MS256-RT-n10 & --- & ---  & 53.0 & 0.91 \\
MS256-n30   & 58.1 & 0.26 & 64.1 & 0.70 \\ 
\enddata
\tablecomments{$t_{\rm res}$ is the time at which the Truelove criterion
is first violated during the course of the run; when no value is given,
this indicates that the criterion was never violated. $t_{\rm f}$ is the
time at which the simulation was stopped.}
\end{deluxetable}

\end{document}